\documentclass[a4paper,fleqn,usenatbib]{mnras}

\usepackage{newtxtext,newtxmath}

\usepackage[T1]{fontenc}
\usepackage{ae,aecompl}
\usepackage{xcolor}
\usepackage{multirow}
\DeclareRobustCommand{\VAN}[3]{#2}
\let\VANthebibliography\thebibliography
\def\thebibliography{\DeclareRobustCommand{\VAN}[3]{##3}\VANthebibliography}

\usepackage{graphicx}	
\usepackage{amsmath}	
\usepackage{amssymb}	
\usepackage{subfig}

\usepackage{hyperref}
\newcommand{\angstrom}{\mbox{\normalfont\AA}}
\usepackage{etoolbox}
\makeatletter
\newcount\c@additionalboxlevel
\newcount\c@maxboxlevel
\AtBeginShipout{%
   \ifnum\value{additionalboxlevel}>\value{maxboxlevel}%
     \typeout{Warning: maxboxlevel might be too small, increase to %
       \the\value{additionalboxlevel}%
     }%
   \fi
   \@whilenum\value{additionalboxlevel}<\value{maxboxlevel}\do{%
     \typeout{* Additional boxing of page `\thepage'}%
     \setbox\AtBeginShipoutBox=\hbox{\copy\AtBeginShipoutBox}%
     \stepcounter{additionalboxlevel}%
   }%
   \setcounter{additionalboxlevel}{0}%
}
\makeatother

\title[Multiwavelength AGN properties]{Forward modeling the multiwavelength properties of Active Galactic Nuclei: application to X-ray and WISE mid-infrared samples} 

\author[A. Georgakakis]{
Antonis Georgakakis$^{1}$\thanks{E-mail: age@noa.gr},  Angel Ruiz$^{1}$, Stephanie M. LaMassa$^{2}$
\\
$^1$Institute for Astronomy \& Astrophysics, National Observatory of Athens, V.  Paulou  \& I.  Metaxa, 11532,  Greece\\ 
$^2$Space Telescope Science Institute, 3700 San Martin Drive, Baltimore, MD 21210, USA\\
}
\date{Accepted XXX. Received YYY; in original form ZZZ}

\pubyear{2020}

\begin{document}
\label{firstpage}
\pagerange{\pageref{firstpage}--\pageref{lastpage}}
\maketitle

\begin{abstract}
An empirical forward-modeling framework is developed to interpret the multi-wavelength properties of Active Galactic Nuclei (AGN) and provide insights into the overlap and incompleteness of samples selected at different parts of the electromagnetic spectrum. The core of the model are observationally derived probabilites on the occupation of galaxies by X-ray selected AGN. These are used to seed mock galaxies drawn from stellar-mass functions with accretion events and then associate them with spectral energy distributions that describe both the stellar and AGN emission components. This approach is used to study the complementarity between X-ray and WISE mid-infrared  AGN selection methods. We first show that the basic observational properties of the X-ray and WISE AGN (magnitude, redshift distributions) are adequately reproduced by the model. We then infer the level of contamination of the WISE selection and show that this is dominated by non-AGN at redshifts $z<0.5$. These are star-forming galaxies that scatter into the WISE AGN selection wedge because of photometric uncertainties affecting their colours. Our baseline model shows a sharp drop in the number density of heavily obscured AGN above the Compton thick limit in the WISE bands. The model also overpredicts by a factor of 1.5 the fraction of X-ray associations in the WISE AGN selection box compared to observations. This suggests a population of X-ray faint sources that is not reproduced by the model. This discrepancy is discussed in the context of either heavily obscured or intrinsically X-ray weak AGN. Evidence is found in favour of the latter.
\end{abstract}




\section{Introduction}

It is widely accepted that most, if not all, local spheroids host at their centres black holes with masses that scale with proxies of the stellar mass component of the parent bulges \citep[e.g.][]{Kormendy_Ho2013}. Understanding the formation history of these relic supermassive black holes and the role they play in the evolution of their host galaxies remain major challenges in current astrophysical research. The class of sources dubbed Active Galactic Nuclei \citep[AGN;][]{Padovani2017} are thought to represent the earlier growth phases of the quiescent black holes we observe today. Therefore, the study of this population can provide clues on both the formation of supermassive black holes and their interplay with galaxies. For example, measurements of the space density of AGN as a function of redshift \citep[e.g.][]{Ueda2003, Ueda2014, Aird2010, Aird2015, Buchner2015}  can be used to reconstruct the formation history of the supermassive black hole population \citep[e.g.][]{Marconi2004, Merloni_Heinz2008, Shankar2013}. Moreover, comparison of the host galaxy properties of AGN (e.g. star-formation, morphology, environment) with those of non-active galaxies can constrain the physical mechanisms that trigger accretion events onto supermassive black holes or provide clues on the impact of AGN outflows on the formation of new stars \citep[e.g.][]{Brandt_Alexander2015}. 

A prerequisite for the above type of investigations is a complete census of the AGN population in the Universe. This is however observationally  challenging. Obscuring clouds of dust and gas along the line-of-sight to the central engine \citep{Hickox_Alexander2018}, dilution of the AGN light by stellar emission \citep{Georgantopoulos_Georgakakis2005, Trump2009, Brandt_Alexander2015}, or variations of the radiative efficiency of the accretion flow among AGN \citep{Cattaneo_Best2009, Heckman_Best2014, Weigel2017} mean that there is no single waveband that can yield complete AGN samples. Instead, multiwavelength approaches that combine information from different parts of the electromagnetic spectrum are essential to address observational biases and compile (nearly) complete samples of active black holes \citep[][]{Padovani2017}. It is argued for example, that mid-infrared wavelengths are sensitive to heavily obscured AGN that are being missed from X-rays or the ultra-violet \citep[e.g.][]{Donley2012, Assef2015}. Moreover, it is suggested that the combination of radio and X-ray AGN surveys better samples the underlying Eddington ratio distribution of the population \citep{Best_Heckman2012, Weigel2017}. 

Although multiwavelength surveys of AGN have become the norm \citep{Brandt_Alexander2015}, synthesising the information from different wavebands into a coherent picture is not trivial. The level of overlap between AGN samples selected at different parts of the spectrum depends critically on the relative flux limits of the different surveys \citep{Mendez2013}, as well as the intrinsic scatter of the wavelength-dependent bolometric corrections. It is therefore often challenging to interpret multiwaveband AGN surveys and separate observational selection effects from intrinsic trends and relationships between physical parameters. 

In this paper we propose a new tool for the interpretation of the multiwavelength observations of AGN, which is based on the principles of forward modeling. They key feature of the method is that it provides a strong handle on observational effects such as photometric uncertainties, flux limits and flux-dependent sample incompleteness. The proposed tool therefore, enables the isolation of selection biases to gain insights into the underlying physical parameters that govern the overlap and incompleteness of AGN samples identified at different parts of the electromagnetic spectrum. Additionally, the forward modelling framework can provide quantitative estimates of how a specific AGN identification method applied to a set of observations sample the underlying population of active black holes. This is important in the case of complex AGN selection functions that are not known apriori and depend on different physical parameters, e.g. accretion properties, host galaxy characteristics, etc. An example is the identification of AGN using mid-infrared colours \cite[e.g.][]{Donley2012, Stern2012, Assef2013}, which essentially depends on the contrast between AGN and stellar light. The properties of AGN hosts (e.g. stellar mass, star-formation rate, dust content) are therefore important for understanding the type of accretion events (e.g. redshift accretion luminosity, Eddington ratio) a certain mid-infrared selection method is sensitive to or not. Moreover, the dominant population in the mid-infrared is star-forming galaxies and hence, contamination of AGN samples is also an important aspect of the selection function \citep[e.g.][]{Georgantopoulos2008, Assef2013}  that needs to be understood and accounted for. 

Motivated by these challenges and the increasing popularity of the mid-infrared for identifying active black holes, we choose to demonstrate the potential of the forward modeling tool by investigating the complementarity of X-ray and mid-infrared AGN samples. We show how the method can quantify in detail the mid-infrared selection function, including contamination, and how the comparison with observations provides critical tests on the assumptions used to construct the model. Throughout this work we adopt a cosmology with $H_0=\rm 70\, km/s$ and $\Omega_\Lambda=0.7$.

\section{Methodology}\label{sec:method}

The construction of a model that describes the multiwavelength properties of AGN is enabled by empirical relations that quantify the incidence of accretion events within the galaxy population. Such relations are used to seed the stellar mass function of galaxies with active nuclei and produce realistic mock catalogues of AGN and their hosts. Template Spectral Energy Distributions (SEDs) that represent the nuclear emission and the stellar component of galaxies are then implemented to add multiwavelength information to the mocks. The key parameter that enables the above methodology, and which can be constrained from observations, is the specific accretion rate, $\lambda$. This is defined to be proportional to the ratio of the AGN luminosity and the host-galaxy stellar mass, $\lambda \propto L_{AGN}/M_{*}$. By construction it measures the level of nuclear activity per unit stellar mass of galaxies. The determination of the specific accretion rates of large samples of galaxies provides an estimate of how likely a galaxy is to experience nuclear activity with amplitude $\lambda$. The specific accretion-rate distribution, $P(\lambda)$, of a population is therefore directly related to the incidence of accretion events within the sample. The $P(\lambda)$ is therefore the fundamental quantity that is used in this work to populate galaxies  with AGN and construct mock catalogues. It is emphasized that the definition of specific accretion rate used in this work is based entirely on observationally derived quantities, e.g. AGN luminosity measured at a given wavelength regime (X-rays in our analysis) and stellar mass. Bolometric corrections or black-hole scaling relations therefore do not enter the calculations. 

It has recently become possible to measure observationally the specific accretion-rate distribution using data from multiwaveband extragalactic survey fields \citep[e.g.][]{Aird2012, Aird2018, Aird2019, Bongiorno2012, Bongiorno2016, Georgakakis2017_plz}. These observational constraints however, are not optimal for the purposes of this work. This is because they do not include corrections for the incidence of obscured AGN in galaxies. It is well established that most supermassive black holes in the Universe grow their masses behind cocoons of gas and dust clouds that block direct view to the accretion process \citep[e.g.][]{Ueda2014,Buchner2015,Aird2015}. The selection of AGN at wavelengths that are sensitive to the effects of absorption by dust or gas (e.g. X-ray, UV, optical) introduces biases against obscured systems. Studies on the specific accretion-rate distribution of AGN that do not account for this effect underestimate the incidence of AGN among galaxies. Since the multi-wavelength demographics of AGN as a function of nuclear obscuration is one of the motivations of the present work, the construction of mocks needs to be based on specific accretion-rate distributions that account for the heavily obscured AGN component of the Universe. In this work we derive new estimates of the specific accretion-rate distributions that match our analysis requirements. Our approach is based on the fact that the specific accretion-rate distribution of AGN convolved with the stellar mass function of galaxies yields the AGN luminosity function \citep[e.g.][]{Georgakakis2017_plz, Grimmett2019}. Knowledge of the evolving stellar mass function and the obscuration-corrected AGN luminosity function can then yield the specific accretion-rate distribution. Details of the derivation are presented in the Appendix \ref{sec:plz} and only the most salient details are discussed in this section. We first choose to model the specific accretion-rate distribution by a three-segment broken power-law. This is motivated by non-parametric studies of the $P(\lambda)$  \citep{Georgakakis2017_plz, Aird2018}, which indicate significant deviations from a single power-law functional form. The most obvious manifestations of these deviations are turnover points at both high and very low specific accretion rates, where the local gradient changes substantially. For the stellar mass function of galaxies we adopt the parametrization of \cite{Ilbert2013}. Finally, the space density of AGN as a function of redshift and luminosity is represented by the obscuration-corrected binned estimates of \citet{Aird2015}. These measurements account for the moderately obscured (Compton thin) AGN population, i.e. those with equivalent hydrogen column density along the line-of-sight $N_H \la \rm 10^{24}\,cm^{-2}$. AGN above this limit (Compton thick) are assumed to represent 34\% of the Compton thin population \citep{Aird2015}, independent of luminosity and redshift \citep[e.g.][]{Buchner2015}. We choose such a simple recipe to describe the space density of Compton thick AGN in the Universe because current observations cannot constrain more complex models. 

The starting point of the analysis is the evolving mass function of galaxies. This observational quantity has been constrained to a satisfactory level of accuracy over a wide redshift baseline using data from recent large multi-wavelength programmes \citep[e.g.]{Ilbert2010, Muzzin2013, Ilbert2013}. Monte Carlo methods are applied to the observationally-derived parametric models of the stellar mass function to draw galaxies in the 2-dimensional space of stellar mass and redshift. It is customary in the galaxy community to parametrise the galaxy stellar-mass function at fixed redshift intervals, rather than using a global model with redshift-dependent terms. The Monte Carlo method interpolates the mass functions between neighbouring redshift intervals to provide  continuous sampling in stellar mass and  redshift  space. Each mock galaxy of stellar mass $M_{*}$ is assigned  a specific accretion rate, $\lambda \propto L_{AGN}/ M_{*}$, which is drawn in a probabilistic way from the distributions derived in Appendix \ref{sec:plz}. The corresponding AGN luminosities of mock galaxies are then estimated as $L_{AGN}= \lambda \times M_{*}$, where in this application $L_{AGN}$ is the 2-10\,keV X-ray luminosity, $L_X(\rm 2-10\,keV)$. This approach generates mock galaxy samples that follow the observed stellar mass function and have AGN painted on them, with a (obscuration-corrected) space density that is consistent with observations.
 
Each mock galaxy is assigned a redshift, a stellar mass and an AGN luminosity. These parameters are combined with empirical relations and statistical properties of AGN and galaxies (e.g. the star-formation main sequence \citealt{Schreiber2015}; UV to X-ray luminosity ratio of QSOs \citealt{Lusso2010}), to generate Spectral Energy Distributions for the AGN emission and the stellar light. Each of these spectral components are described in detail in the following sections. The total flux of mock sources in a given spectral band is the sum of the two components.  

\subsection{AGN Spectral Energy Distribution}

One of the properties of AGN that has a strong impact on their observed Spectral Energy Distributions is the level of obscuration along the line-of-sight. At X-ray wavelengths this is parametrised by the atomic hydrogen column densit y, $N_H$, which measures the amount of intervening gas (including metals) that absorbs or scatters X-ray photons. The $N_H$ distribution of AGN has been studied extensively as a function of redshift and accretion luminosity \citep[e.g.][]{Akylas2006, Ueda2014, Buchner2015, Aird2015} to constrain the fraction of absorbed systems in the Universe. In this work we adopt the parametrisation of \citet{Aird2015}. They model independently the luminosity function of unobscured ($\log N_H/\rm cm^{-2} = 20 - 22$) and moderately obscured ($\log N_H/\rm cm^{-2} = 22 - 24$; Compton thin) AGN assuming they are described by a double power-law with parameters that are allowed to evolve differently with redshift for each of the two populations. These luminosity functions are then partitioned to column densities assuming that the fraction of AGN is constant in the each of the logarithmic intervals  $\log N_H/\rm cm^{-2} = 20-21$, $21-22$, $22-23$, $23-24$. A population of Compton-thick AGNs with atomic hydrogen columns $\log N_H/\rm cm^{-2} = 24-26$ is also allowed. The luminosity function of the latter population is tied to that of moderately obscured AGN via a scaling factor, $\beta_{Thick}$, which represents the fraction of Compton thick AGN relative to the moderately obscured population. In our work we adopt the flexible double power-law model for the evolution of the AGN X-ray luminosity function using the best-fit parameters listed in Table 9 of \citet{Aird2015}. The fraction of Compton thick AGN in the model is set to $\beta_{Thick}=0.34$. 

\begin{figure}
\begin{center}
\includegraphics[height=0.9\columnwidth]{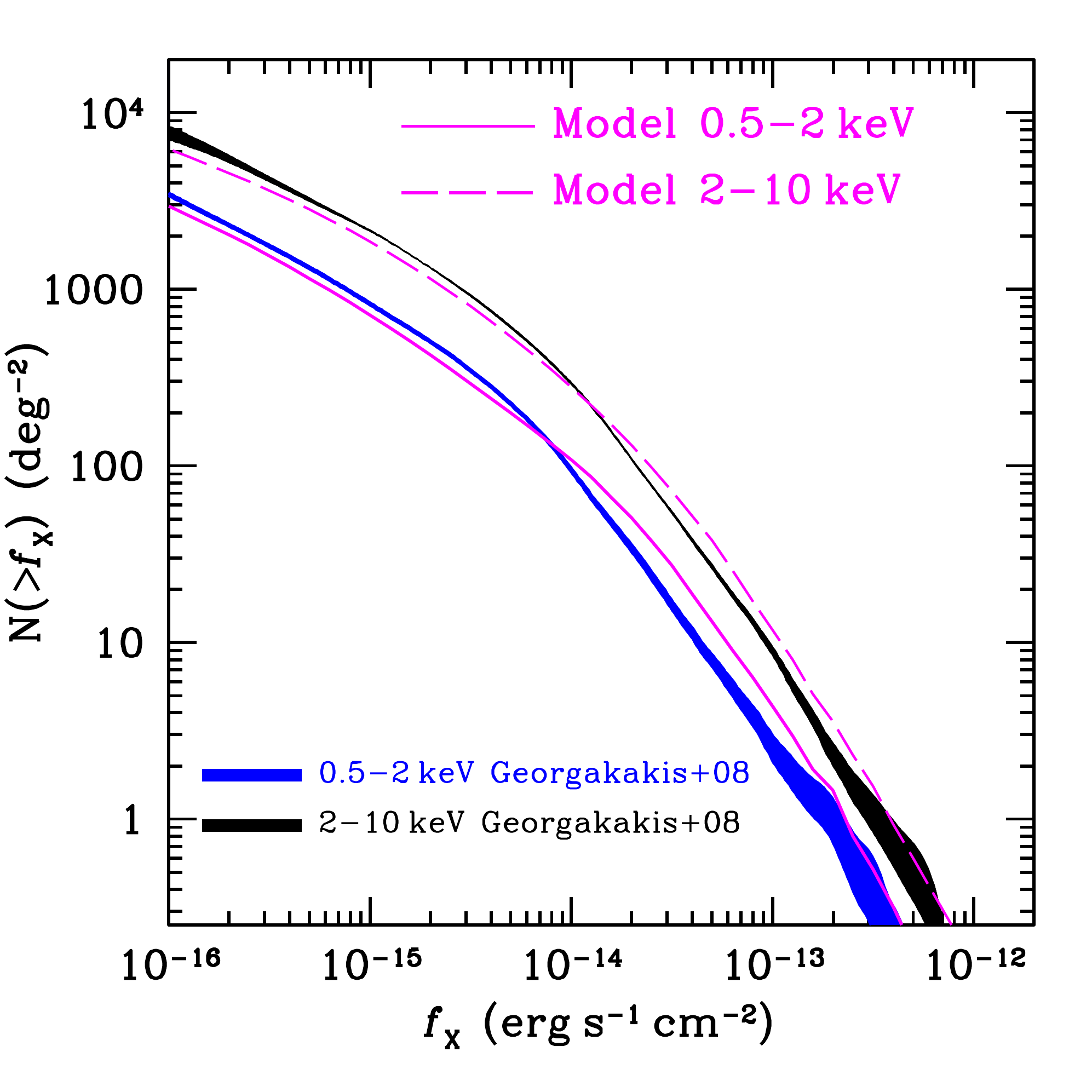}
\end{center}
\caption{Cumulative AGN number counts per square degree as a function of X-ray flux in the soft (0.5-2\,keV) and hard (2-10\,keV) bands. The observational results of  \protect\cite{Georgakakis2008_sense} are plotted as blue (0.5-2\,keV) and black (2-10\,keV) shaded regions. The width of these regions corresponds to the Poisson uncertainty in the number count distribution. The predictions of the model presented in this paper are shown with the magenta curves for the 0.5-2\,keV (solid) and 2-10\,keV (dashed) energy bands.}\label{fig:lgNlogS}
\end{figure}

The above assumptions on the AGN $N_H$ distribution are consistent with the adopted total (i.e. corrected for obscuration effects) X-ray luminosity function of \cite{Aird2015}, which, as described in the Appendix \ref{sec:plz}, is used to derive the corresponding specific accretion-rate distributions.  The important implication is that the mocks reproduce the observed sky density of AGN as a function of X-ray flux. We demonstrate this point by constructing the cumulative number-count distribution of X-ray sources. First, each mock AGN at redshift $z$ with luminosity $L_X$ is assigned in a probabilistic manner column densities based on the above assumptions for the $N_H$ distribution. The observed flux of each source is then estimated in the spectral bands 0.5--2 and 2--10\,keV. For the calculation of fluxes an X-ray spectral model should be adopted. We assume that the X-ray spectrum of AGN consists of an intrinsic power-law with spectral index $\Gamma$ that is transmitted through an obscuring medium with column density $N_H$ that photoelectrically absorbs or Compton-scatters the X-ray photons. We use the torus model of \citet{Brightman_Nandra2011} to describe this process and produce the resulting X-ray spectrum. This model assumes a sphere of constant density with two symmetric conical wedges with vertices at the centre of the sphere removed. The opening angle of the cones is fixed to $45\deg$ and the viewing angle of the observer is set to $87\deg$ , i.e. nearly edge on. The power-law index of the intrinsic power-law is set to $\Gamma=1.9$. In addition to the transmitted photons we also assume an additional soft component. This is often observed in the X-ray spectra of obscured AGN and may represent Thomson scattering of the intrinsic X-ray emission off ionised material within the torus opening angle \cite[e.g.][]{Guainazzi_Bianchi2007}.  This soft component is approximated  by a power-law model with spectral index $\Gamma=1.9$ and normalisation that is fixed to 5\% of the intrinsic power-law spectrum. The resulting $\log N - \log S$ in the 0.5-2 and 2-10\,keV spectral bands is compared with observations in Figure \ref{fig:lgNlogS}.

\begin{figure}
\begin{center}
\includegraphics[height=0.9\columnwidth]{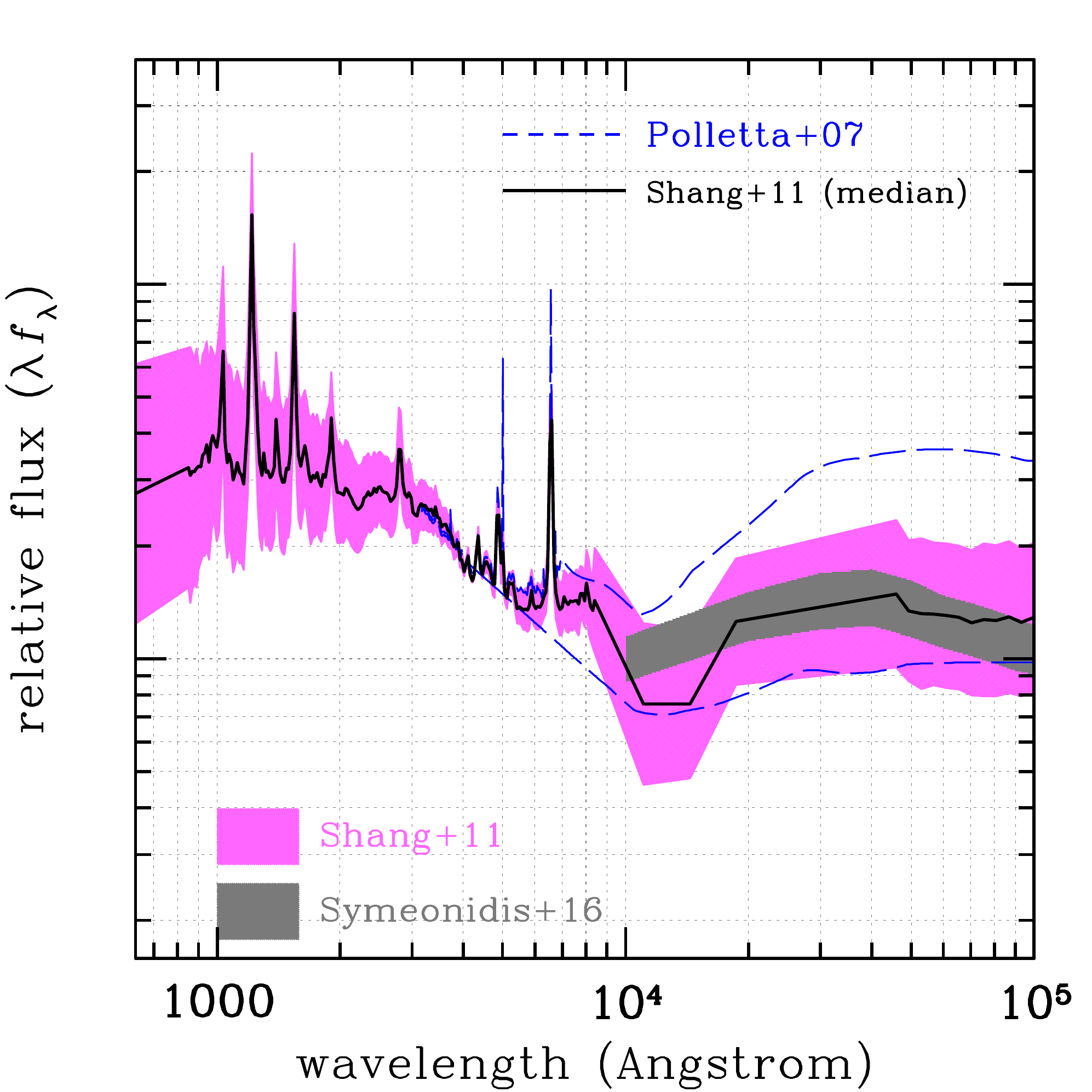}
\end{center}
\caption{Observationally derived AGN intrinsic SEDs. The black line shows the mean composite SED constructed by \protect\cite{Shang2011}, which is the one adopted in this work. The pink region shows the $1\sigma$ scatter around the mean. The grey shaded region shows the $1\sigma$ confidence interval of the mid-infrared SED estimated by  \protect\cite{Symeonidis2016}. Both these templates are normalised at $1\mu m$. The blue dashed curves provide a measure of the diversity of the infrared QSO SEDs using two templates from \protect\cite{Polletta2007} with different infrared-to-optical flux ratios. Both templates are normalised to be identical in the optical regime \protect\citep[SDSS quasar composite spectrum][]{VandenBerk2001}. The infrared data are from a sample of 35 SDSS/SWIRE quasars \protect\citep{Hatziminaoglou2005}. The upper curve  is obtained by averaging in each wavelength bin the 25\% brighter fluxes from these sources and corresponds to the TQSO1 template of \protect\cite{Polletta2007}. The lower curve is  obtained by averaging in each wavelength bin the  25\% fainter fluxes of the sample and corresponds to the BQSO1 template of \protect\cite{Polletta2007}.}\label{fig:agnsed}
\end{figure}

Next we describe the steps to generate multiwavelength AGN SEDs from the basic parameters of the mock catalogue, i.e. redshift, $z$, X-ray luminosity in the 2-10\,keV band, $L_X$, and atomic hydrogen column density, $N_H$. This requires assumptions on the overall shape of the intrinsic AGN SED from the UV to the infrared, its absolute normalisation relative to the accretion luminosity at X-ray wavelengths, the amount of dust along the line-of-sight and the wavelength dependence of the extinction law. 

In this work we adopt the composite intrinsic AGN SED constructed by \cite{Shang2011} using a sample of radio-quiet QSOs. Figure \ref{fig:agnsed} compares this composite with other observationally derived AGN SEDs in the literature. The normalisation of the SED, which is needed to estimate fluxes at different wavebands, is tied to the X-ray luminosity. The correlation between monochromatic X-ray [$L_\nu (\rm 2\,keV)$] and UV [$L_\nu(\rm 2500\angstrom)$] luminosities  \citep[e.g.][]{Steffen2006, Just2007, Lusso2010} is used to link the 2-10\,keV luminosity of mock AGN with the UV/optical regime. We adopt the \citep{Lusso2010} bisector best-fit relation $\log L_\nu ({\rm 2\,keV}) = 0.760\, \log L_\nu({\rm2500\angstrom})+3.508$. At a given monochromatic optical luminosity we assume that the data points scatter around the above relation following a Gaussian distribution with standard deviation $\sigma=0.4$ (logarithm base 10). The normalisation of the template SED at rest-frame $2500\angstrom$ is used to calculate the optical fluxes of mock AGN. At longer wavelengths however, this approach may introduce systematics because of variations in the intrinsic shape of the AGN SEDs. For the near- and mid-infrared fluxes we therefore normalise the template AGN SED at $6\,\mu m$ using the correlation between $L_X(\rm 2-10\,keV)$ and  $\nu L_\nu(6\,\mu m)$ luminosities of \cite{Stern2015} with an adopted scatter around the mean of 0.4\,dex. This relation deviates from linearity at high luminosities and is consistent with recent findings for intrinsically low $L_X/\nu L_\nu(6\,\mu m)$ ratios for luminous QSOs \citep{Martocchia2017}. Figure \ref{fig:agnsed} compares the adopted intrinsic AGN SED with constraints from studies that combine photometric/spectroscopic measurements for individual sources from the literature to estimate the average flux and the corresponding scatter as a function of wavelength.

In this work we choose to link the obscuration measured at X-rays with the extinction at longer wavelengths, UV, optical and near-inrfared. The challenge in this approach is that the X-ray absorber may be spatially distinct from the dusty medium that extincts the bulk of the UV/optical photons, because of e.g. dust sublimation.  Nevertheless, observations suggest a broad correlation between the hydrogen column density of the cold gas that absorbs the X-ray photons and the optical reddening $E(B-V)$ at the optical part of the SED inferred from e.g. broad emission line ratios \citep[e.g.][]{Burtscher2016}. We adopt a mean gas-to-dust ratio $N_H/E(B-V)=6\times10^{22}\rm cm^{-2} \, mag^{-1}$, which is similar to the median  value of \citet[e.g.][]{Maiolino2001} and significantly lower, by about 1\,dex, than the Galactic value. It is cautioned that large variations relative the above mean relation are observed for individual sources \citep[e.g.][]{Burtscher2016}. This may indicate differences in the physical conditions of the absorber, e.g. dust-grain sizes, variations in the physical scales of the obscuring medium, e.g. the presence of galactic-scale dust lanes along the line-of-sight, the impact of variability due to no concurrent observational measurements at different wavelengths. We attempt to partly mitigate these effects by introducing a $\log_{10}$ scatter of 0.5\,dex in the above dust-to-gas ratio relation when assigning mock AGN optical redennings, $E(B-V)$, based on their atomic hydrogen column densities. The $E(B-V)$ is then translated to extinction at different wavelengths by adopting the methodology described in \citet{Assef2010}. At wavelengths $\lambda<3300$\AA\, the extinction law is represented by a Small Magellanic Cloud curve based on the functional form of \citet{Gordon_Clayton1998} for the star AzV\,18. At longer wavelengths the Galactic extinction curve of \citet{Cardelli1989} is adopted. The ratio of the $V$-band extinction to the reddening, $R_V=A_V/E(B-V)$, is fixed to $R_V=3.1$. 

\begin{figure}
\begin{center}
\includegraphics[width=0.9\columnwidth]{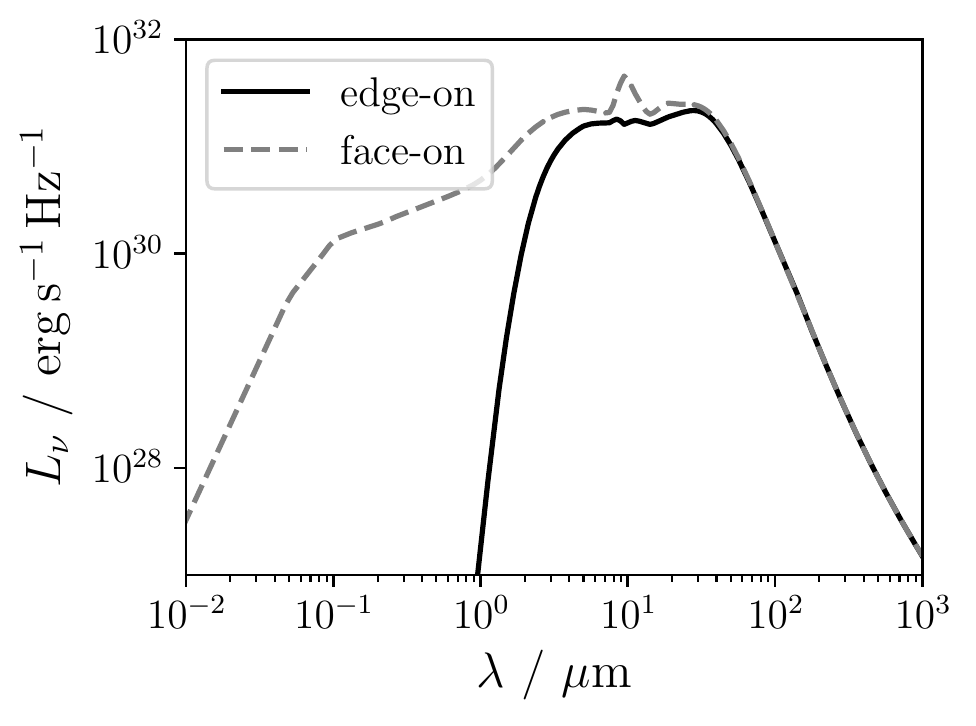}
\includegraphics[width=0.9\columnwidth]{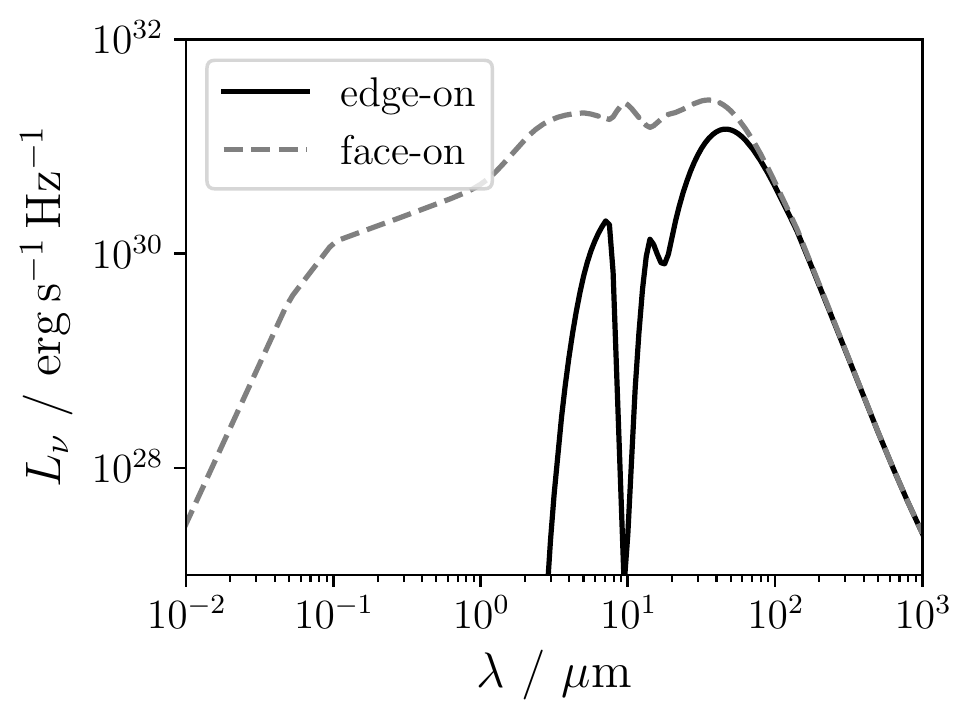}
\end{center}
\caption{The SEDs of AGN transmitted through a toroidal geometry of dust and gas clouds under different assumptions for the line-of-sight optical depth and the viewing angle. The results are based on the radiative transfer models of \protect\cite{Fritz2006}. The top panel shows the case of a torus with an (equatorial) optical depth $\tau(9.7\mu m)=1$, which corresponds to hydrogen column densities of $N_H \rm \approx9 \times 10^{22}cm^{-2}$. The dashed and solid curves are for face-on and edge-on viewing angles respectively.  The bottom panel is for a torus with optical depth $\tau(9.7\mu m)=10$ ( $N_H \rm \approx9 \times 10^{23}cm^{-2}$). }\label{fig:fritz}
\end{figure}

The extrapolation of the extinction law to the rest-frame mid-infrared warrants some discussion as it directly affects the detectability of heavily obscured AGN, i.e. those above the Compton thick limit, in the WISE bands. It is traditionally thought that the mid-infrared SED of AGN is dominated by thermal radiation from a hot dusty medium and hence may not be subjected to obscuration effects. This is supported by the observational fact that many Compton thick AGN in the local Universe closely follow the tight correlation between intrinsic X-ray (2-10\,keV) and mid-infrared ($12\rm \mu m$) luminosities of less obscured systems \citep{Gandhi2009}. At the same time however, radiative transfer calculations that describe how the intrinsic SED of AGN is modified when it is transmitted through an obscuring medium \citep[][]{Silva2004, Fritz2006} suggest that dense dust clouds along the line-of-sight can absorb part of the emitted radiation at the rest-frame mid-infrared. Figure \ref{fig:fritz} demonstrates this point using the \citet{Fritz2006} radiative-transfer models for the transmitted spectrum of AGN in the case of toroidal geometries of the dust clouds. We assume a torus opening angle of 60\,deg, a ratio between  external and internal radii of the torus of 30 and a constant dust density distribution in the torus \citep[][parameters $\beta=0$ and $\gamma=0$]{Fritz2006}. For this demonstration two different values of the (equatorial) optical depth of the above torus geometry are adopted, $\tau(9.7\mu m)=1$ and 10, which correspond to hydrogen column densities of $N_H \rm \approx9 \times 10^{22}cm^{-2}$ and $\rm 9 \times 10^{23} \, cm^{-2}$, respectively. Figure \ref{fig:fritz} shows the resulting SEDs in the case of a face-on \citet[][parameter $\psi=90$]{Fritz2006} and an edge-on ($\psi=0$) viewing angles. High levels of line-of-sight obscuration can significantly suppress the emerging radiation  in the mid-infrared. The discovery of heavily obscured (Compton thick) AGN that appear sub-dominant in the mid-infrared for their intrinsic (i.e. obscuration corrected) X-ray luminosity \citep{Krabbe2001, Gandhi2015} supports the above radiative-transfer model results. It therefore appears that for at least a subset of the Compton thick AGN population, absorption of the mid-infrared photons by an obscuring medium affects their observed SEDs. 

Based on the evidence above we choose to proceed with two distinct model incarnations that correspond to different assumptions on the form of the extinction law at rest-frame mid-infrared,  $\lambda>\rm 3\mu m$, in an attempt to capture the diversity of the observational results on the mid-infrared SEDs of Compton thick AGN. The baseline model assumes that the extinction in the rest-frame mid-infrared is zero, i.e. $A_\lambda/E(B-V) =0$ for $\lambda>\rm 3\mu m$. This assumption is favourable to Compton thick AGN. The limit $\lambda= \rm 3\mu m$ is chosen because it represents the approximate wavelength beyond which the thermal radiation dominates the SED of type-1 AGN. At shorter wavelengths the accretion disk component (which should be subjected to extinction) has a non-negligible contribution to the total emitted radiation and becomes dominant at about $\rm 1\mu m$ \citep{Hernan_Caballero2016, Hernan_Caballero2017}. Our AGN  SED model does not discriminate between disk and thermal components and therefore it is not possible in the current implementation to apply distinct extinction laws to each of them. Lowering below $3\mu m$ the wavelength limit where $A_\lambda/E(B-V)=0$  means that all the mock obscured AGN have a non-negligible contribution from unattenuated accretion-disk radiation in the rest-frame near-infrared. In addition to the evidence above, observations of nearby Seyferts further indicate that dust extinction suppresses even the thermal SED component of type-2 systems at rest-frame near-infrared wavelengths \citep[$\rm 2.5\mu m$][]{Burtscher2015}. This further supports the choice of  $\lambda= \rm 3\mu m$ as the limit  beyond which $A_\lambda/E(B-V) =0$. Nevertheless, the results and conclusions are insensitive to variations of this parameter by about 20\%, i.e. for wavelength in the approximate range $\rm 2.5-3.5\mu m$.
The second model version extrapolates the extinction law described in the previous paragraphs to the mid-infrared. Under this assumption the extinction in the rest-frame 3.6 and $\rm 4.5\mu m$ are $A_{3.6\rm \mu m}/E(B-V) =0.16$ and  $A_{4.5\rm \mu m}/E(B-V) =0.11$. It should be emphasised that the two models differ primarily in the predicted fraction of Compton thick AGN in the WISE bands. For less obscured sources, $N_H < \rm 10^{24} \, cm^{-2}$, the differences between the two models are minimal. For that purpose in the following sections we present results only for the baseline model (i.e. $A_\lambda/E(B-V) =0$ for $\lambda>\rm 3\mu m$) and only discuss the comparison of the two models in relation to Compton thick AGN only.

\subsection{Galaxy Spectral Energy Distribution}

The estimation of the Spectral Energy Distribution of a galaxy requires knowledge of its star-formation history, i.e. the time variations of its instantaneous star-formation rate. This record can be translated to a stellar population mix at a given time. The luminosity of individual classes of stars within the population can then be synthesised to determine the overall energy emitted by the galaxy as a function of wavelength. The star-formation history of galaxies however, can be complex and erratic \citep{Ciesla2015} depending on e.g. their position on the cosmic web, the frequency of galaxy encounters and interactions, the supply of gas from cosmological scales, feedback mechanisms and stochastic processes related to the secular evolution of galaxies. Because of the complexity of the physics involved it is hard to make predictions on the star-formation history of individual galaxies.  This is even more challenging in the case of the mock galaxies considered here, for which the only information available is their total stellar masses and redshifts. We therefore resort to assumptions and empirical relations to infer spectral energy distributions based on the information at hand. 

The first step is to assign instantaneous star-formation rates to mock galaxies by exploiting the empirically found correlation between star-formation rate and stellar mass, often referred to as the Main Sequence of Star Formation. Galaxies are split into star-forming and quiescent using the analytic relations of \cite{Brammer2011}, which estimate the fraction of passive galaxies as a function stellar mass and redshift. For stellar masses below $M= 10^{10} M_{\odot}$, which corresponds to the lower-limit of the parameter space studied by  \cite{Brammer2011}, we simply extrapolate their analytic relation. The  \citet{Schreiber2015} Main Sequence relation is then used to assign star-formation rates to the star-forming subset of the mock galaxy population. In this exercise it is assumed that the star-formation rate at fixed mass and redshift is normally distributed with scatter of 0.2\,dex.  The quiescent mock galaxies are simply assumed to have star-formation rates 1 to 2\,dex below the Main Sequence expectation. 

Next we assume an analytic form for the star-formation rate to infer the SEDs of galaxies using their assigned stellar masses and star-formation rates as boundary conditions. This approach is inverse of that usually followed by observers, whereby the broad-band SEDs of galaxies are fit with analytic star-formation history models to infer physical parameters, such as stellar masses and star-formation rates. We use an exponentially declining star-formation history law of the form

\begin{equation}\label{eq:sfr}
SFR(t) = A \, \exp^{-t/\tau},
\end{equation}

\noindent where $A$ is the normalization, $\tau$ is the characteristic time-scale of the exponential decay and $SFR(t)$ is the instantaneous star-formation rate at time $t$ after the formation of the galaxy at $t=0$. The stellar mass that corresponds to the above star-formation history model is

\begin{equation}\label{eq:mstar}
M_{\star}(t) = A \, \tau \, \big( 1 - \exp^{-t/\tau} \big).
\end{equation}

\noindent The goal is to infer for each mock galaxy the parameters $A$, $\tau$, $t$ using Equations \ref{eq:sfr}, \ref{eq:mstar} and the allocated SFR and $M_{\star}$ of the galaxy at the redshift $z$. Clearly this is an ill-posed problem since there are 3 unknown parameters to be determined from two equations. We simplify this exercise by fixing $\tau$. It is then possible to analytically determine $t$ and $A$ that reproduce the  assigned SFR and $M_{\star}$ of each mock galaxy. We also require that the assigned $t$ does not exceed the age of the Universe at the redshift of the mock galaxy, i.e. $t<t_{\rm Univ}(z)$. This constraint is important in the case of quiescent galaxies with very low star-formation rates. If the above relation is violated, then the $\tau$ is incrementally reduced and the estimation of $t$, $A$ is repeated until the age constraint above is fulfilled. Once the star-formation history parameters are constrained for each mock galaxy we pass Equation \ref{eq:sfr} to a stellar population synthesis code to generate the corresponding SEDs and estimate fluxes at different wavebands. For this latter step the {\sc cigale} code \citep{Ciesla2015, Boquien2018} is used. 

\begin{figure}
\begin{center}
\includegraphics[height=0.7\columnwidth]{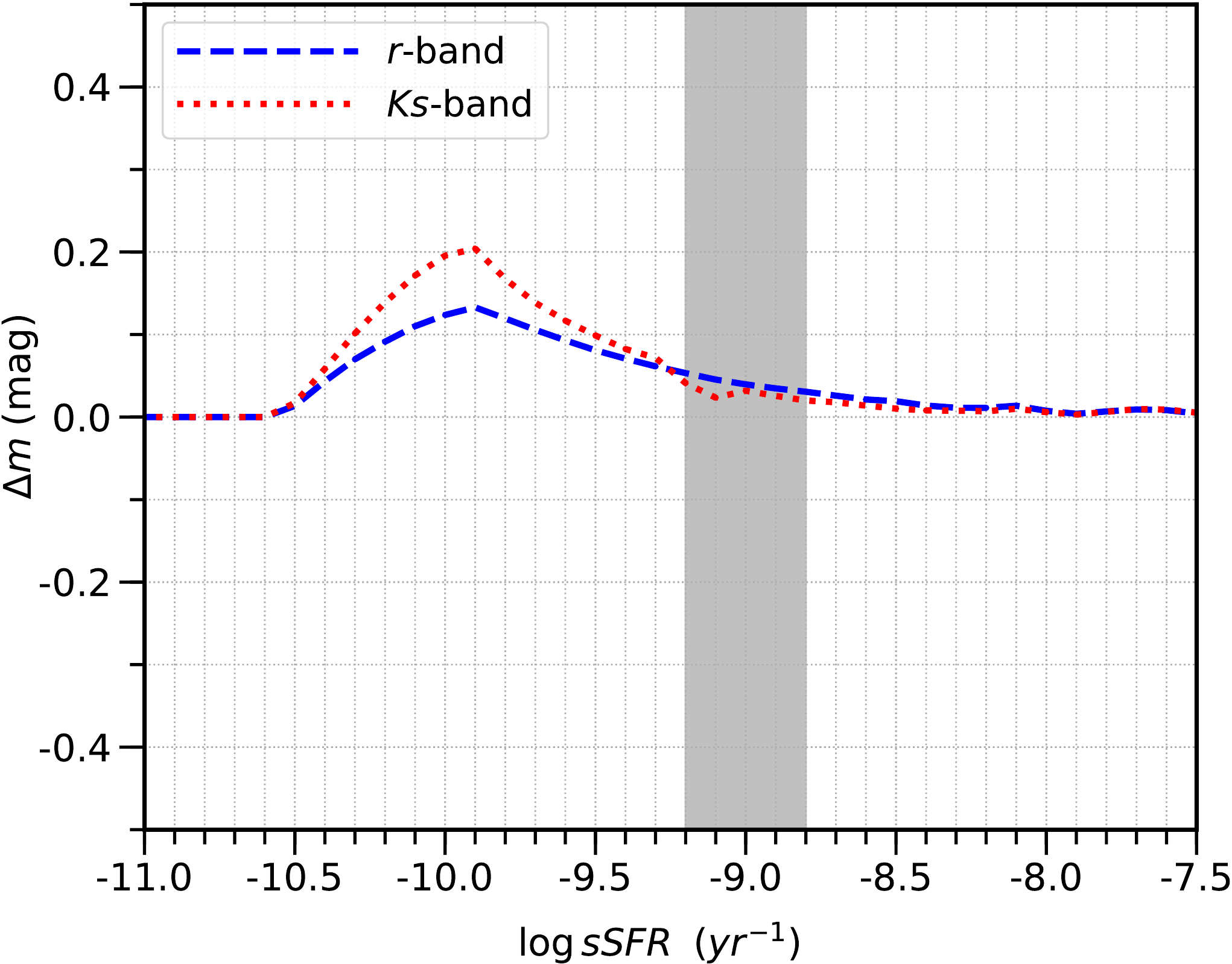}
\end{center}
\caption{Difference in the magnitudes (normalised to stellar mass) estimated by the stellar population synthesis model described in the text for an exponential decay star-formation history with two different values of $\tau$, 2 and 10\,Gyr. The magnitude difference is plotted as a function of the specific star-formation rate.  The red dotted and blue dashed lines correspond to the near-infrared $Ks$-band and the optical $r$ filter respectively. These curves are estimated for a galaxy at a redshift $z=1$. The grey-shaded rectangle shows the $1\sigma$ extent of the Main Sequence (log scatter 0.2) for a galaxy with stellar mass $\log M_\star/M_\odot=10$.}\label{fig:deltaM}
\end{figure}

It is recognized that an issue of the approach outlined above is the non-uniqueness of the inferred star-formation histories. Diverse values of $\tau$ lead to different ($A$, $t$) pairs that reproduce the SFR and $M_{\star}$ of a mock galaxy. Nevertheless, there are strong aliases between the star-formation history parameters and the resulting SEDs. Very different combinations of $A$, $\tau$, $t$ can lead to very similar SEDs. Figure \ref{fig:deltaM} demonstrates this point by comparing the magnitudes (normalised to stellar mass) estimated by fixing  $\tau$ to two distinct values, 2 and 10\,Gyr, in the methodology outlined above. This figure shows that for the application described in this paper the well-established degeneracies that plague attempts to age-date galaxies from their SEDs work in our favour. 

In practice, for the derivation of fluxes in different wavebands we fix $\tau=10$\,Gyrs. A 3-dimensional grid in redshift, stellar mass and star-formation rate is generated and fluxes are estimated for each ($z$, $M_{\star}$, SFR) point. Redshifts range between 0.1 and 4.1 in steps of 0.1. Stellar masses take values in the logarithmic interval $\log M_{\star} = (9,13)$ with a step $\delta log M_{\star} =0.1$. Star-formation rates are defined relative to the Main Sequence \citep{Schreiber2015} at a given ($z$, $\log M_{\star}$) pair and vary between -2\,dex below and +2\,dex above it in logarithmic bins of 0.1. For each ($z$, $M_{\star}$, SFR) grid point the {\sc cigale} stellar synthesis code estimates fluxes in the $ugrizJHKs$ filters as well as the WISE and IRAC mid-infrared bands. The \cite{BC03} stellar library is used to synthesize stellar populations based on the star-formation histories parametrised by Equation \ref{eq:sfr}. The \cite{Chabrier2003} initial mass function is adopted, the metallicity is fixed to solar and the \protect\cite{BC03} stellar libraries are used. The stellar light is absorbed by dust that follows the \cite{Calzetti2000} law with extinction $E(B-V)=0.4$\,mag  for the star-forming galaxies and zero extinction for the passive galaxies. These choices are motivated by the galaxy SED-fitting results in the COSMOS field \citep{Scoville2007} presented by \cite{Laigle2016}. They find that the dust extinction distribution has a mode in the range $E(B-V)=0.3-0.5$\,mag for star-forming galaxies with stellar mass $\log M_\star/M_\odot>10$ and that passive galaxies typically have  $E(B-V)\approx0.0$\,mag. We acknowledge that these values are model dependent. They nevertheless provide some guide on the choice of extinction for the mock galaxies. We also choose not to adopt more complex dust extinction models that may include for example a dependence on redshift or stellar mass, because of the lack of appropriate empirical prescriptions.   
 
\begin{figure}
\begin{center}
\includegraphics[height=0.75\columnwidth]{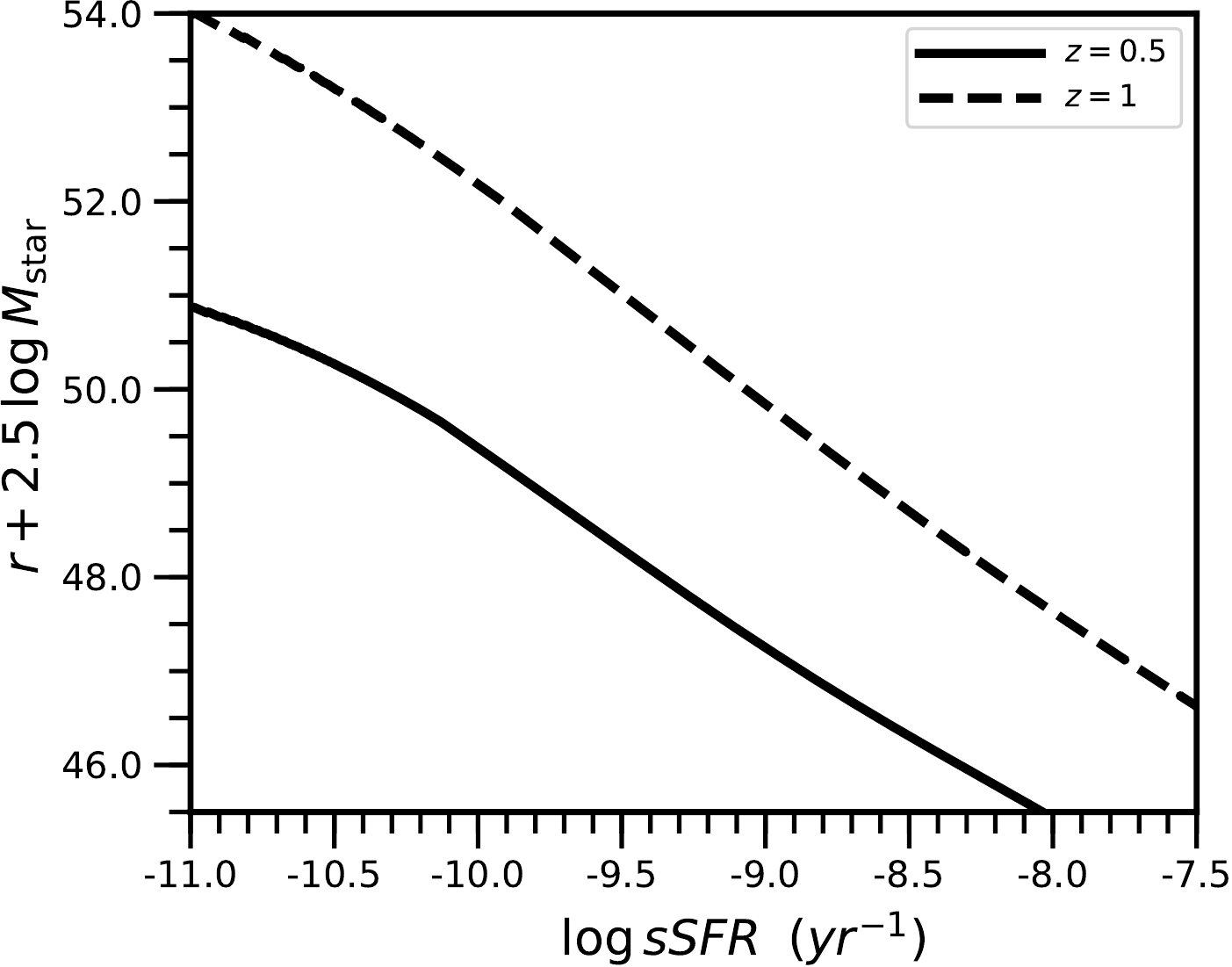}
\end{center}
\caption{$r$-band magnitude normalised to stellar mass plotted as a function of specific star-formation rate for redshifts $z=0.5$ and $z=1$. The curves correspond to the models described in the text. They assume an exponential declining star-formation history to determine the fluxes in different bands using the instantaneous stellar mass and star-formation rate as boundary conditions. Both curves include reddening $E(B-V)=0.4$\,mag.}\label{fig:mag_vs_ssfr}
\end{figure}

\begin{figure}
\begin{center}
\includegraphics[height=0.8\columnwidth]{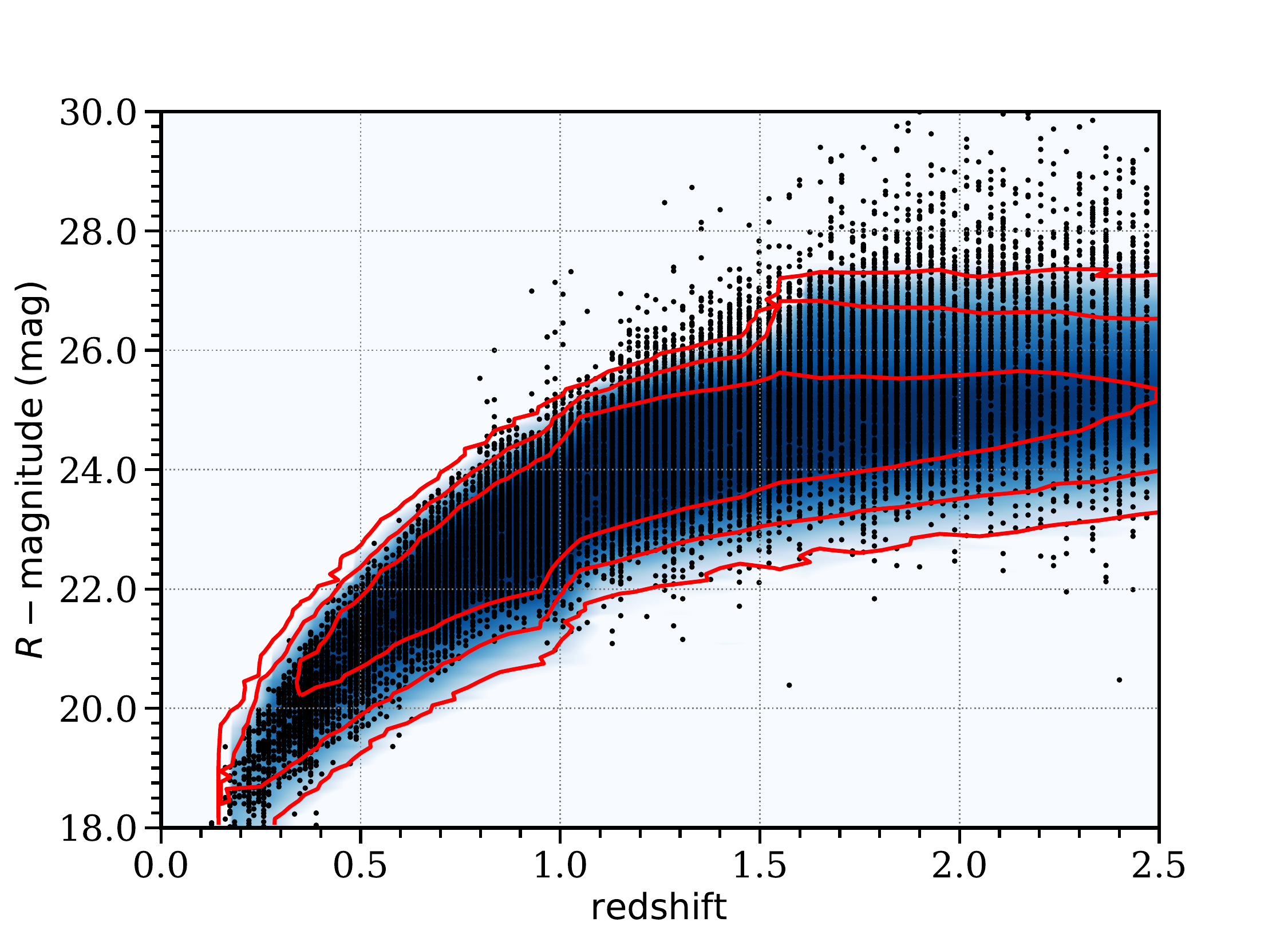}
\includegraphics[height=0.8\columnwidth]{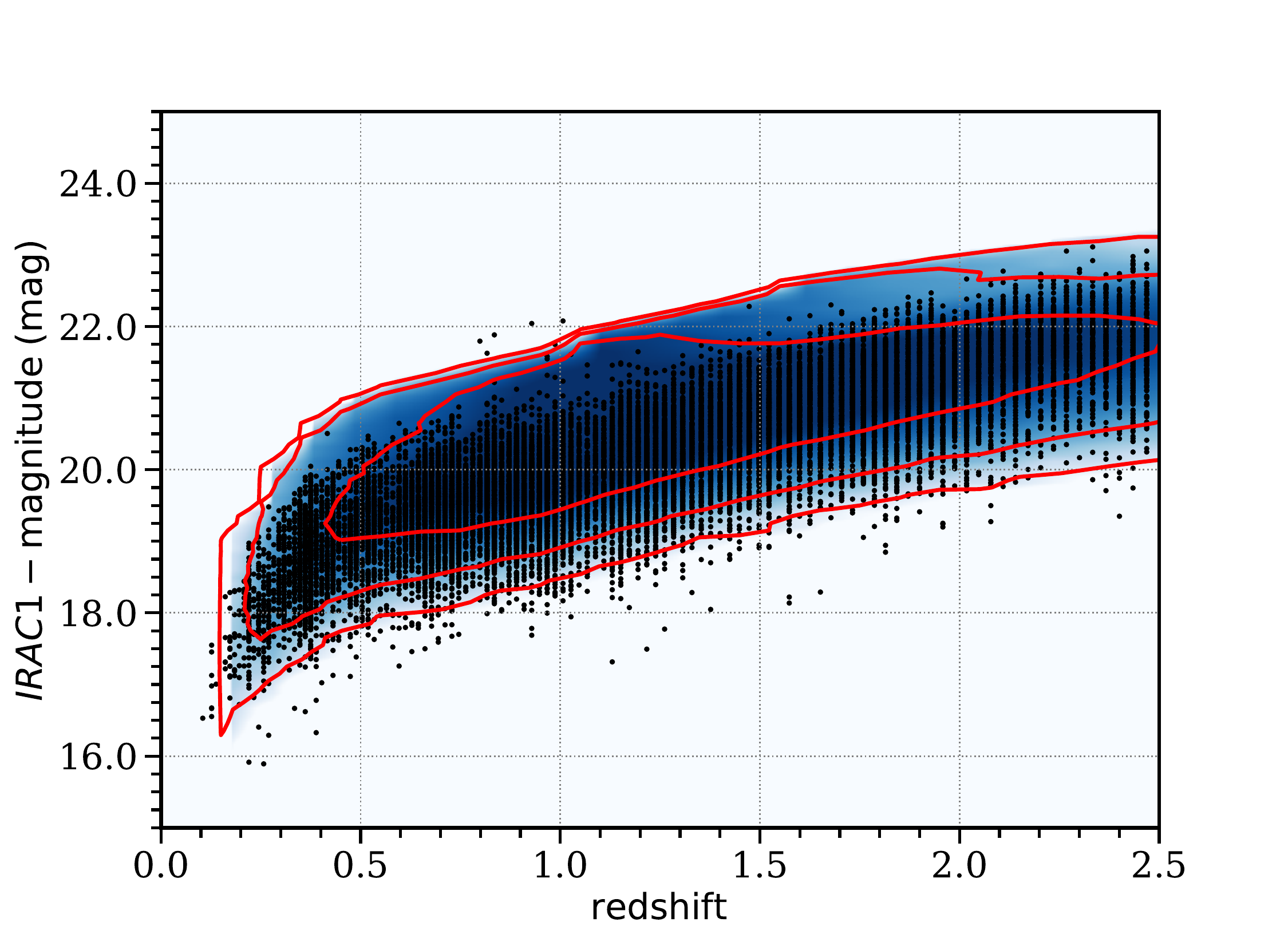}
\end{center}
\caption{The distribution of mock and real galaxies on the apparent magnitude vs redshift plane. The top panel plots the optical $r$-band magnitude. The bottom panel shows results for the mid-infrared IRAC\,$3.6\mu m$ band. In both panels the blue shaded regions and the red contours show the distribution of mock galaxies with stellar masses $\log M/M_\odot>10.5$. The contours encompass 68, 95 and 99.7\% of the mock galaxies. The black dots are galaxies in the COSMOS field with photometry, redshifts and stellar mass estimates from \protect\citet{Muzzin2013}. A stellar mass cut of $\log M/M_\odot>10.5$ has also been applied to this data. We use the galaxy stellar masses of \protect\citet{Muzzin2013} determined for a \protect\cite{Chabrier2003} initial mass function and the \protect\cite{BC03} stellar libraries.}\label{fig:galRZ}
\end{figure}

The end-product of the SED-estimation step are look-up tables that list for each redshift bin the mass-normalised fluxes in the filters above as a function of specific star-formation rate. Examples of such curves are shown in Figure \ref{fig:mag_vs_ssfr}. These curves are then used to assign fluxes via nearest-neighbor interpolation to each mock galaxy with redshift $z$, stellar mass $M_{\star}$, star-formation rate SFR and hence, specific star-formation rate $sSFR=SFR/M_{\star}$. The performance of this approach is demonstrated in Figure \ref{fig:galRZ}. It shows the distribution of model galaxies on the apparent magnitude vs redshift plane at the stellar mass cut $\log M/M_{\odot}>10.5$ in comparison with the observational results of \cite{Muzzin2013} to the same stellar mass limit.  Results are shown for apparent magnitudes in the optical $r$-band and IRAC\,$3.6\mu m$ filter. The mock galaxies broadly follow the observed distribution of magnitudes at a given redshift interval. Small systematic offsets between the observations and the model predictions or differences in the broadness  of the distribution at fixed redshift are also present. These discrepancies may be associated with systematic offsets of $0.1-0.2$\,dex in the model vs the observationally determined stellar masses, uncertainties in the photometric aperture corrections applied to the observations, or differences in the mass-to-light ratios of the stellar-population models, which are associated  to the adopted star-formation history parametrisation. Nevertheless, to the zero order our approach of assigning SEDs to galaxies produces mock samples that are broadly consistent with observations.   

\section{Results}

\subsection{Observed propertied of X-ray selected AGN}

In this section the performance of the model described in the previous section is assessed by making predictions on the multiwavelength photometric properties of AGN detected at X-rays and then comparing with observations. For this exercise we use two recent  extragalactic X-ray survey fields with different depth and area characteristics, the XMM-XXL \citep{Pierre2016} and the COSMOS-Legacy \citep{Civano2016}. The former is one of the largest contiguous surveys ($\rm 2\times25deg^2$) carried out by the XMM-{\it Newton} to a depth of $f_X \approx 10^{-15}\rm \, erg \, s^{-1} \,cm^{-2}$. The latter has a limiting flux of $\approx 10^{-16} \rm   erg \, s^{-1} \,cm^{-2}$  and represents relatively deep surveys over smaller areas ($\approx \rm 2\,deg^2$). For each field we construct the magnitude distribution of the counterparts of X-ray AGN in different wavebands, e.g. optical, near-infrared, mid-infrared. The comparison of these distributions with the model expectations requires the application of observational selection effects to the mock catalogue. These include the X-ray sensitivity of each survey field and the depth of the corresponding photometric observations. 

The X-ray selection function quantifies the probability of detecting sources of a given flux. Because of instrumental effects, such as vignetting, the detection probability increases smoothly from faint to bright fluxes. These variations are well understood and can be estimated to a good level of accuracy to generate X-ray sensitivity curves. For the XMM-XXL and COSMOS-Legacy surveys we use the X-ray source catalogues and corresponding sensitivity curves presented by \citet{Liu2016} and \citet{Georgakakis2015}, respectively. For both fields the X-ray sensitivity curves are constructed using the methods described in \cite{Georgakakis2008_sense}. For each mock AGN of a given X-ray flux the detection probability is estimated from the sensitivity curve. A random number between zero and one is generated and if it is larger than the detection probability the mock AGN is rejected. This methodology generates mock samples that mimic the XMM-XXL or Chandra-Legacy X-ray selection functions.

\begin{figure}
\begin{center}
\includegraphics[height=0.9\columnwidth]{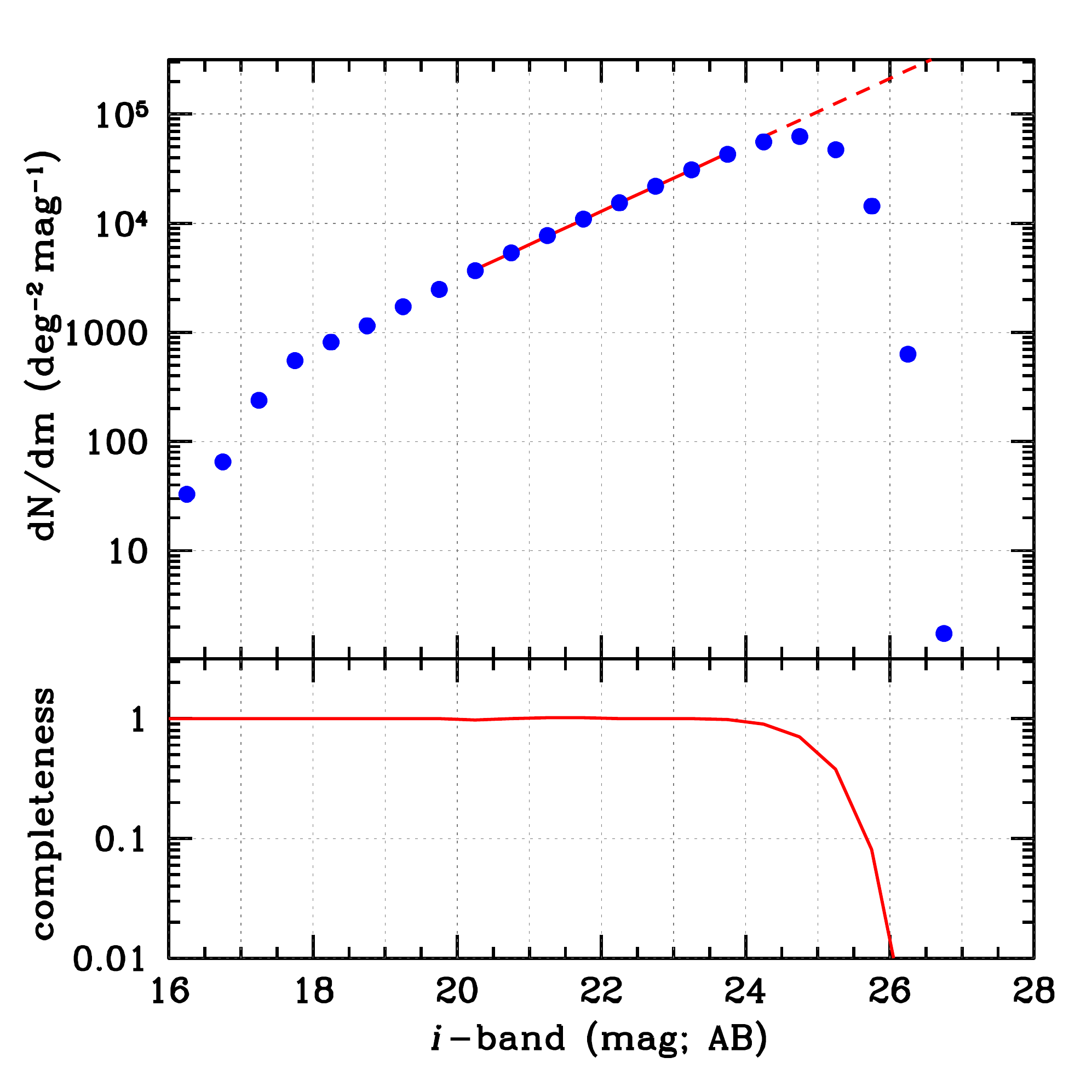}
\end{center}
\caption{The top panel plots the differential number counts in the $i$-band (blue points). These are constructed using the Canada-France Hawaii Telescope Lensing Survey \protect\citep{Heymans2012, Erben2013} optical photometric catalogue in the XMM-XXL field. The sample is affected by incompleteness at magnitudes fainter than $i\approx25$\,mag. This is manifested by the turnover in the differential counts (blue points) beyond this magnitude limit. The solid red line shows the best-fit power-law relation to the observed number counts between magnitudes $i=20$ and 24\,mag, i.e. in a range unaffected by incompleteness. The red-dashed lines shows the extrapolation of this relation to faint magnitudes. It provides an estimate of the expected unbiased number density of $i$-band selected sources as a function of magnitude. The ratio between the best-fit power-law relation and the observed counts (blue points) at faint magnitudes measures the completeness of the optical observations, which can be used to approximate the selection function of the specific dataset. This is shown in the lower panel with the red curve.}\label{fig:counts-example} 
\end{figure}

The multiwavelength properties of X-ray AGN can be studied by associating them in a statistical manner with photometric catalogues selected at different wavebands, e.g. optical, infrared etc. The fraction of associations primarily depends on the depth of the multiwavelength observations and the choice of the photometric filters. Such selection effects need to be applied to the model before comparing the predicted  magnitude distribution of X-ray AGN to observations. It is therefore important to be able to quantify the incompleteness of  a photometric catalogue as a function of magnitude. One approach to address this point is by comparing the observed incomplete number count distribution of sources in a given sample (sky density of sources per magnitude bin) with an unbiased expectation. The ratio between the two distributions provides an estimate of the selection function of the sample. In this paper we approximate the unbiased (complete) count-rate distribution by exploiting the fact that the observed logarithmic galaxy number counts can be represented by a power-law. The slope of the power-law may vary slowly with magnitude, but at least within relatively small intervals it is assumed to remain constant. For a given photometric survey we therefore construct the number counts as a function of magnitude. An example is plotted in Figure  \ref{fig:counts-example}) and shows the typical power-law form followed by a turn-over at the faint-end, which is because of incompleteness, i.e. the photometric limit of the specific survey. A power-law functional form is fit to the number count distribution (e.g. see Fig \ref{fig:counts-example}) for magnitudes in the range $m_{to}-4$ and $m_{to}-1$, where $m_{to}$ is the turnover magnitude beyond which incompleteness dominates. The choice of magnitude interval is to avoid incompleteness and provide an acceptable representation of the faint-end slope of the logarithmic number counts (e.g. avoid Euclidean-slope regime). The extrapolation of the power-law fit to faint magnitudes that are affected by incompleteness provides an estimate of the unbiased expectation. The ratio between this and the observed number counts approximates the selection function of a photometric catalogue and is applied to the mocks to mimic observational biases. In practice each mock galaxy is weighed by the inferred selection function of a photometric catalogue. The weighted histogram of magnitudes is then constructed and compared with the corresponding photometric observations.   

The identification of X-ray sources in the XMM-XXL field with optical counterparts is presented in \citet{Georgakakis2017xxl}. They used the Canada-France Hawaii Telescope Lensing Survey \citep[CFHTLenS][]{Heymans2012, Erben2013} optical photometric catalogue ($ugriz$-bands) selected in the $i$-band.  The X-ray sources in the COSMOS-Legacy survey are matched to the multiwavelength catalogue of \citet{Laigle2016} that includes photometry from the ultraviolet to the far-infrared. For this field it is therefore possible to test the model performance in reconstructing the magnitude distribution of X-ray AGN in near- and mid-infrared wavebands. Also,  at these longer wavelengths AGN are typically associated with bright sources and therefore incompleteness corrections are irrelevant. They do play a role however, at shorter wavelengths, i.e. optical. The identification of the COSMOS-Legacy X-ray fields with multiwavelength counterparts is described in Appendix \ref{appendix:cosmos}.

Figures \ref{fig:xxl}--\ref{fig:cosmos-hard} compare the observed and simulated magnitude distributions of X-ray selected AGN in the XMM-XXL and COSMOS-Legacy survey fields. The soft (0.5-2\,keV) and hard (2-10\,keV) band selected samples are plotted separately. For the XMM-XXL the optical magnitude distributions ($r$, $i$ filters) are shown. Near- and mid-infrared bands ($Ks$, {\it Spitzer} IRAC1) are also included in the case of the COSMOS-Legacy field. In these figures the simulated magnitude distributions are split into type-I and type-II AGN contributions. The boundary between the two classes is the column density limit $\rm \log [N_H/cm^{-2}]=22$. As expected type-I AGN are offset to bright optical magnitudes, where they dominate in the case of shallow fields like the XMM-XXL. The type-II AGN contribution increases toward fainter magnitudes. The interplay between the two populations yields a broad, nearly bimodal, $r$-band magnitude distribution in the case of shallow X-ray fields like the XMM-XXL. In the near- and mid-infrared the type-I and II AGN have nearly identical magnitude histograms. Overall there is reasonable agreement between the observed and simulated magnitude distributions of X-ray selected AGN. The basic characteristics of the observations, e.g. broadness, peak of the histograms, are roughly reproduced by the model. 

\begin{figure*}
\begin{center}
\includegraphics[height=0.9\columnwidth]{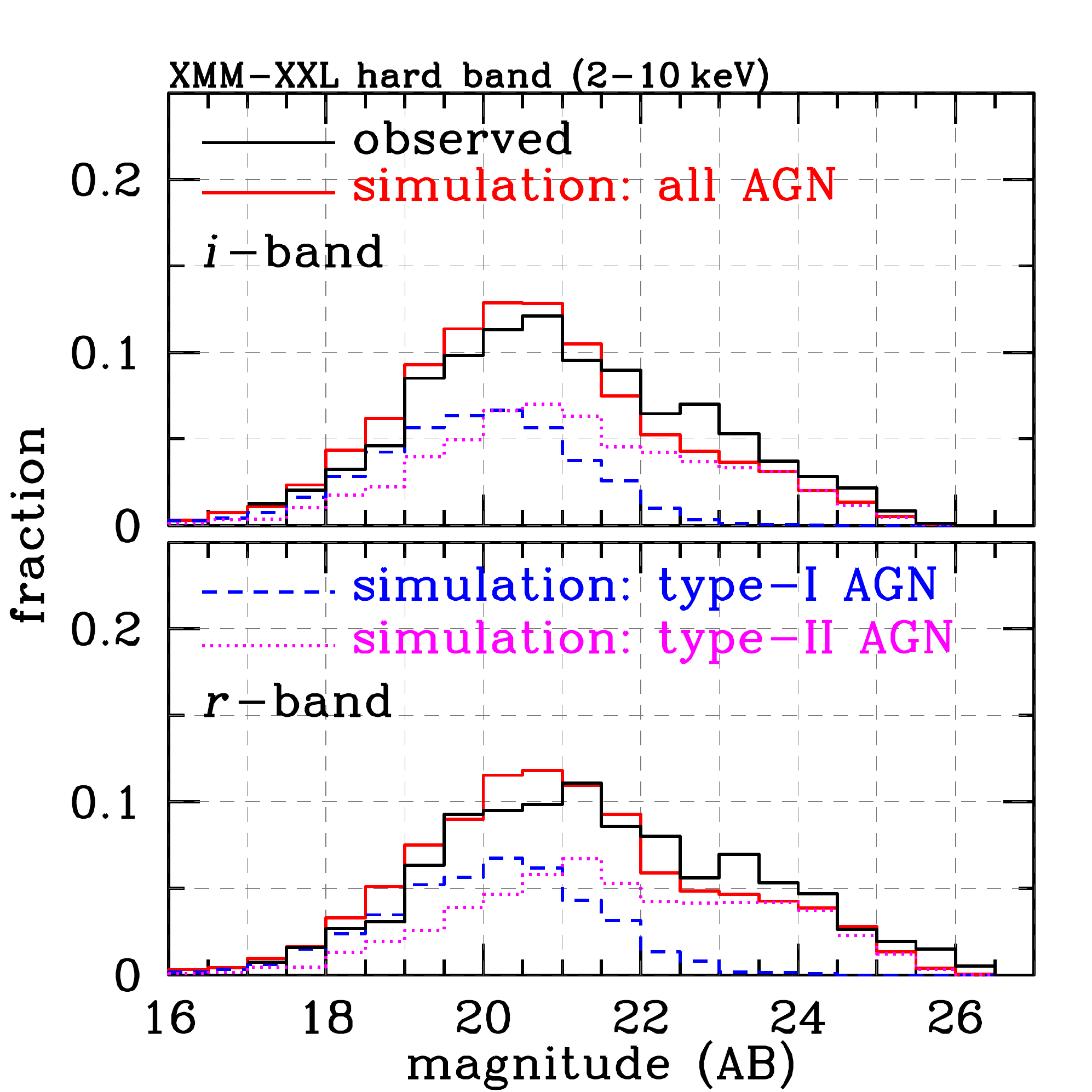}
\includegraphics[height=0.9\columnwidth]{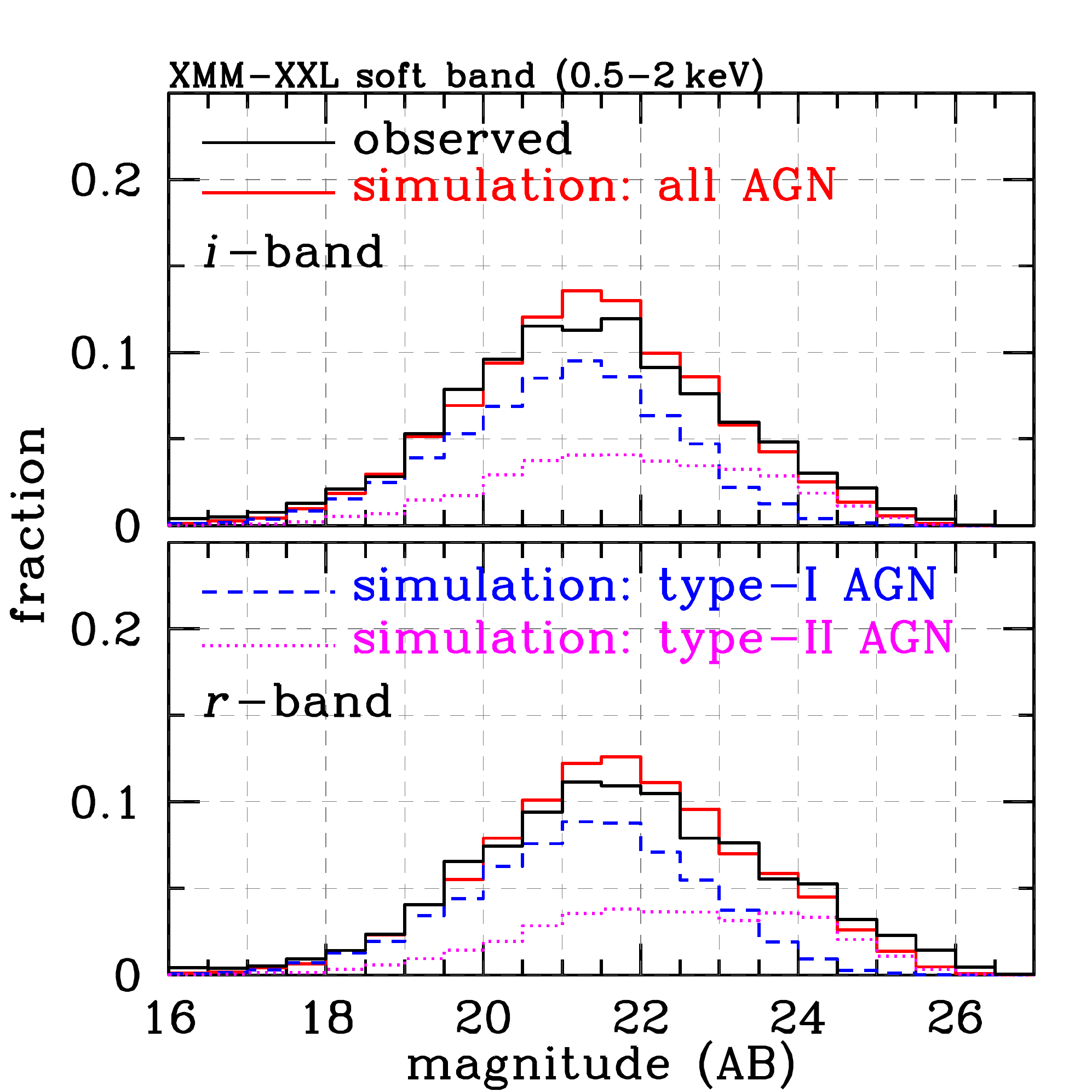}
\end{center}
\caption{Magnitude distribution of X-ray AGN in the XMM-XXL field. The left panel shows the hard-band (2-10\,keV) selected sample. The right panel corresponds to XMM-XXL AGN selected in the soft (0.5-2\,keV) band. In both panels the observations  \protect\citep{Georgakakis2017xxl} are shown with the black solid histogram, while the red solid line corresponds to the simulations described in text after applying the appropriate observational selection effects (X-ray sensitivity, photometric completeness). The red solid histogram is further broken down into the type-I (blue dashed) and type-II (magenta dotted) AGN contributions. Type-I or unobscured AGN are defined as those with $\log [N_H \rm /cm^{-2}] < 22$. AGN with hydrogen column density above this limit are type-II or obscured.}\label{fig:xxl}
\end{figure*}

\begin{figure*}
\begin{center}
\includegraphics[height=0.9\columnwidth]{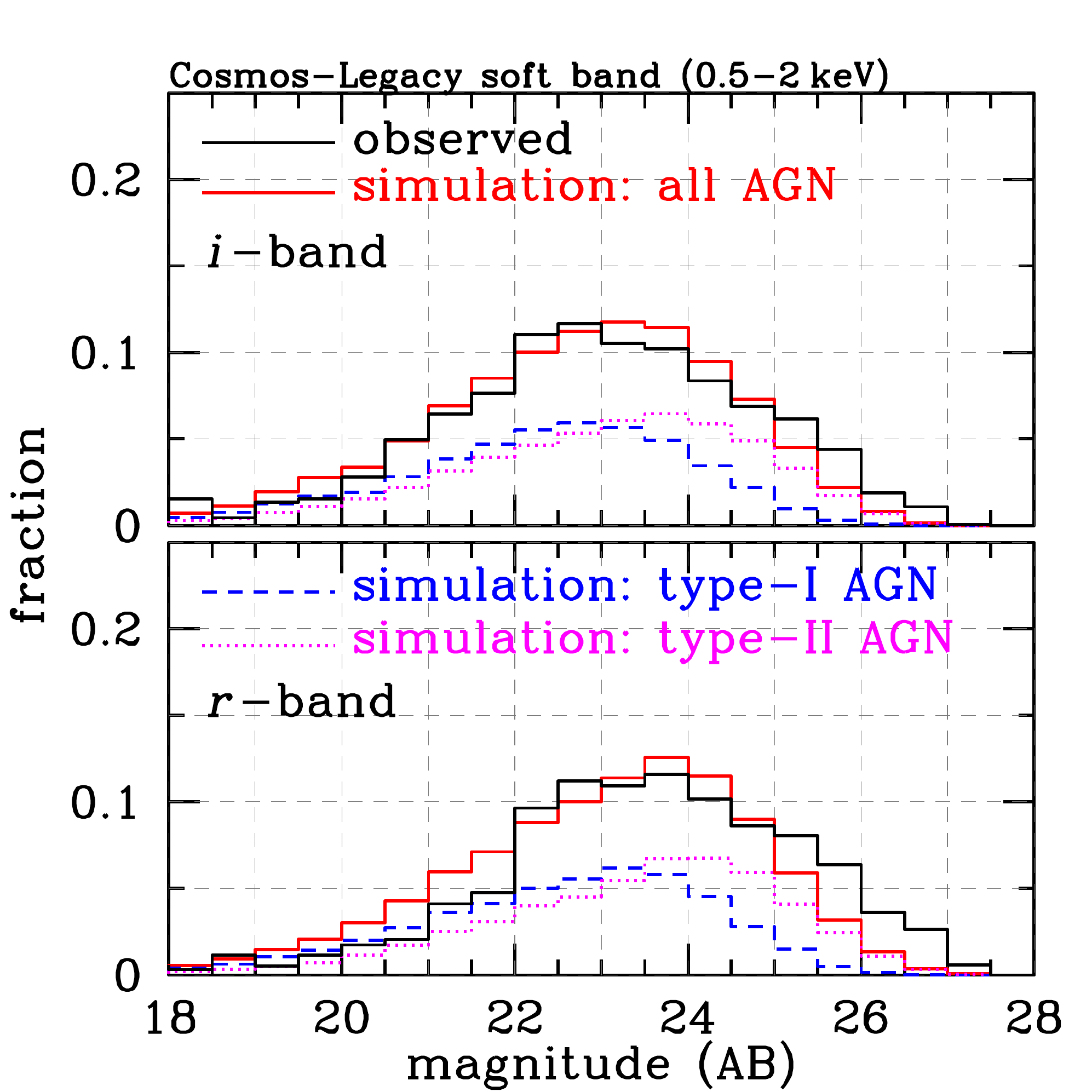}
\includegraphics[height=0.9\columnwidth]{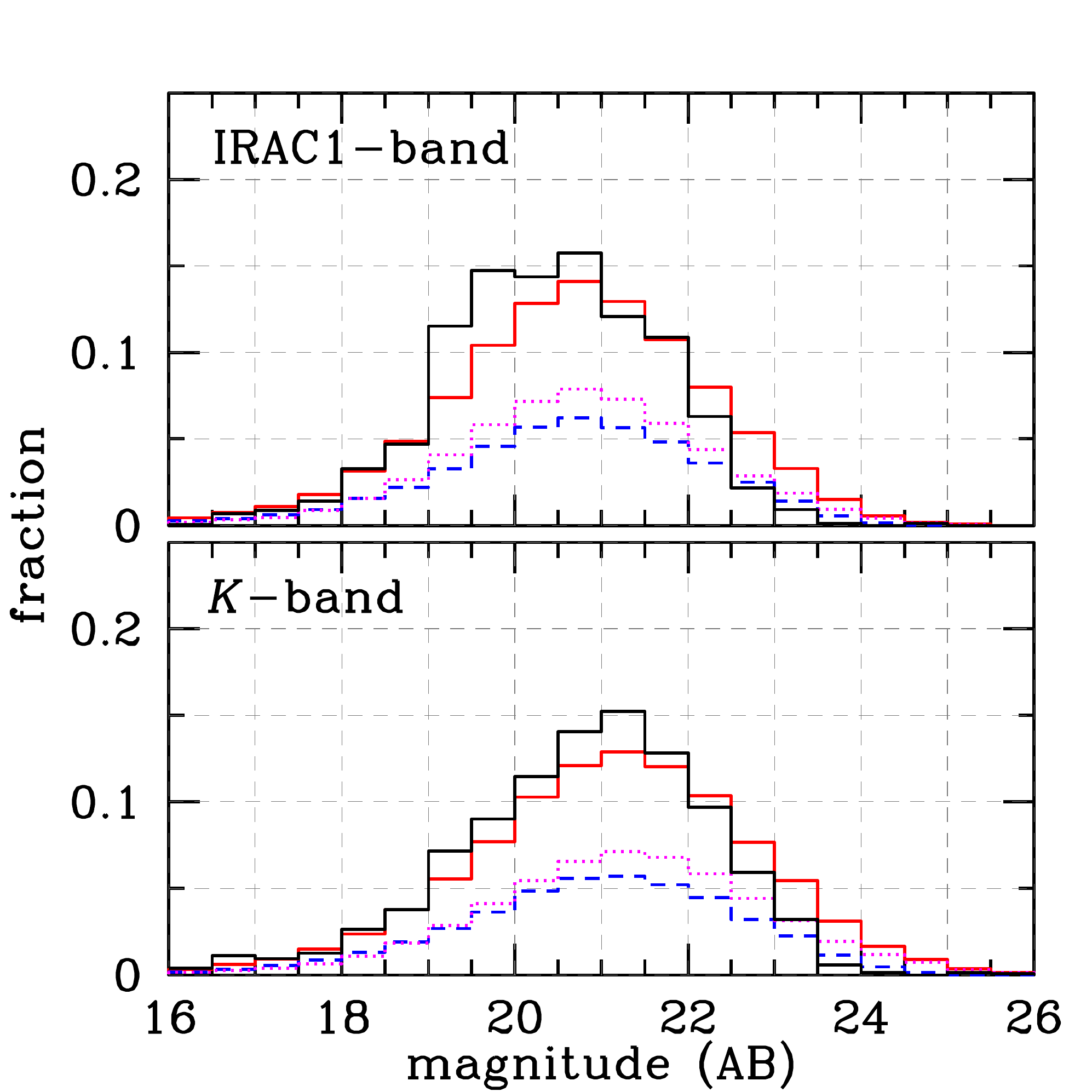}
\end{center}
\caption{Magnitude distribution of X-ray AGN selected in the soft-band (0.5-2\,keV) of the COSMOS-Legacy field. The left panel shows the $r$ and $i$ optical band  histograms. The right panel plots the near-infrared ($Ks$) and mid-infrared ({\it Spitzer} IRAC1\,3.6$\mu$m) wavebands. In both panels the observations are shown with the black solid histogram. The red solid line corresponds to the simulations described in the text after applying the appropriate observational selection effects (X-ray sensitivity, photometric completeness). The red solid histogram is further broken down into the type-I (blue dashed) and type-II (magenta dotted) AGN contributions. Type-I or unobscured AGN are defined as those with $\log [N_H \rm /cm^{-2}] < 22$. AGN with hydrogen column density above this limit are type-II or obscured.}\label{fig:cosmos-soft}
\end{figure*}

\begin{figure*}
\begin{center}
\includegraphics[height=0.9\columnwidth]{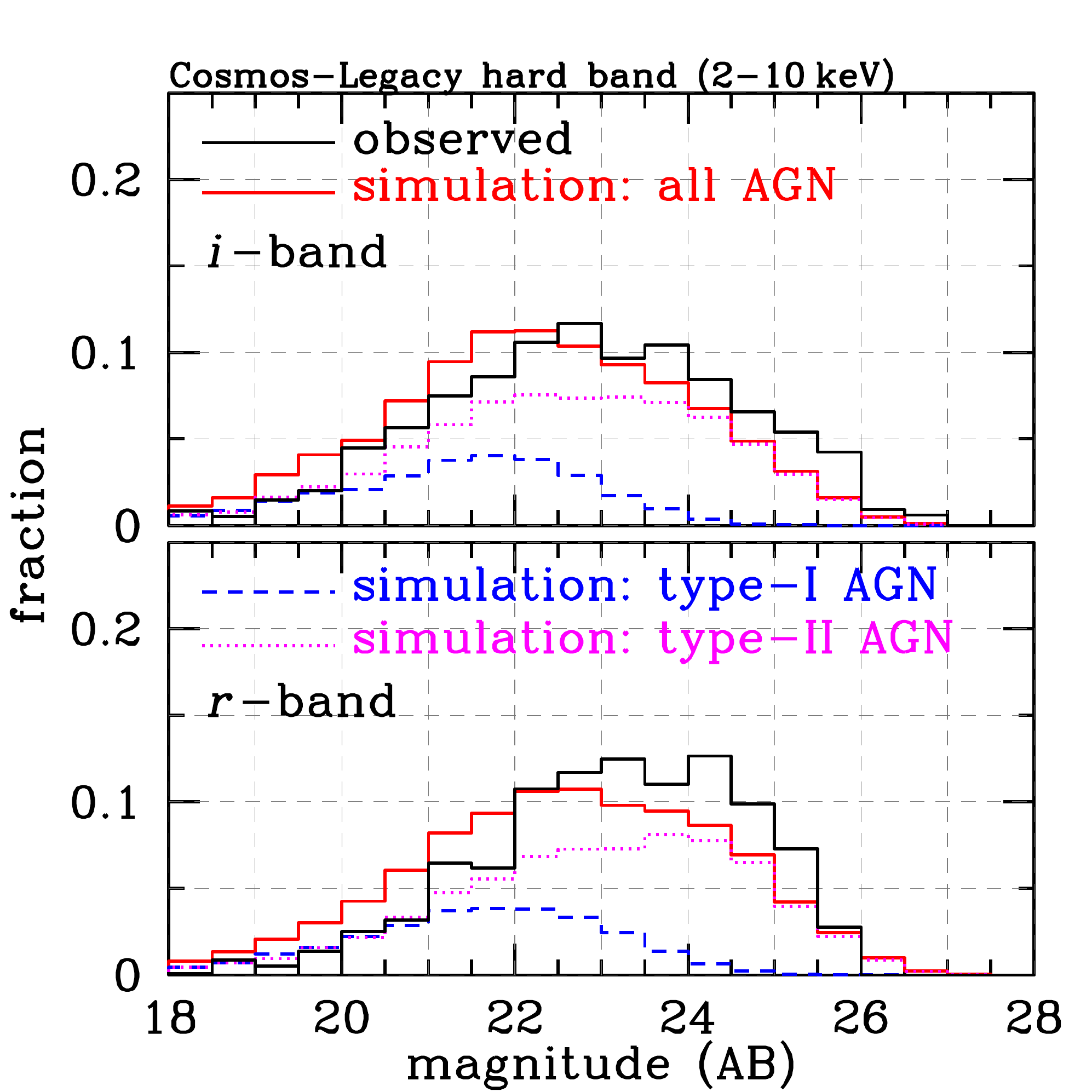}
\includegraphics[height=0.9\columnwidth]{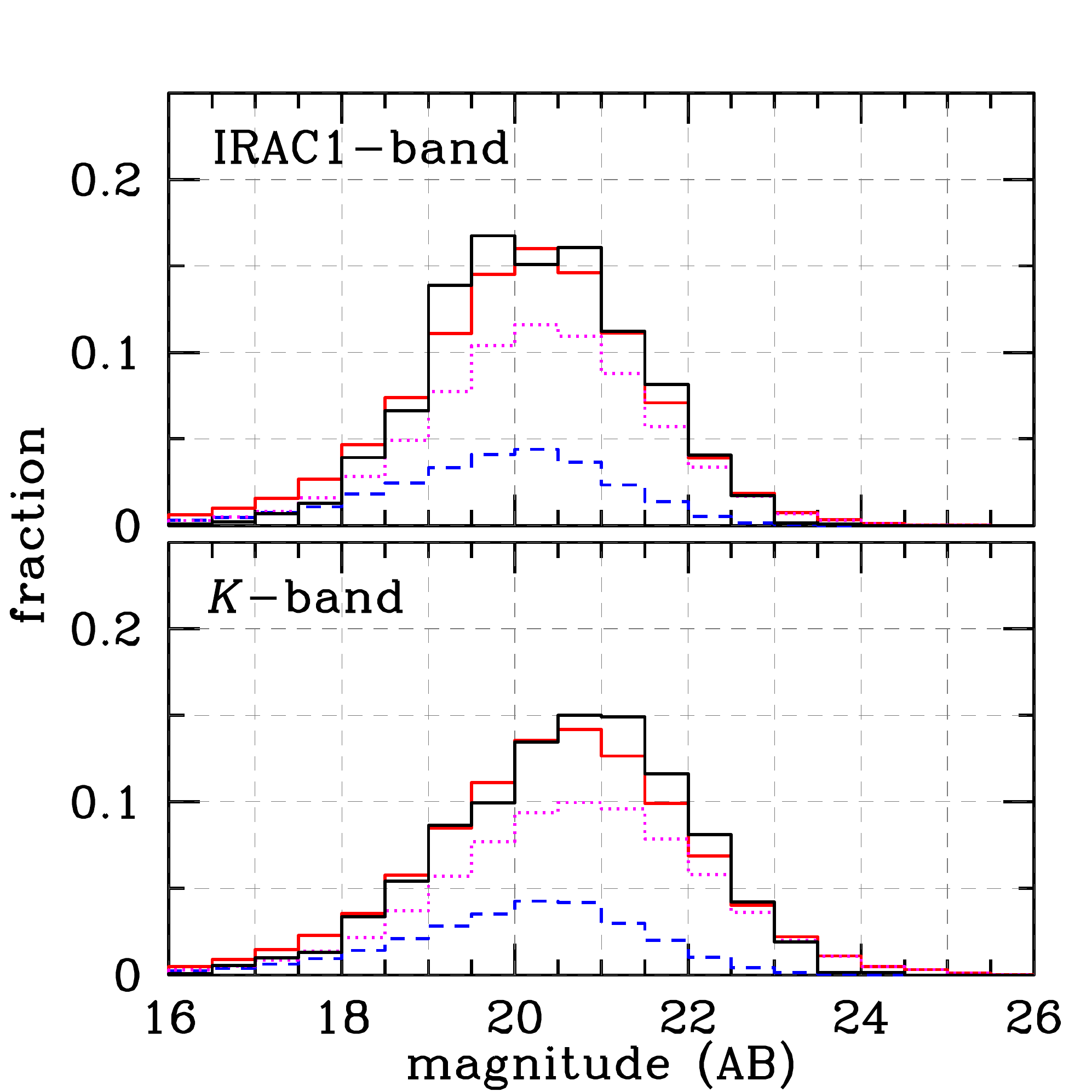}
\end{center}
\caption{Magnitude distribution of X-ray AGN selected in the hard-band (2-10\,keV) of the COSMOS-Legacy field. The left panel shows the $r$ and $i$ optical band  histograms. The right panel plots the near-infrared ($Ks$) and mid-infrared ({\it Spitzer} IRAC1\,3.6$\mu$m) wavebands. In both panels the observations are shown with the black solid histogram. The red solid line corresponds to the simulations described in the text after applying the appropriate observational selection effects (X-ray sensitivity, photometric completeness). The red solid histogram is further broken down into the type-I (blue dashed) and type-II (magenta dotted) AGN contributions. Type-I or unobscured AGN are defined as those with $\log [N_H \rm /cm^{-2}] < 22$. AGN with hydrogen column density above this limit are type-II or obscured.}\label{fig:cosmos-hard}
\end{figure*}

\subsection{Observed properties of the R75 WISE selected AGN}

In this section the predictions of the empirical model is tested against the observed properties of AGN selected in the WISE mid-infrared bands. We choose sources from this mission because of the rich observational data available for them \citep[e.g. photometry, spectroscopy;][]{Mateos2013, Assef2013, Hainline2014, Assef2018, LaMassa2019}, which is a result of their apparent brightness that facilitates follow-up studies. Among the different photometric criteria proposed in the literature for isolating AGN among the WISE source population, we adopt the one proposed by \citet{Assef2013} for compiling samples with reliability 75\% (R75 sample). This choice is motivated by the availability of observational data that can be compared with the model predictions. In particular, the SDSS-IV special-plate spectroscopic programme in the  Stripe82 field \citep{LaMassa2019} includes R75 WISE-selected AGN as prime targets. These observations resulted in the largest spectroscopic sample of R75 WISE AGN to date to the optical magnitude limit $r\approx22.5$\,mag. This is therefore an excellent observational resource for model comparisons. 

\begin{figure*}
\begin{center}
\includegraphics[height=0.7\columnwidth]{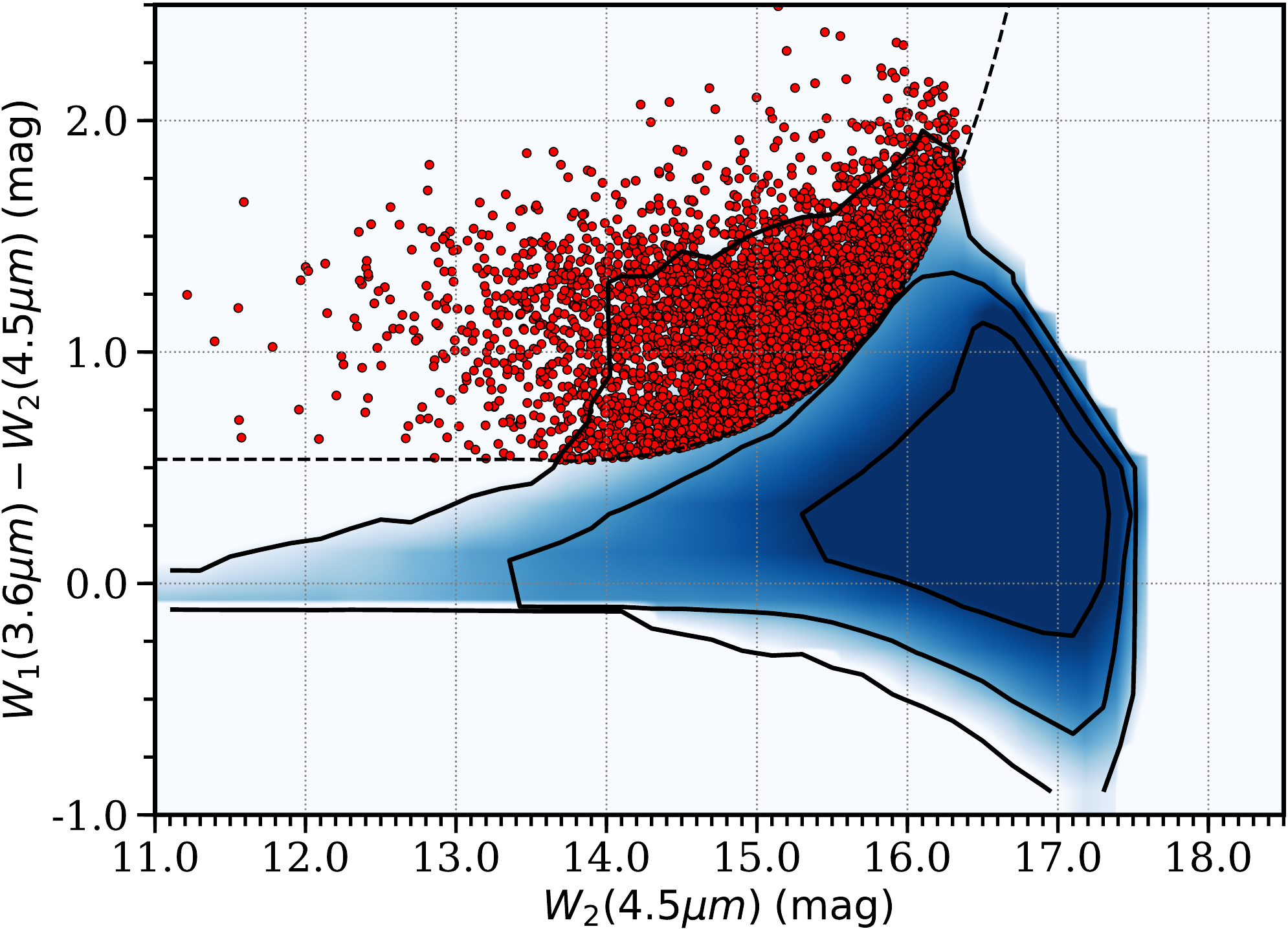}
\includegraphics[height=0.7\columnwidth]{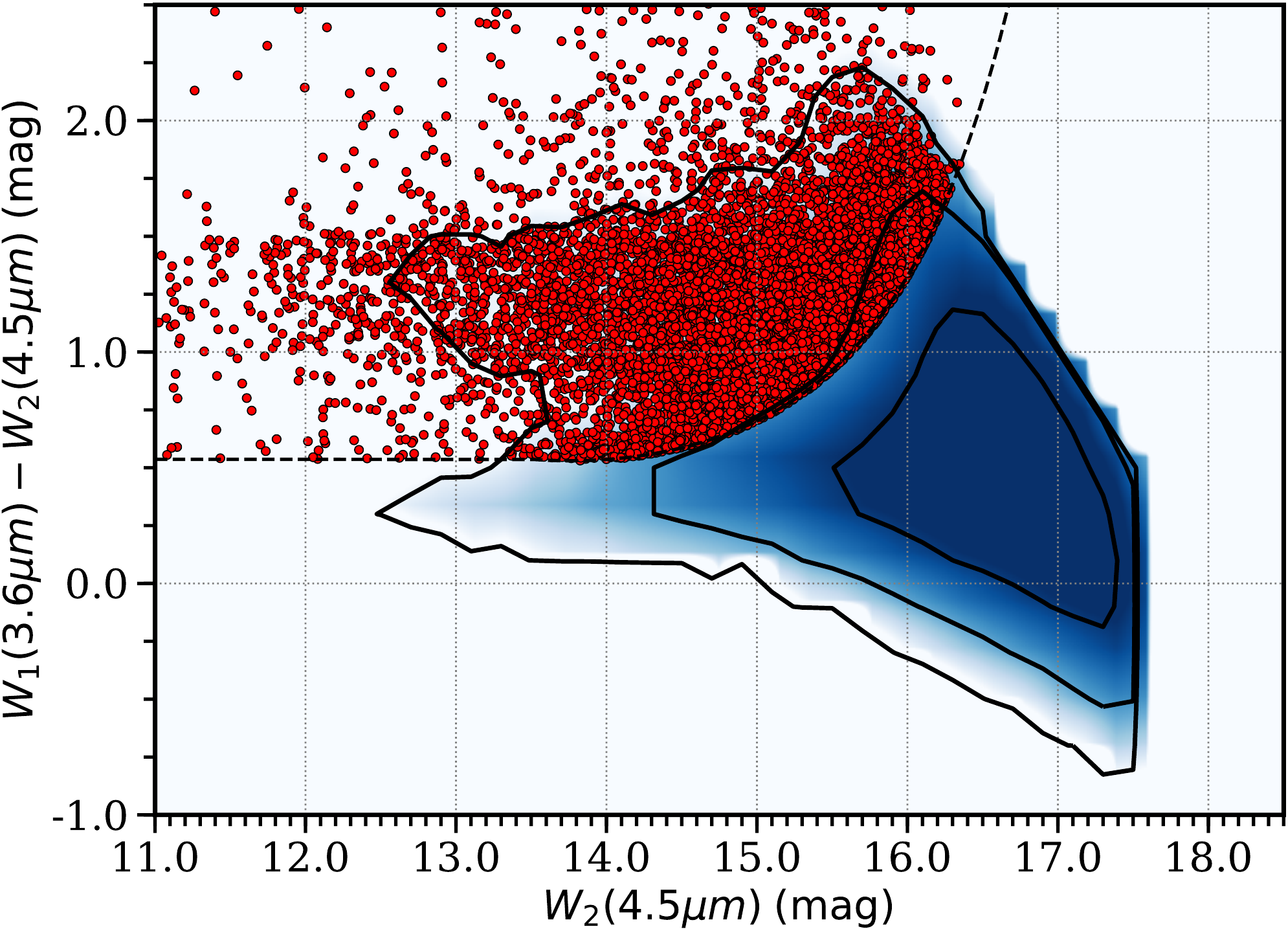}
\end{center}
\caption{WISE $W2$ vs $W1-W2$ colour-magnitude plot. The panel on the left corresponds to the ALLWISE observations in the Stripe82 field presented by \protect\citet{LaMassa2019}. The panel on the right shows the distribution of mock sources on the colour-magnitude space after applying the WISE selection function to the empirical model as described in the text.}\label{fig:wise-swedge}
\end{figure*}

Figure \ref{fig:wise-swedge}-left  demonstrates the R75 selection wedge of \citet{Assef2013} based on the $W1$, $W2$ photometric bands of the ALLWISE catalogue. We reproduce this selection in the model by first applying to the mock catalogue the ALLWISE photometric characteristics. This enables mimicking the ALLWISE source detection and hence approximate the selection function of the \citet{Assef2013} sample. The first step to achieve this is to assign uncertainties to the mock WISE photometry that are representative of the observed ones. The ALLWISE data processing pipeline includes a noise model that yields the $1\sigma$ rms photometric error, $\sigma_{Wi}$, in each of the four WISE bands, $i=1,2,3,4$. This quantity also defines the signal-to-noise ratio of individual sources

\begin{equation}\label{eq:wise-snr}
{\rm SNR}_{Wi}=\frac{\delta f_{Wi}}{f_{Wi}} = \frac{2.5}{\sigma_{Wi}\,\ln(10)}, 
\end{equation}

\noindent where $f_{Wi}$, $\delta f_{Wi}$ are the flux and corresponding error of the $Wi$ band, respectively. The  $\sigma_{Wi}$ is therefore directly related to the source detection process. ALLWISE detections are required to have ${\rm SNR}_{Wi}>5$ in at least one band ($\sigma_{Wi}<0.22$\,mag). In practice this cut translates to  detection magnitude limits of $W1=18$ and $W2=16$\,mag (Vega system) in the ALLWISE catalogue. Detections in a given filter have photometric measurements in other WISE bands if the corresponding signal-to-noise ratio is  ${\rm SNR}_{Wi}>2$ or equivalently $\sigma_{Wi}<0.54$\,mag. This threshold defines secondary limits, $W1=19$ and $W2=17.5$\,mag (Vega system), for which the $W1-W2$ colour can be measured in the ALLWISE catalogue. The $1\sigma$ rms photometric error produced by the ALLWISE noise model is a monotonic function of source magnitude. This is shown in Figure \ref{fig:errorwise}, which plots the median $\sigma_{W1}$, $\sigma_{W2}$ ({\tt w1sigmpro}, {\tt w2sigmpro},  parameters of the ALLWISE catalogue) as a function of the $W1$, $W2$  magnitude ({\tt w1mpro}, {\tt w2mpro} ALLWISE catalogue parameters) respectively. The errorbars provide a measure of the scatter in each magnitude bin. We use the magnitude dependence of the $\sigma_{Wi}$ in Figure \ref{fig:errorwise} to assign signal-to-noise ratios and photometric errors to mock sources and hence, approximate the ALLWISE catalogue selection. 

We first assign $1\,\sigma$ rms errors, $\sigma_{W1}$, $\sigma_{W2}$, to the model magnitudes $W1$, $W2$. These are estimated by fitting relations of the form $\sigma_{Wi} = A + B \, e^{-Wi/C}$ to the data points of Figure \ref{fig:errorwise}. For each mock AGN with magnitudes $W1$, $W2$ the corresponding standard deviations $\sigma_{W1}$, $\sigma_{W2}$ are estimated using the above parametric fits. Equation  \ref{eq:wise-snr} is then used to define the signal-to-noise ratio of mock sources. By thresholding the model catalogue to ${\rm SNR}_{W1}>5$ or ${\rm SNR}_{W2}>5$ it is possible to mimic the ALLWISE source detection process. The standard deviations $\sigma_{W1}$, $\sigma_{W2}$ are also used to generate Gaussian deviates, which represent the photometric uncertainty in individual bands for each model source.  These are added to the mock photometric magnitudes $W1$, $W2$ of a given source to supplement the model with noise characteristics similar to those of the ALLWISE catalogue. For the estimation of the $W1-W2$ color only sources with ${\rm SNR}_{W1}>2$ or ${\rm SNR}_{W2}>2$ should be used. This essentially translates to filtering the catalogue to the magnitude limits $W1<19$ and $W2=17.5$\,mag (Vega system). Figure \ref{fig:wise-swedge}-right plots the resulting mock sample on the $W1-W2$ vs $W2$ plane. The overall distribution of the model sources in that figure resembles the real observations on the left. There are however, differences. Firstly, the mock catalogue does not include Galactic stars. As a result the observed ALLWISE population with $W1-W2\approx0$ that extends to bright $W2$ magnitudes is absent from the model sample. Secondly the R75 selection wedge in the mocks includes a higher fraction of apparently bright ($W2<13$\,mag) AGN compared to the observations. This is related to the bright-end [$L_X(\rm 2-10\,keV) \ga 10^{44} -  10^{45}\, erg \, s^{-1}$] slope of the X-ray luminosity function and the $6\mu m$/X-ray  luminosity relation used to construct the model. A lower space density of luminous X-ray AGN or a flattening of the  $6\mu m$/X-ray luminosity relation (less mid-infrared emission at fixed X-rays) at the bright-end \citep[e.g.][]{Mateos2015}, both affect the number of mock AGN with apparent magnitude $W<13$\,mag. 

\begin{figure}
\begin{center}
\includegraphics[height=0.7\columnwidth]{PLOTS/errorswise.pdf}
\end{center}
\caption{Measured ALLWISE photometric errors ($1\sigma$ rms) in the $W1$ and $W2$ filters as a function of magnitude. The data-points represent the median photometric uncertainty within 0.4\,mag intervals. The errorbars represent the 16th and 84th percentile of the photometric error distribution in each magnitude bin. The blue dotted and red solid lines correspond to the best fit relation of the form $\sigma_{Wi} = A + B  e^{-Wi/C}$ to the  $W1$ and $W2$ bands respectively. 
}\label{fig:errorwise}
\end{figure}

\begin{figure}
\begin{center}
\includegraphics[height=0.9\columnwidth]{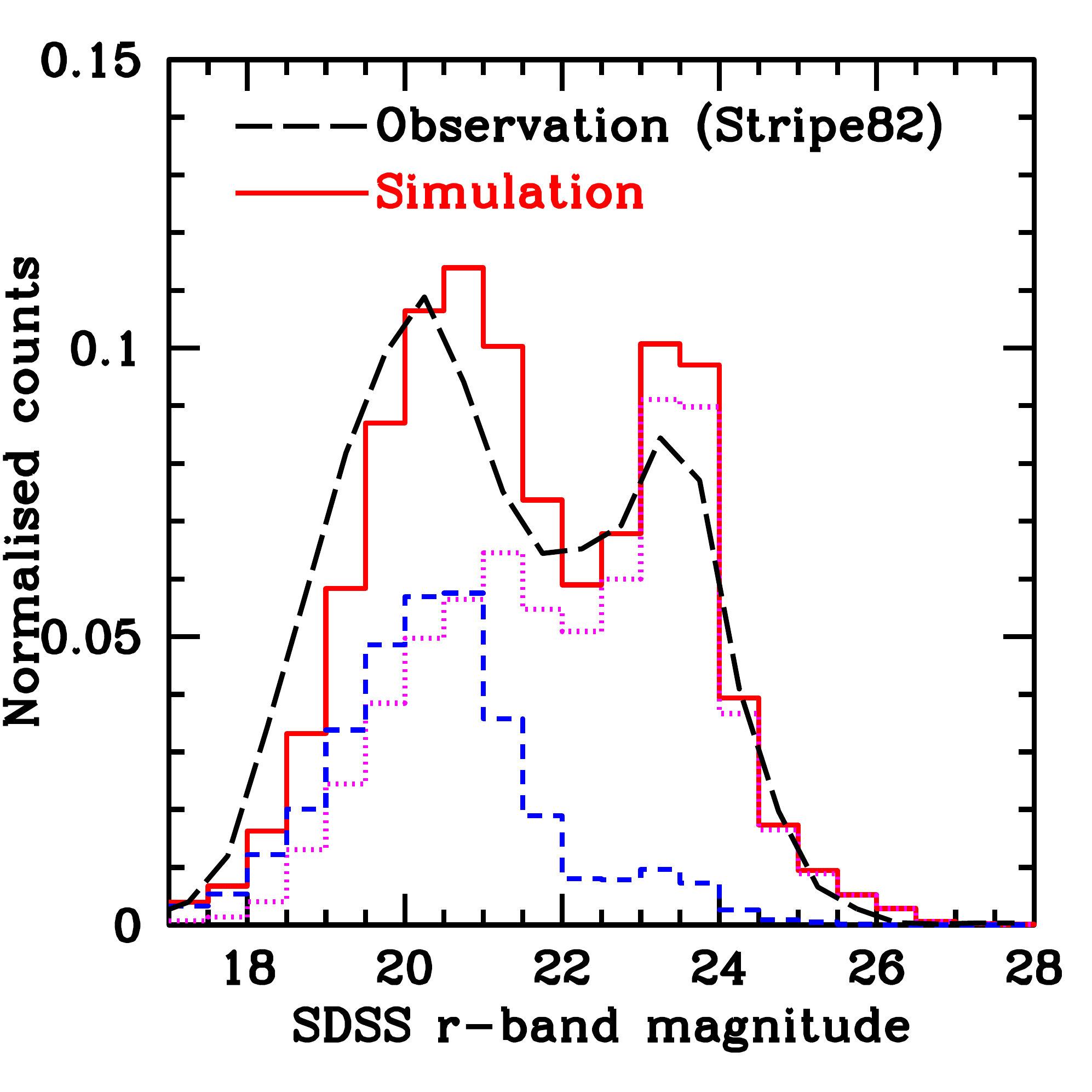}
\end{center}
\caption{Optical $r$-band magnitude distribution of R75 WISE AGN. The dashed black curves shows the observations in the Stripe82X field presented by \protect\cite{LaMassa2019}. The red histogram corresponds to the simulations described in text. This is further broken down into the type-I (blue dashed) and type-II (magenta dotted) AGN contributions. Type-I or unobscured AGN are defined as those with $\log [N_H \rm /cm^{-2}] < 22$. AGN with hydrogen column density above this limit are type-II or obscured.}\label{fig:wise-rdist}
\end{figure}

Having a mock R75 WISE AGN sample we can next explore its  optical magnitude and redshift distributions in comparison with the observations presented in \citet{LaMassa2019}. Figure \ref{fig:wise-rdist} presents the observed and model $r$-band magnitude histograms of the R75 AGN. The observations use the \citet{Jiang2014} coadded photometry in the Stripe82 area that reaches depths of $r\approx25$\,mag. The empirical model predicts a bimodal $r$-band distribution similar to the observed one. Figure \ref{fig:wise-rdist}  shows that the faint-magnitude peak in the model consists  primarily of obscured type-II AGN, for which the optical bands are dominated by stellar emission from the host galaxy. There is a transition from Type-II to Type-I AGN with increasing apparent brightness. 



\begin{figure}
\begin{center}
\includegraphics[height=0.9\columnwidth]{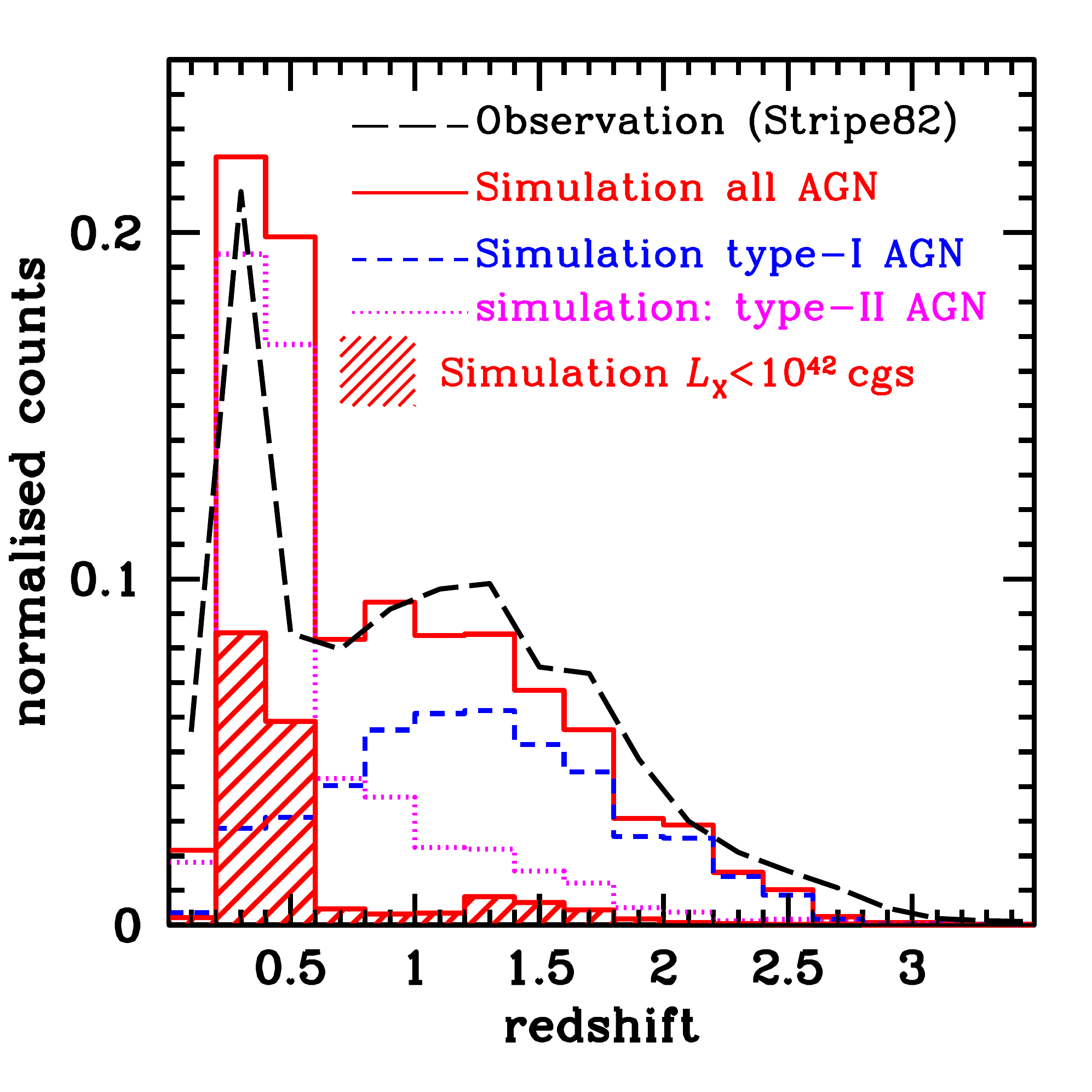}
\end{center}
\caption{Redshift distribution of R75 WISE AGN. The black dashed line shows the observations in the Stripe82X field presented by \protect\cite{LaMassa2019}. The red solid-line histogram corresponds to the simulations described in text after weighing each mock AGN with the $r$-band dependent spectroscopic completeness of the Stripe82X observations. The red solid histogram is further broken down into the type-I (blue dashed) and type-II (magenta dotted) AGN contributions. Type-I or unobscured AGN are defined as those with $\log [N_H \rm /cm^{-2}] < 22$. AGN with hydrogen column density above this limit are type-II or obscured.}\label{fig:wise-zdist}
\end{figure}

The redshift distribution of the mock AGN sample is plotted in  Figure \ref{fig:wise-zdist}. This is compared with spectroscopic observations from the SDSS-IV special-plate QSO programme presented by \citet{LaMassa2019}. These data are limited to the optical magnitude of $r\approx23.5$\,mag. At fainter fluxes the signal-to-noise ratio of the SDSS spectra is too low for reliable redshift estimates. This bias has to be applied to the model before comparing with the observations. One approach to quantify this selection effect is to determine the success rate of reliable redshift estimates from the SDSS spectra as a function of magnitude. \citet{LaMassa2019} have empirically measured this success rate defined as the ratio of sources with reliable redshift estimates at a given $r$-band magnitude and the total number of targeted sources at the same magnitude (see their Figure 8). They find a smooth transition from 100\% redshift-success rate at $r\approx21$ to zero at $r>23.5$. This empirically determined relation provides efficiency factors that are used to weigh mock AGN based on their model $r$-band magnitudes. The resulting weighted histogram is plotted in Figure \ref{fig:wise-zdist}. Both observations and model show a prominent peak at $z<0.5$ followed by a smooth decline to higher redshifts. The low redshift peak of the model is dominated by type-II AGN while type-Is extend to higher redshifts. This is a combined effect of the optical spectroscopic magnitude limit (see for example Fig. \ref{fig:wise-rdist}) and dust extinction that makes obscured AGN fainter that the ALLWISE magnitude limits. The latter is because the WISE spectral bands sample the rest-frame near-infrared and optical with increasing redshift, where the impact of dust reddening is substantial in the case of heavily obscured systems.

\begin{figure}
\begin{center}
\includegraphics[height=0.7\columnwidth]{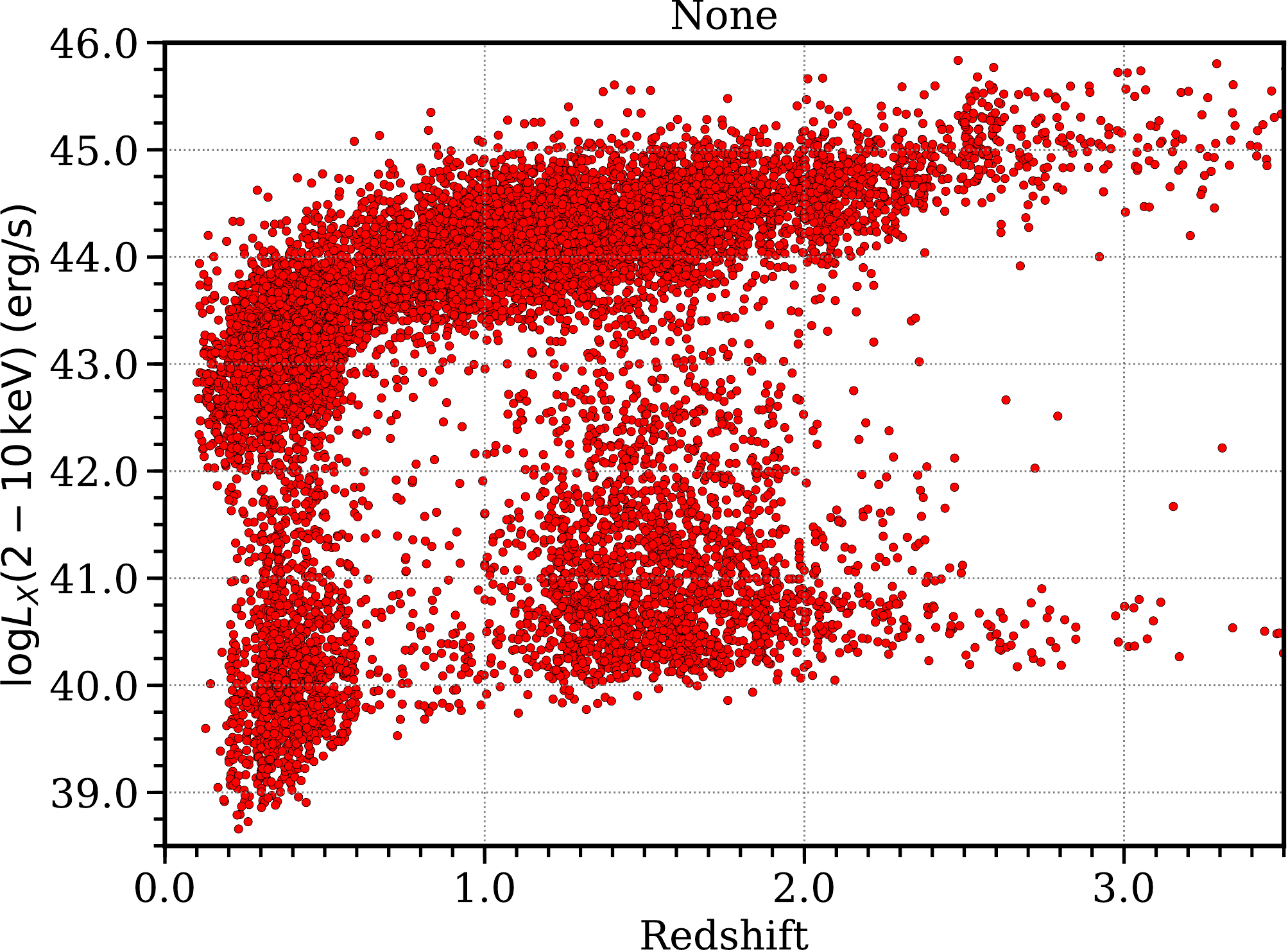}
\end{center}
\caption{Distribution of mock R75 AGN on the X-ray luminosity vs redshift plane. Red dots correspond to individual mock sources.}\label{fig:wise-lxz}
\end{figure}

\begin{figure}
\begin{center}
\includegraphics[height=0.7\columnwidth]{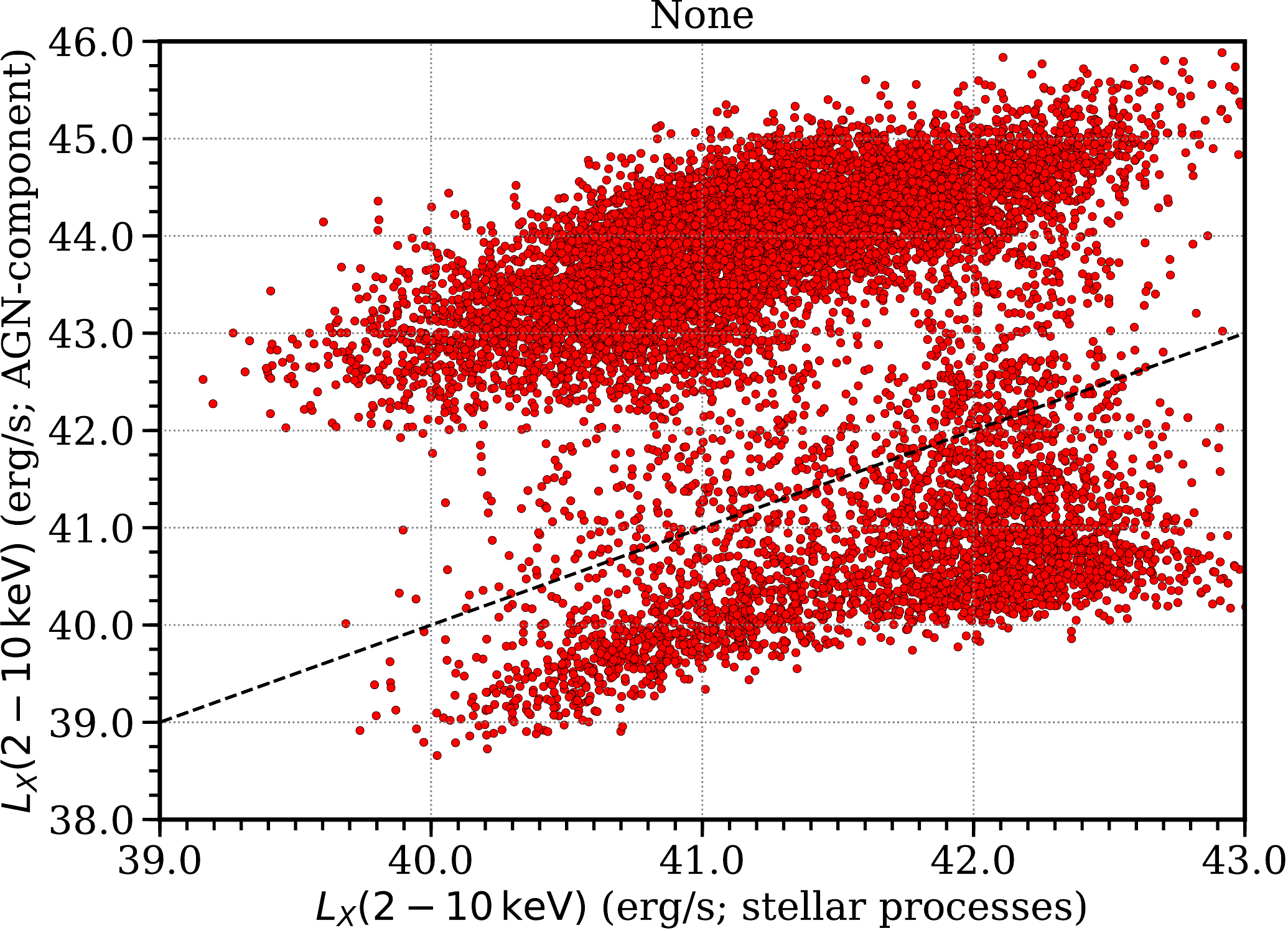}
\end{center}
\caption{X-ray AGN luminosity of the mock R75 WISE AGN as a function of the expected X-ray luminosity from stellar processes. The latter is estimated using the \protect\citet{Lehmer2016} empirical relation between X-ray luminosity, stellar mass, star-formation rate and redshift. The dashed black line shows the one-to-one luminosity relation. The majority of low luminosity AGN in the  R75 selection wedge of \protect\cite{Assef2013} lie below the dashed line. Their expected X-ray emission from stellar processes (mostly star-formation) exceeds the luminosity produced by the accretion flow onto the central supermassive black hole.}\label{fig:AGNvsSFR}
\end{figure}

The model also predicts a sizeable contribution (about 30\%) of low luminosity AGN, $L_X(\rm 2-10\,keV)<10^{42} \, erg \, s^{-1}$ to the low-redshift peak of Figure \ref{fig:wise-zdist}. These systems, mostly star-forming galaxies, scatter into the R75 selection wedge as a result of the photometric uncertainties affecting their colours. If we had not added photometric noise to the model $W1$, $W2$ magnitudes, these sources would not have made it into the R75 wedge. They are therefore contamination to the R75 AGN selection. Figure \ref{fig:wise-lxz} shows that the typical AGN luminosities of these sources are $L_X(\rm 2-10\,keV)\approx10^{40} \, erg \, s^{-1}$, i.e. comparable to the X-ray emission that stellar processes in galaxies can produce \citep[e.g. low and high-mass X-ray binaries,][]{Tremmel2013}. Figure \ref{fig:AGNvsSFR} demonstrates this point by plotting for the mock galaxies in the WISE R75 selection wedge the expected X-ray luminosity from stellar processes based on the empirical relations of \citep[][]{Lehmer2016}.  The majority of low-luminosity AGN in the WISE R75 wedge are expected to have X-ray emission from stellar processes in excess of their assigned AGN radiative output. The AGN emission component in these mock galaxies is therefore subdominant and cannot explain their position on the $W_1-W_2$ vs $W_2$ colour-magnitude plane. Figure \ref{fig:wise-lxz} also reveals a second cloud of low luminosity AGN that contaminate the R75 wedge at redshifts $1\la z\la 2.5$. In terms of fraction relative to the total R75 population  at these redshifts however, this second cloud is not as significant (see Figure \ref{fig:wise-zdist}). The low-luminosity AGN population [$L_X(\rm 2-10\,keV)<10^{42} \, erg \, s^{-1}$] at all redshifts corresponds to about 20\% of the mock R75 sources. It is also worth highlighting in Figure \ref{fig:wise-zdist} the transition from Type-II AGN at $z<1$ to Type-I AGN at higher redshifts.   

An important observational constrain on the multi-wavelength properties of AGN is the overlap of samples selected independently at different parts of the electromagnetic spectrum. Of particular interest in that respect is the comparison between mid-infrared and X-ray selected AGN samples, since these wavelengths are believed to provide complementary views of the active black hole population. Observationally, it is shown that the level of overlap between X-ray and mid-infrared samples depends on the relative flux limits at the two wavelength regimes \cite[e.g.][]{Park2010, Donley2012, Mendez2013}. We choose to compare the empirical model predictions with the Stripe82X sample presented by \citet{LaMassa2019}. We find that 11\% (339/3325) and 20\% (651/3325) of the R75 WISE-selected AGN in that sample have X-ray counterparts in the hard (2-10\,keV) and soft (0.5-2\,keV) bands of the Stripe82X survey respectively (see Appendix \ref{app:stripe82x} for details).  This fraction should be compared with 17\% (2-10\,keV band) and 30\% (0.5-2\,keV band) X-ray overlap predicted by the empirical model. For this estimate the observationally derived X-ray sensitivity curve of the Stripe82X (Appendix \ref{app:stripe82x}) in the hard (2-10\,keV) and soft (0.5-2\,keV) bands have been applied to the model to select a mock X-ray sample that resembles the Stripe82X detection. It is then possible to estimate the fraction of R75 AGN in the model that have X-ray counterparts in the Stripe82X-like sample. The model predicts a factor of about 1.5 more X-ray/WISE AGN associations compared to the observations. This discrepancy suggests a population of X-ray--faint sources within the WISE R75 selection wedge that cannot be reproduced by the current parametrisation of the empirical model. These could be either intrinsically X-ray--faint \citep[e.g.][]{Martocchia2017} or heavily obscured AGN. We explore each of these possibilities separately. 

 A population of intrinsically X-ray faint AGN that was not included in our baseline empirical model would have a measurable effect on the predicted vs observed fraction of X-ray associations among unobscured (Type-I) AGN. In the Stripe82X survey, type-I AGN are defined as those with broad optical emission lines based on the visual classification of their SDSS optical spectra described in \citet{LaMassa2019}. The fraction of X-ray (0.5-2keV) associations for the type-I subsample of the R75 WISE selected AGN in the Stripe82X survey is 48\% (513/1058). This fraction should be compared with the empirical model prediction of 72\% (0.5-2\,keV band), i.e. a factor of 1.5 higher than the observations. This estimate assumes that type-I mock AGN are those with $\log N_H/\rm cm^{-2}<22$. Lowering the threshold to $\log N_H/\rm cm^{-2} <21.5$ \citep[e.g][]{Merloni2015} has no effect on the measured fraction. In the model type-I R75 WISE AGN are dominated by optically bright ($r\la22.5$\,mag) sources where the optical spectroscopic completeness of the SDSS-IV special-plate QSO programme has close to 100\% completenees. We therefore choose not to apply such corrections to the model predictions. This comparison shows that the difference in the fraction of X-ray/WISE associations between observation and model pertains to type-I AGN. This can be interpreted as evidence for a population of intrinsically X-ray faint AGN beyond the baseline model assumptions on the scatter in the $L_X({\rm 2-10\,keV})-\nu L_{\nu}(6\,\mu m)$ and $L_\nu (\rm 2\,keV)-L_\nu(\rm 2500\angstrom)$ correlations.


An alternative possibility for reducing the fraction of X-ray/WISE associations in the R75 wedge is to allow heavily obscured and hence X-ray faint AGN to be selected in the mid-infrared. The baseline model does predict a (small) fraction of Compton-thick (X-ray-faint) AGN within the WISE R75 selection wedge (see next section). The number of these sources however, has to be increased by a factor of 1.5 to accommodate the observed (low) fraction of X-ray/WISE associations.  This can be accomplished for example, if the parent population of Compton think AGN is increased by that factor compared to the baseline model assumption, $\beta_{Thick}=34$\%. 

\begin{figure}
\begin{center}
\includegraphics[height=0.95\columnwidth]{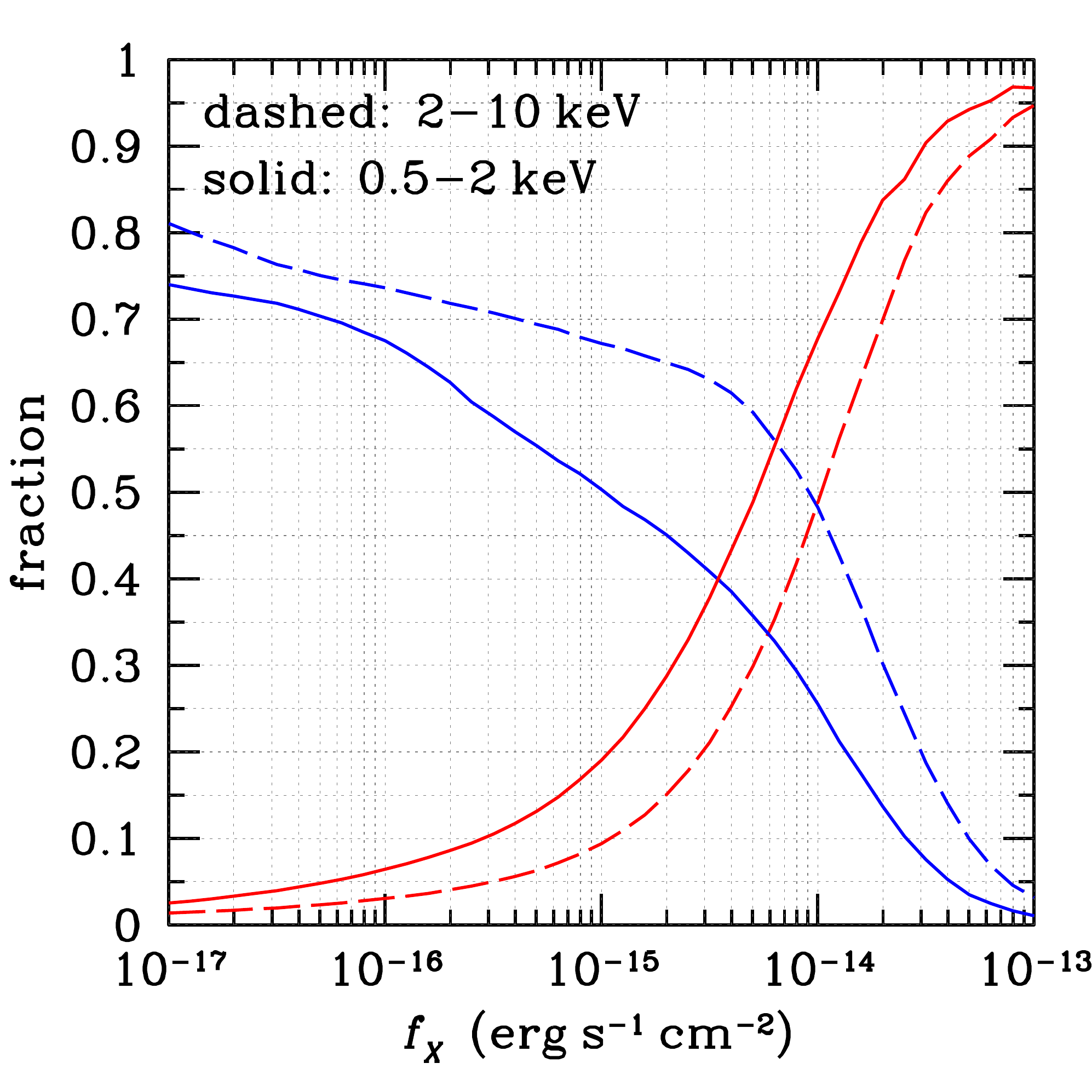}
\end{center}
\caption{Fraction of X-ray/WISE associations as a function of flux limit in the hard (2-10\,keV; dashed curves) and soft (0.5-2\,keV; solid line) energy bands. The blue set of curves correspond to the fraction of X-ray sources associated with WISE R75 AGN above a given flux limit. This fraction increases with decreasing X-ray flux.  The red set of curves show the fraction of R75 WISE sources among the X-ray selected AGN at a given flux limit. The model predicts that nearly 100\% of the X-ray AGN lie in the WISE R75 wedge at bright fluxes, $f_X \ga 10^{-12}\rm \, erg \, s^{-1} \, cm^{-2}$.}\label{fig:wise-xray}
\end{figure}

Despite the issues discussed above we use the baseline parametrisation of the empirical model to predict the fraction of R75 WISE sources that have X-ray counterparts above a given flux limit, as well as the fraction of X-ray sources at a given limit that lie within the R75 wedge. These curves are plotted as a function of X-ray flux in Figure  \ref{fig:wise-xray}. Even at very faint X-ray fluxes not all the R75 WISE AGN have X-ray counterparts. This is related to contamination of the R75 wedge by non-AGN or low luminosity AGN. Also, the fraction of X-ray selected AGN that lie within the R75 wedge decreases rapidly from about 100\% at $f_X(\rm 0.5-2keV ) > 3 \times 10^{-14} \, erg \, s^{-1} \, cm^{-2}$ to about 5\%  at $f_X(\rm 0.5-2keV ) > 10^{-16} \, erg \, s^{-1} \, cm^{-2}$. 

\subsection{Inferred properties of the R75 WISE selected AGN}

In this section the baseline parametrisation of the empirical model is used to explore the intrinsic properties of the R75 WISE mock AGN population, including line-of-sight obscuration, Eddington ratios and host galaxy characteristics. 

\begin{figure}
\begin{center}
\includegraphics[height=0.65\columnwidth]{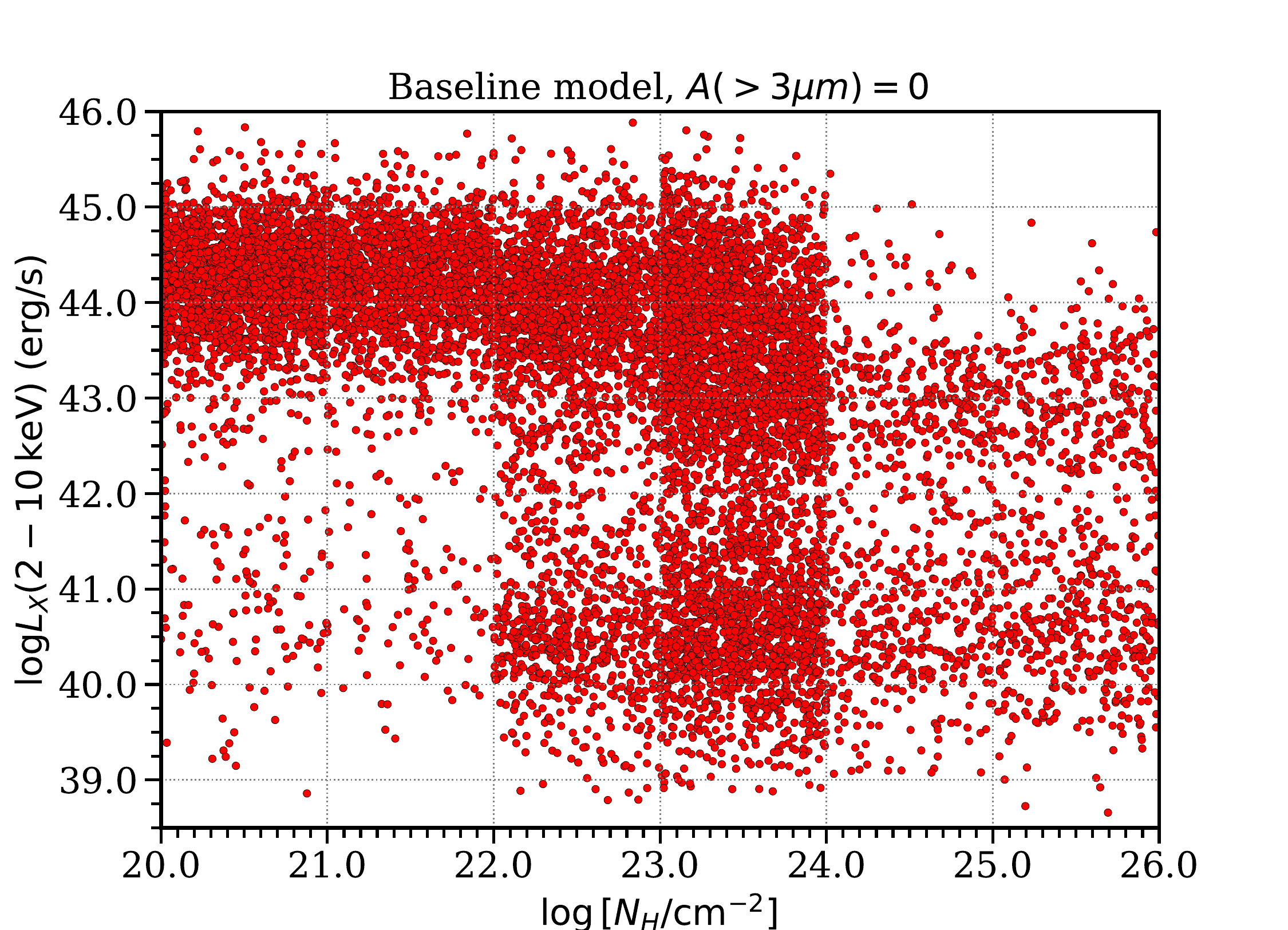}
\includegraphics[height=0.65\columnwidth]{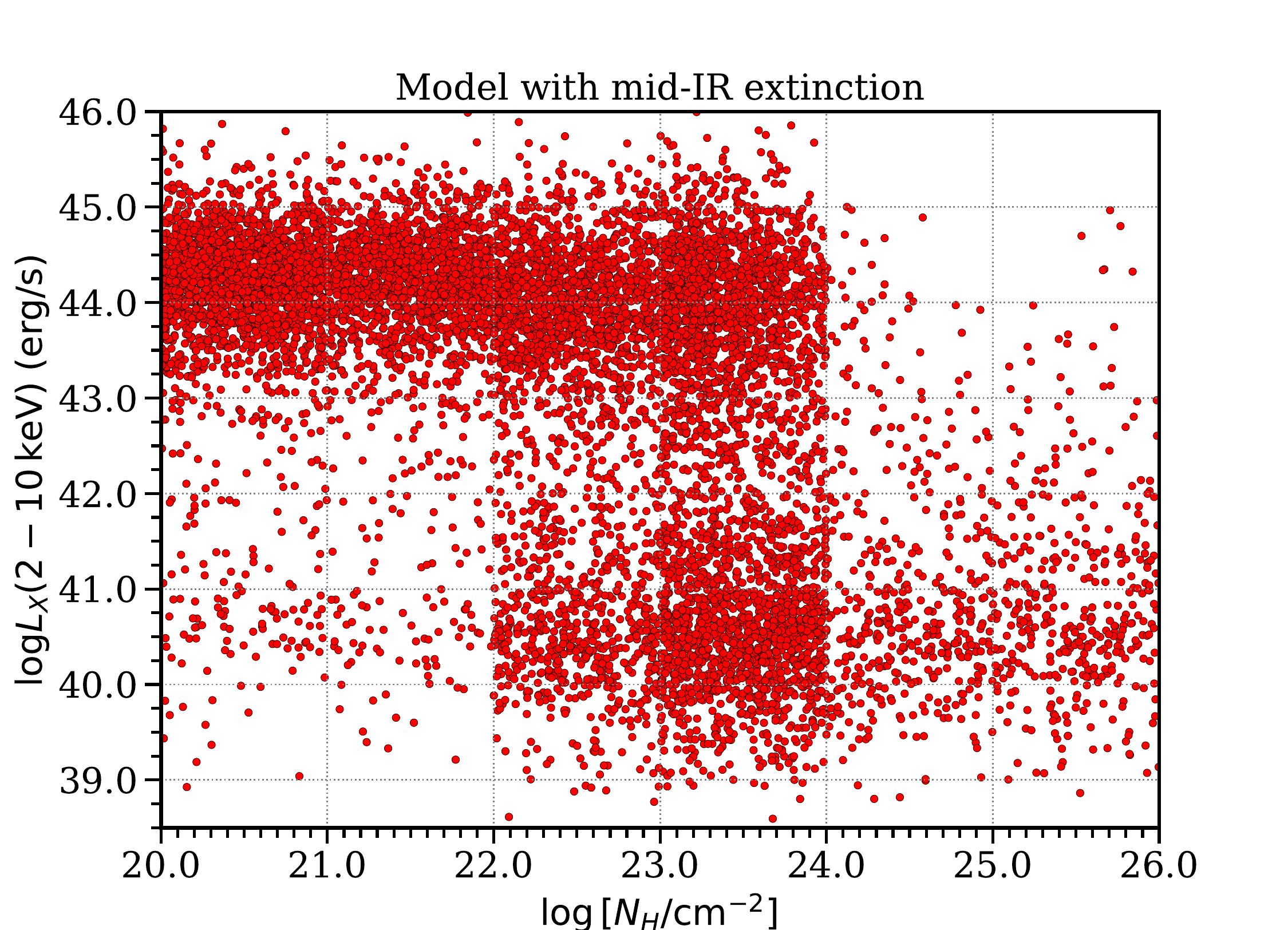}
\includegraphics[height=0.65\columnwidth]{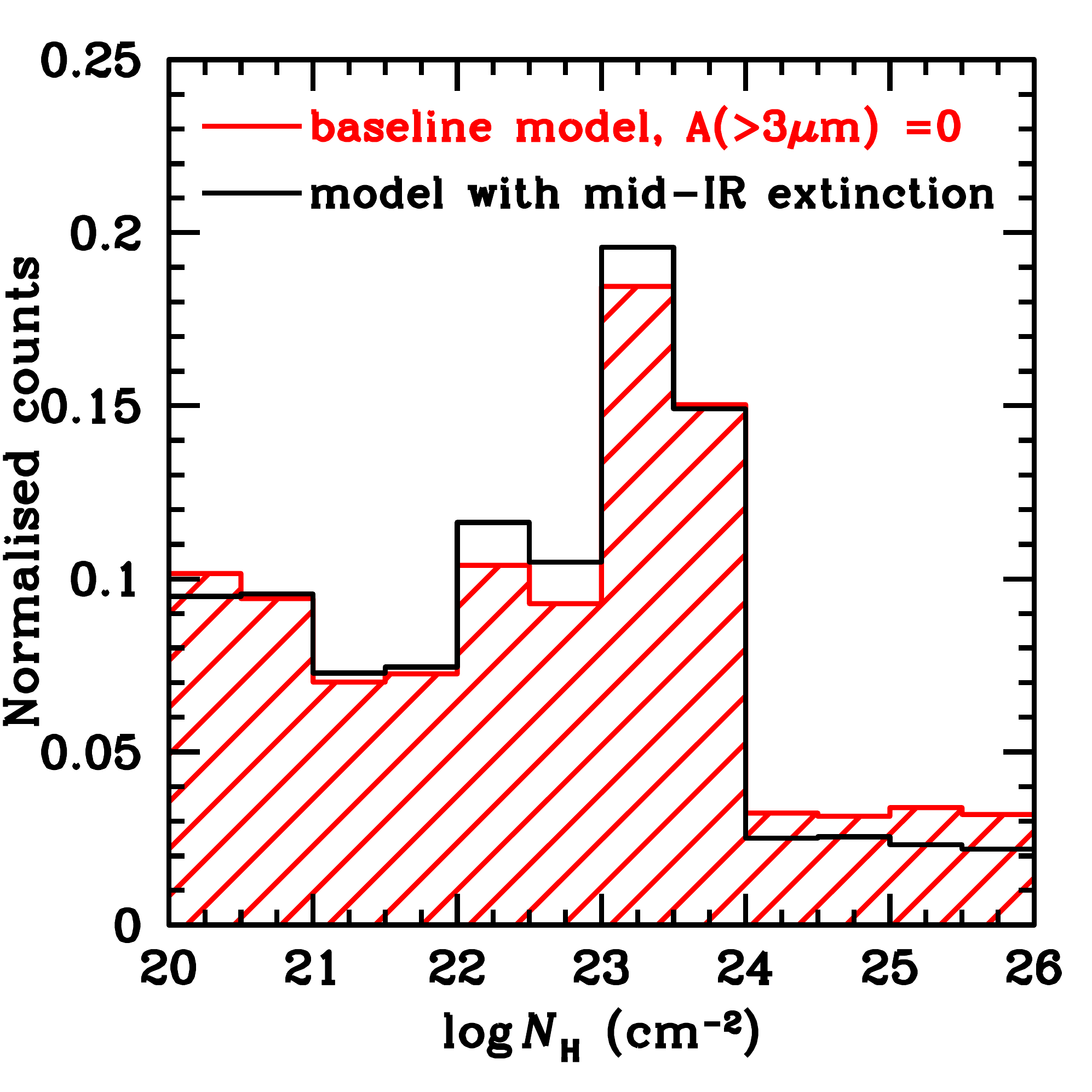}
\end{center}
\caption{The {\bf top panel} shows the distribution of mock WISE R75 sources in the 2-dimensional space of hydrogen column density and intrinsic X-ray luminosity in the 2-10\,keV band for the baseline model with $A(\rm > 3\mu m ) =0$. Each dot on this diagram corresponds to a source in the model that fulfills the WISE R75 selection criteria. The {\bf middle panels} corresponds the same parameter space but in the model version that extrapolates the extinction curve to the mid-infrared.  The  {\bf bottom panel} compares the corresponding 1-dimensional $N_H$ distribution of the two models by collapsing the top and middle diagrams along the luminosity axis.}\label{fig:wise-nhdist}
\end{figure}

A motivation for the development of the empirical model presented in this work is to explore the incidence of obscured sources among the WISE mid-infrared selected AGN population. The top panel of Figure \ref{fig:wise-nhdist} plots for the baseline model ($A_\lambda=0$ for $\lambda>3\mu m$) the distribution of WISE R75 AGN on the $L_X$ vs $N_H$ plane. There is a sharp drop in the density of sources at $N_H\rm >10^{24}\, cm^{-2}$. This is because the WISE $W1$-band corresponds to the rest-frame mid-infrared ($2.2\mu m$) at redhifts $z\ga0.7$. At these rest-frame wavelengths the dust extinction is substantial in the case of Compton thick levels of obscuration [$A(2.2\mu m )/E(B-V) = 0.35$]. In our model these sources are therefore expected to have mid-infrared colours/magnitudes dominated by the host galaxy and hence, lie outside the R75 AGN selection wedge and/or fainter that the WISE limits. The paucity of sources at $N_H\rm >10^{24}\, cm^{-2}$ is more striking for the version of the semi-empirical model in which the extinction curve is extrapolated to the mid-infrared. This is demonstrated in the middle panel of Figure \ref{fig:wise-nhdist}. In this case the extinction in the $W1$, $W2$ bands is large at all redshifts for Compton thick levels of obscuration. The sharp cut in the number of sources with $N_H\rm > 10^{24}\, cm^{-2}$ for both model versions is further demonstrated in the bottom panel of Figure \ref{fig:wise-nhdist}. It plots the $N_H$ distribution of the baseline model vs the one with mid-infrared extinction. We caution that the low number density of Compton thick AGN predicted by the model is a direct consequence of the choice to link the hydrogen column density $N_H$, which affects the X-ray spectrum, with the extinction at longer wavelengths via the linear relation $N_H/E(B-V)=6\times10^{22}\rm cm^{-2} \, mag^{-1}$. Relaxing this requirement by e.g. introducing (substantial) scatter, will impact the detectability of heavily obscured AGN in the WISE bands. A feature  of the top and middle panels of Figure \ref{fig:wise-nhdist} is the large number of low luminosity AGN [$L_X(\rm 2-10\,keV)<10^{42} \, erg \, s^{-1}$] that extend into the Compton thick regime. These are contaminating sources that spuriously enter the R75 selection wedge as a result of photometric uncertainties affecting their colours/magnitudes. The obscuration therefore does not play any role on whether these sources lie in the R75 selection region. The fact that there are less Type-I sources among the contaminating low luminosity AGN population simply reflects the low fraction of such systems at the faint-end of the X-ray luminosity function of \citet{Aird2015}.

\begin{figure*}
\begin{center}
\includegraphics[height=0.85\columnwidth]{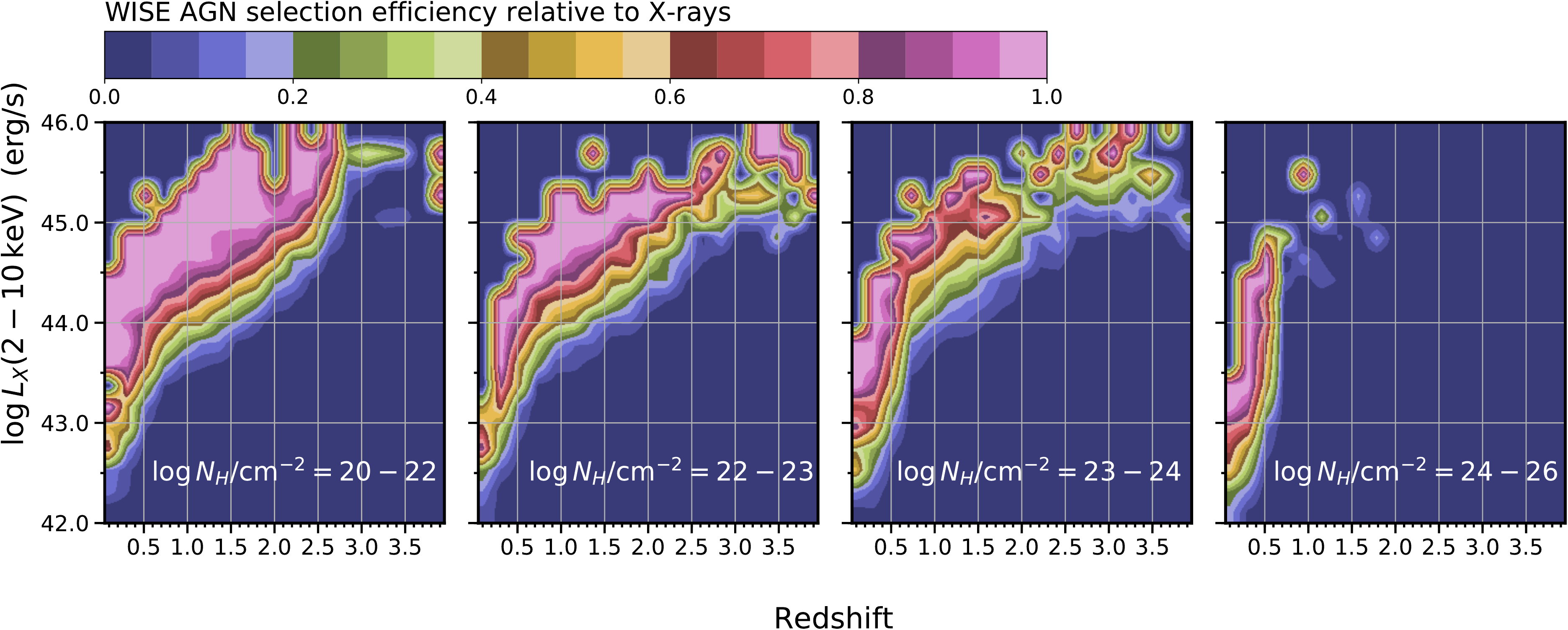}
\end{center}
\caption{Completeness of the WISE R75 AGN selection in the 2-dimensional space of X-ray luminosity (2-10\,keV band) and redshift. The completeness is defined as the ratio of the number AGN in bins of $L_X$, $z$ and $N_H$ that lie within the WISE R75 selection wedge and the total number of AGN in the simulation within the same $L_X$, $z$ and $N_H$ bins. Each panel corresponds to a different hydrogen column density interval $\log N_H/\rm cm^{-2}=20-22$, $22-23$, $23-24$ and $24-26$. The colours correspond to different levels of completeness fractions as indicated in the colour bar.}\label{fig:wise-lxznh}
\end{figure*}

Next we explore the WISE R75 AGN selection function and its dependence on X-ray luminosity, redshift and hydrogen column density. Figure \ref{fig:wise-lxznh} shows the distribution of the R75 completeness on the X-ray luminosity vs redshift plane. Each panel corresponds to a different hydrogen column density interval $N_H=20-22$, $22-23$, $23-24$ and $\rm 24-26\,cm^{-2}$. The completeness is defined as the fraction of the total AGN population at a given $z$, $L_X$ and $N_H$ bin that lies within the R75 selection wedge. The somewhat erratic behaviour of the upper bound (bright $L_X$) of the completeness regions in Figure \ref{fig:wise-lxznh} is because of shot noise associated with the low number of AGN at such bright luminosities predicted by the X-ray luminosity function model. Figure \ref{fig:wise-lxznh} shows that the WISE R75 completeness is a complex function of redshift and AGN physical parameters. A general remark is that the R75 selection is sensitive to powerful AGN with typical luminosities $L_X (\rm 2 - 10\,keV) \ga 10^{44}\, erg \, s^{-1}$. In detail however, the completeness at fixed luminosity decreases with either increasing column density or increasing redshift. It is nevertheless possible to identify regions of the parameter space that are highly complete (>80\%) to high column densities, $N_H\rm \approx 10^{24} \, cm^{-2}$, e.g. $\log L_X(2-10)>44.5$ (erg/s) and $z<1$. In the case of Compton thick AGN the baseline model predicts that the vast majority lie at low redshift, $z\la0.7$. This is the limit where the WISE $W1$-band moves into to the near-infrared ($\rm \approx 2.2\mu m$) and the observed AGN radiation is affected by dust extinction. The low-redshift Compton thick AGN ($N_H \rm > 10^{24} \, cm^{-2}$, $L_X(\rm 2-10\,keV)>10^{42}\, erg \, s^{-1}$) predicted by the baseline model are optically bright (median $r \approx 21$\,mag) and have a sky number density of about $\rm 12\,deg^{-2}$. This population of Compton thick AGN is essentially absent from the  model version in which the extinction law is extrapolated to the rest-frame mid-infrared.

\begin{figure}
\begin{center}
\includegraphics[height=0.95\columnwidth]{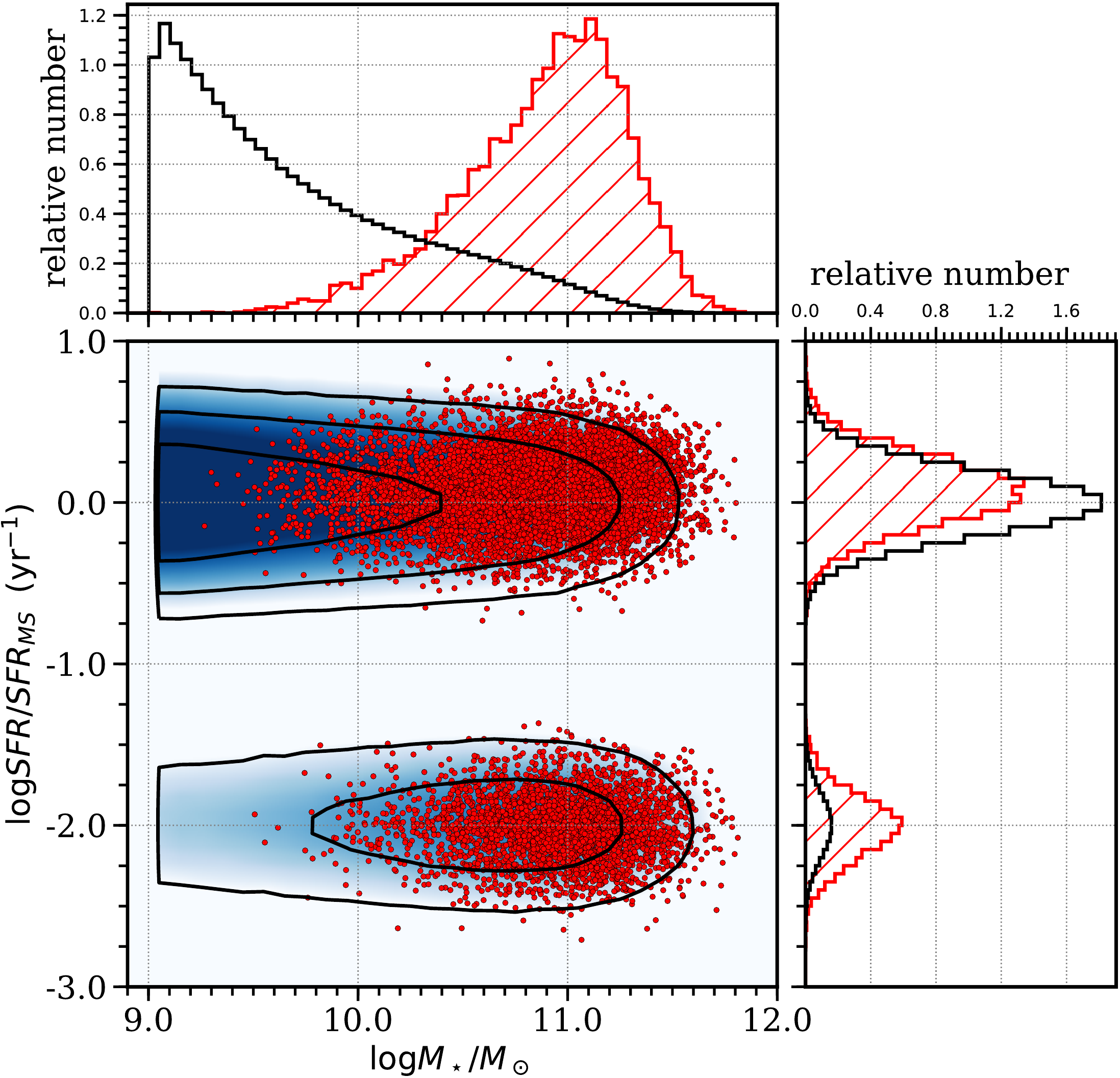}
\end{center}
\caption{Distribution of mock WISE R75 sources in the 2-dimensional space of  stellar mass and normalised specific star-formation rate in comparison to the overall galaxy population in the simulation. For a given stellar mass and redshift the sSFR is normalised to the Main Sequence value defined by \protect\cite{Schreiber2015}. Star-forming galaxies are scattered around a mean normalised sSFR of zero. Passive galaxies are offset to 2\,dex below the main sequence. The contours and blue-shaded regions show the galaxy distribution.  Darker colours correspond to a higher density of sources. The contours enclose 68, 95 and 99.7 per cent of the population. The red dots are mock galaxies within the WISE R75 selection wedge. The histograms above and on the right of the central panel show 1-dimensional slices through the parameter space of stellar mass and normalised star-formation rate. The black solid-line is for the overall galaxy population. The red-hatched histogram corresponds to WISE R75 selected sources.}\label{fig:sSFR_Mstar}
\end{figure}

Finally we explore the physical properties of the WISE selected AGN predicted by the mock to get insights into the type of galaxies that host such systems. Figure \ref{fig:sSFR_Mstar} plots the distribution of R75 WISE AGN on the specific SFR (sSFR) vs stellar mass plane. The sSSFR is normalised to the main sequence expectation for the redshift and stellar mass of each galaxy based on the empirical relation of \citealt{Schreiber2015}. For comparison also plotted in Figure \ref{fig:sSFR_Mstar} is the overall galaxy population in the mock. The R75 WISE AGN are skewed to massive galaxies (distribution peaks at $10^{11}\,M_{\odot}$) relative to the general galaxy population. This is a selection bias. At fixed specific accretion rate, AGN in less massive galaxies are not sufficiently bright to make it above the R75 flux limits. There is clear division between star-forming and passive galaxies in Figure \ref{fig:sSFR_Mstar}, which is imposed by our modeling assumptions. Also by construction AGN are drawn from the general galaxy population and therefore share the star-formation properties of this parent sample. The apparent higher fraction of passive galaxies among the R75 WISE AGN sources relative to galaxies in Figure \ref{fig:sSFR_Mstar} is because of the different mass distribution of the two populations. Therefore in our modeling the R75 WISE AGN are not skewed to star-forming galaxies. However, star-forming galaxies is the main source of contamination, especially at low redshift, $z<0.5$.

\begin{figure}
\begin{center}
\includegraphics[height=0.95\columnwidth]{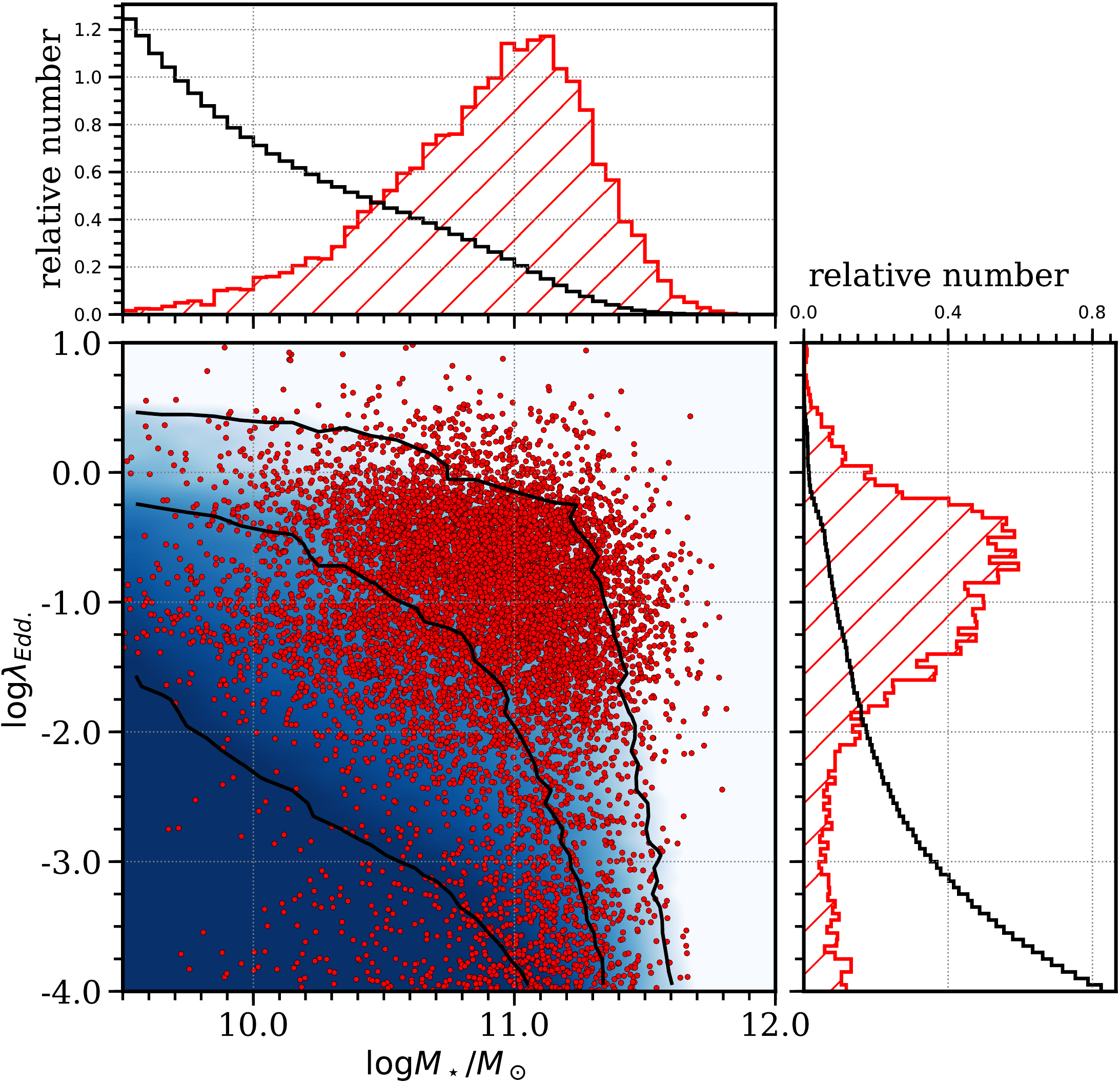}
\end{center}
\caption{Distribution of mock WISE R75 sources in the 2-dimensional space of  stellar mass and specific accretion rate in comparison to the overall galaxy population in the simulation. The specific accretion rate ($\lambda = L_X/M_{*}$) is converted to Eddington ratio via Equation \ref{eq:lambda}. The contours and blue-shaded regions show the galaxy distribution.  Darker colours correspond to a higher density of sources. The contours enclose 68, 95 and 99.7 per cent of the population. The red dots are mock galaxies within the WISE R75 selection wedge. The histograms above and on the right of the central panel show 1-dimensional slices through the parameter space of stellar mass and specific accretion rate. The black solid-line is for the overall galaxy population. The red-hatched histogram corresponds to WISE R75 selected sources.}\label{fig:sBHAR_Mstar}
\end{figure}

Figure \ref{fig:sBHAR_Mstar} plots the distribution of the R75 WISE AGN on the specific accretion rate (defined in Equation \ref{eq:lambda}) vs stellar mass plane. There a clear preference for high-specific accretion rates, $\log \lambda_{Edd}\approx -0.5$. Also evident in this figure is the tail extending to very low specific accretion rates. This corresponds to the contamination of the R75 selection wedge by star-forming galaxies.

\section{Discussion}

A forward-modeling approach is presented to explore the multiwavelength properties of AGN and study the selection function of samples identified at different parts of the electromagnetic spectrum. The starting point of the model development is the X-ray luminosity function of AGN and the assumption that X-rays provide an unbiased view of the active black-hole population at both low luminosities and high levels of obscurations.  It is further assumed that accretion events populate galaxies in a probabilistic way with duty cycles that depend on the specific accretion rate (accretion luminosity normalised by stellar mass). These basic assumptions allow seeding the stellar mass function of galaxies with AGN X-ray luminosities. Empirical relations are then used to associate the physical properties of the mock AGN (e.g. accretion luminosity) and galaxies (e.g. stellar mass) with spectral energy distributions and hence construct the multi-wavelength properties of the population. The various assumptions adopted to build the model can then be tested by comparing the model predictions with the observed properties of AGN at different parts of the electromagnetic spectrum. 

The predictive power of this generic approach is demonstrated on the specific problem of the selection of AGN in the mid-infrared using photometric observations from the WISE mission. Among the different criteria proposed in the literature to identify AGN within the WISE photometric bands we choose to test the R75 selection of \citet{Assef2013} because of the rich set of supporting observations available for this class of sources \citep[e.g.][]{LaMassa2019}. For this specific exercise the model can reproduce two key observational properties of the population, the bimodal optical magnitude distribution in the $r$-band and the prominent peak of the redshift distribution at $z<0.5$. It is then possible to use the model to get insights on the origin of these observational results. 

The bimodal optical magnitude distribution is the result of the transition from type-I (unobscured) AGN at bright magnitudes to type-II (obscured) sources at faint fluxes. For the latter population the observed optical light is predominantly stellar emission from the host galaxy.  The clear distinction between obscured and unobscured WISE AGN in their optical properties, as opposed to e.g. X-ray AGN (see Figures \ref{fig:xxl}, \ref{fig:cosmos-soft} and \ref{fig:cosmos-hard}), is because of the relatively bright accretion luminosities of the selected systems. The R75 WISE sample is dominated by AGN with $L_X \ga 10^{44} \rm \, erg \, s^{-1}$ (e.g. see Figures \ref{fig:wise-lxz}, \ref{fig:wise-lxznh}).  This results in a clear separation between type-I AGN, which appear optically bright, and obscured, for which only  stellar emission is observed. In contrast, X-ray selected AGN samples include a large fraction of moderate and low-luminosity AGN, with the net effect being a smooth optical magnitude distribution with no obvious distinction between Type-Is and Type-IIs.  

A prediction of the model is the high level of contamination of the WISE R75 selection wedge by star-forming galaxies at low redshifts. This is manifested by the prominent peak in the redshift distribution of the population at $z<0.5$ (see Figure \ref{fig:wise-zdist}), which is also present in the sample of \citet{LaMassa2019}. In the model about 35\% of the R75-wedge sources at $z<0.5$ are associated with star-forming galaxies that host low-luminosity AGN, $L_X<<10^{42} \rm \, erg \, s^{-1}$.  This fraction should be compared with the optical spectroscopic analysis of \citet{LaMassa2019} that showed that about 50\% of the $z<0.5$ WISE sources within the R75 wedge have line ratios consistent with those of star-forming galaxies. From the remaining half, about 2/3 are classified as AGN/star-forming composites and 1/3 have line ratios typical of Type-II Seyferts. In our empirical model, the source of contamination is photometric uncertainties affecting the $W1-W2$ colours of star-forming galaxies thus, making them scatter into the R75 selection wedge. Because of the large number of star-forming galaxies within the WISE population, there is a large pool of sources that can potentially scatter into the wedge. Switching off the photometric incertaities in the model eliminates this source of contamination. The redshift distribution of the contaminating population, which is skewed to $z<0.5$, is related to relatively bright magnitude cut of the WISE sample. 

The empirical model presented in this work produces a higher fraction (factor of 1.5) of X-ray associations among the WISE R75 AGN population compared to observations. This inconsistency questions some of the assumptions on which the model is built and suggests a relatively large population of apparently X-ray weak AGN within the WISE R75 wedge. This can be achieved either by allowing a fraction of intrinsically X-ray faint active black holes or by increasing the fraction of Compton thick (and hence apparently X-ray faint) AGN in the WISE R75 selection wedge.

The first option is supported by the fact that the observed fraction of X-ray associations within the WISE broad-line AGN population (i.e. Type-Is) is also lower than the empirical model prediction.  The discrepancy between model and observations for the type-I class of sources is similar to that for the total population, i.e. factor of 1.5. This suggests that the disagreement is related to the scatter and overall shape of the $L_X-L_{6\mu m}$ relation. There is indeed increasing evidence for significant deviations from linearity for the most powerful QSOs, in the sense that the X-ray luminosity increases slower than the mid-infrared one \citep{Stern2015, Chen2017, Martocchia2017}. A non-linear $L_X-L_{6\mu m}$ relation is adopted in this work and therefore the X-ray weakness of AGN at bright accretion luminosities is already accounted for in the analysis. This leaves the possibility of increasing the fraction of intrinsically X-ray faint AGN by increasing the scatter of the  $L_X-L_{6\mu m}$ relation. It is found that a scatter of $\approx 0.8$ could reduce the fraction of X-ray associations in the WISE R75 wedge close to the observed one, at least in the case of the Stripe82X field. This is however, much larger than the value of 0.3-0.4 typically quoted in the literature \citep[e.g.][]{Mateos2015}, although that does not exclude the possibility of a population of intrinsically faint AGN that are underrepresented in current X-ray flux-limited samples.  Such a class of sources are for example, the Broad Absorption Line (BAL) QSOs that represent about 15\%  of optically selected quasars. They are believed to include intrinsically X-ray weak systems relative to their UV luminosity \citep[e.g.][]{Luo2014, Kollatschny2016}, and are also likely to be outliers in the $L_X-L_{6\mu m}$ relation \citep[e.g.][]{DelMoro2016}. Another class of sources that often appear X-ray weak relative to their UV luminosity  as a result of extreme X-ray variability are high Eddington-ratio sources, such as narrow-line Seyferts 1 \citep{Miniutti2012, Liu2019}. It is unclear however, whether the fraction of super-Eddington AGN or BAL QSOs within the ovarall population is sufficiently high to substantially increase the scatter in the $L_X-L_{6\mu m}$ relation. Nevertheless, WISE selected AGN are likely to include a non-negligible fraction of high-Eddington ratio sources, e.g. Figure \ref{fig:sBHAR_Mstar}. A detailed investigation of the X-ray properties of broad-line (i.e. not type-2) WISE selected AGN can provide constraints on their $L_X-L_{6\mu m}$ and $L_\nu ({\rm 2\,keV})-L_\nu({\rm 2500\angstrom})$ relations.

\begin{figure}
\begin{center}
\includegraphics[height=0.9\columnwidth]{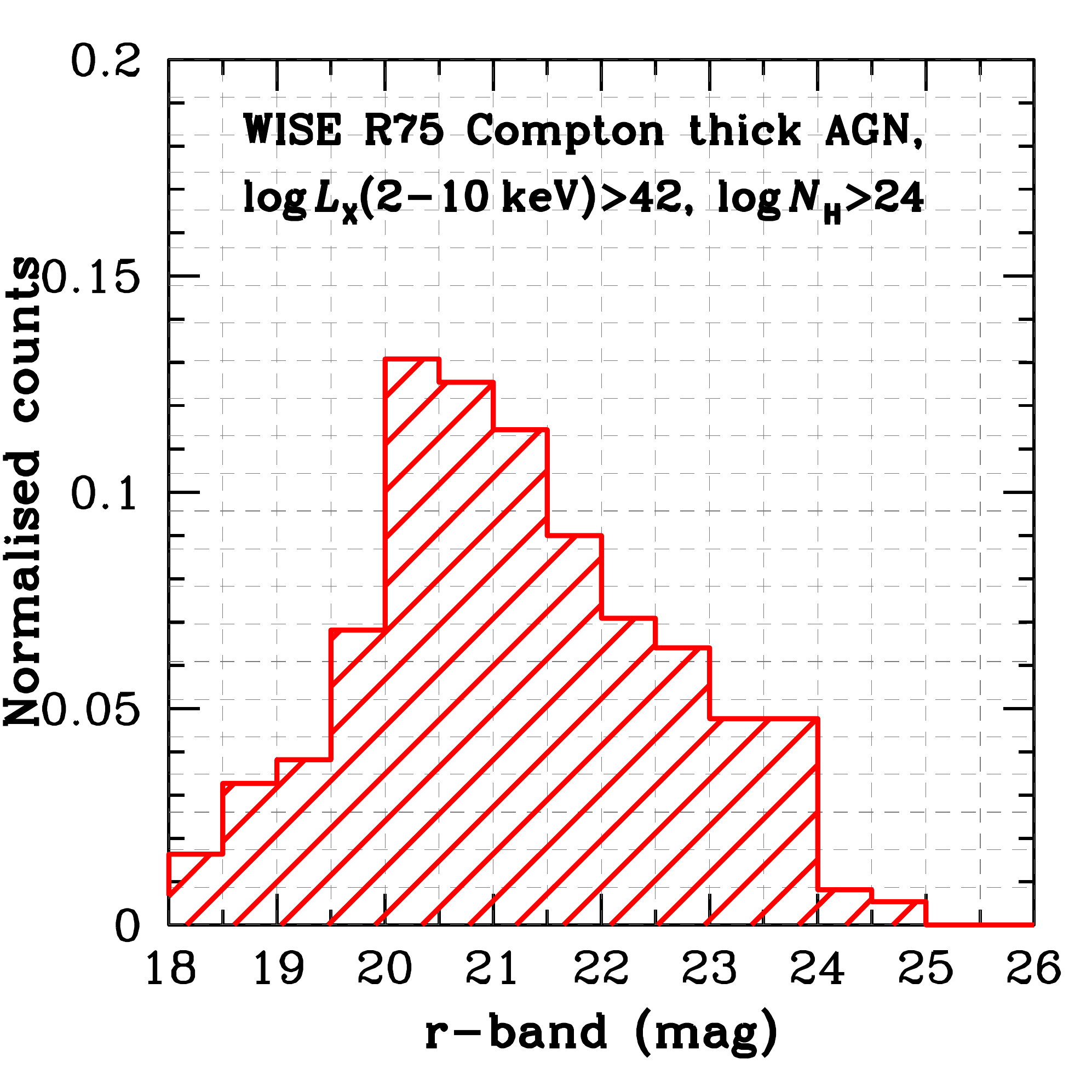}
\includegraphics[height=0.9\columnwidth]{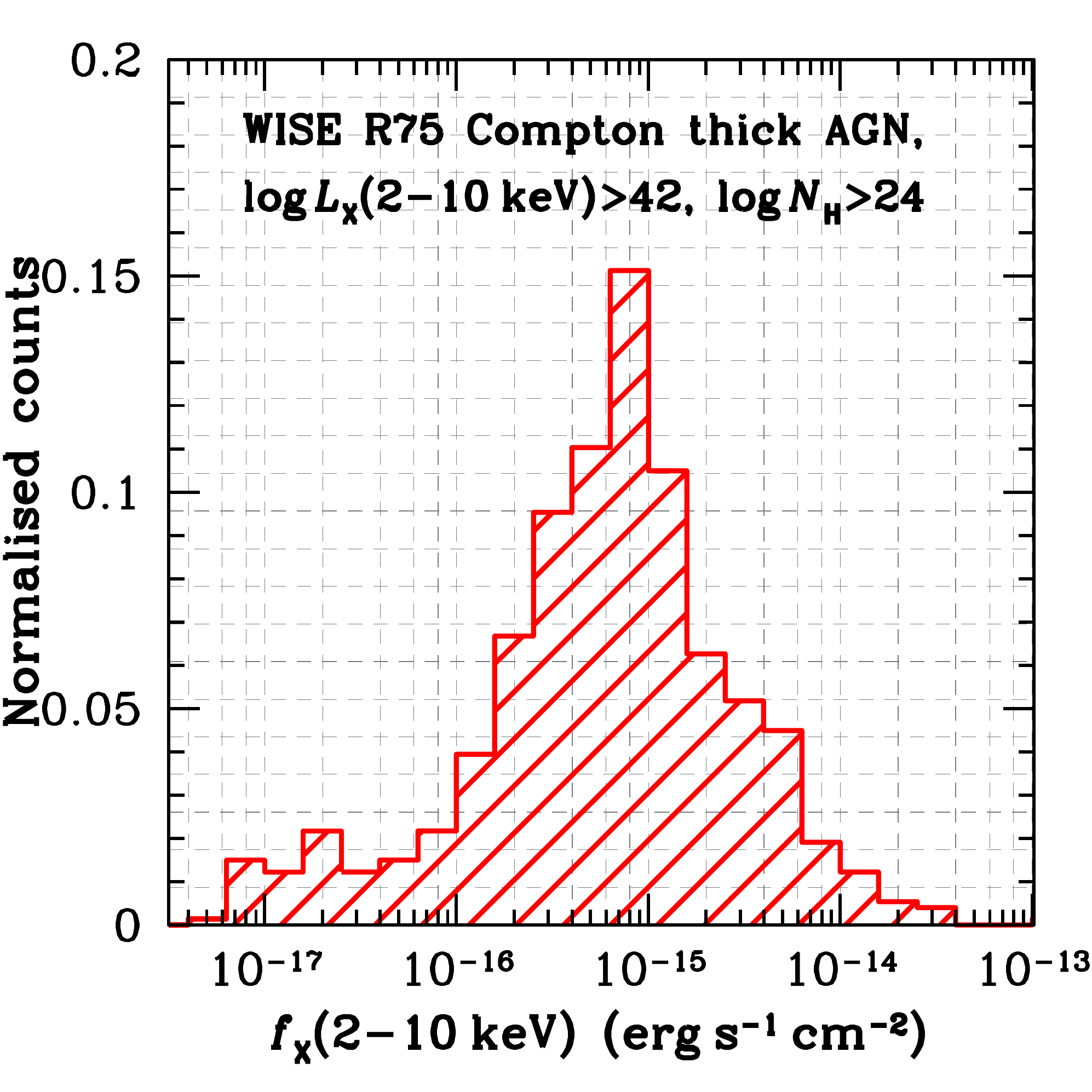}
\end{center}
\caption{Baseline model prediction for the $r$-band (top panel) and 2-10\,keV flux (bottom panel) distributions of Compton thick AGN that lie within the WISE R75 wedge and have X-ray luminosity $L_X(\rm 2-10\,keV) > 10^{42} \, erg \, s^{-1} \, cm^{-2}$.}\label{fig:CT}
\end{figure}

An alternative possibility to reduce the fraction of X-ray sources in the WISE R75 wedge is to allow a large fraction of heavily obscured AGN. This can be achieved for example, by increasing the fraction of Compton thick AGN in the X-ray luminosity function above the current assumption of 34\%, or by increasing the scatter in the relation that links X-ray obscuration to optical extinction beyond the adopted value of 0.5\,dex. Relaxing the above model assumptions would allow heavily obscured and hence, X-ray faint, sources to be selected by the WISE R75 criteria. There is indeed evidence for a potentially large population of heavily obscured, possibly Compton Thick AGN, among  the WISE population \citep[e.g.][]{Assef2015,Mountrichas2017, Yan2019}. The SDSS spectroscopic follow-up programme presented  by \cite{LaMassa2019} also revealed a non-negligible number of WISE R75 sources that are optically faint ($r\ga22$\,mag), lie at redshifts $z\la1$ and are spectroscopically identified by their prominent [OII]\,3727 emission lines. These sources are prime candidates for heavily obscured AGN. Our baseline model predicts that the most heavily obscured, Compton thick, WISE AGN are at low redshift, $z\la0.6$ (see Fig. \ref{fig:wise-lxznh}) and relatively optically bright. The top panel of Fig. \ref{fig:CT} demonstrates the latter point by plotting the $r$-band distribution of the Compton thick AGN population predicted by the model. Observationally, the identification of such AGN needs to account for the relatively high level of contamination of the WISE R75 AGN selection by star-forming galaxies at redshifts $z\la0.6$ (s1ee Fig. \ref{fig:wise-zdist}). One approach to achieve this is via diagnostic optical emission-line ratios \citep[e.g.][]{Kewley2001} to separate star-forming galaxies from Seyfert-2s. Observations at hard X-rays can also provide useful information on the nuclear activity of a galaxy and the level of line-of-sight obscuration. The bottom panel of Figure \ref{fig:CT} shows the expected 2-10\,keV X-ray flux distribution of the Compton thick AGN predicted by the model. The expected fluxes have already been reached in deep X-ray survey fields, e.g. COSMOS-Legacy \citep{Civano2016}. Study of the X-ray spectral properties of WISE selected AGN in such fields can test the baseline model predictions for the demographics of Compton thick AGN in the WISE R75 wedge.

\section{Acknowledgements}
The authors thank the anonymous referee for their careful reading of the paper and their insightful comments. 

\section{Data availability }
The data underlying this article are available in Zenodo, at https://zenodo.org/record/3906341





\bibliographystyle{mnras}
\bibliography{mybib} 

\appendix 

\section{Derivation of Specific Accretion-Rate Distributions}\label{sec:plz}

The determination of specific accretion rate distribution of AGN can be treated as a deconvolution problem. Given a set of measurements of the AGN luminosity function [$\phi (L_\mathrm{X},  z)$] at different redshifts, $z$,  and knowledge of the stellar mass function of galaxies [$\psi(M, z)$], one can recover the specific accretion rate distribution [$P(\lambda, z)$] via the relation 

\begin{equation}\label{eq:lambda2phi}  
\phi(L_\mathrm{X},  z)  =  \int \psi(M, z) \, P(\lambda, z) \, \mathrm{d} \log M,
\end{equation}

\noindent where $M$ is the stellar mass of galaxies, $L_\mathrm{X}$ is the X-ray luminosity produced by the accretion event. The AGN luminosity function in the equation above is represented by measurements at X-ray wavelengths. The $P(\lambda,  z)$  is  a probability density function, i.e., integrates to unity at a given redshift 

\begin{equation}\label{eq:plznorm}  
\int_{\lambda_{min}}^{\lambda_{max}}  P(\lambda, z) \, \mathrm{d}  \log \lambda =1,
\end{equation}

\noindent  where $\lambda_{min}$, $\lambda_{max}$ is the range of specific accretion-rate distributions within which the  $P(\lambda, z)$ is defined. The quantity $\lambda$ is a scaled version of the specific accretion rate designed to resemble the Eddington ratio under the assumptions that the black hole mass is related to the stellar  mass of the AGN  host galaxy and the X-ray luminosity is proxy of the bolometric AGN luminosity. It is emphasized that this scaling is done to provide a qualitative link to the Eddington ratio and help visualise the overall shape of the $P(\lambda, z)$ relative to the Eddington limit \citep[e.g.][]{Georgakakis2017_plz}. It has no impact on the generation of mocks because the quantity used is the X-ray specific accretion rate, $L_X/M_{*}$, which is inferred observationally independent of any bolometric corrections or black-hole scaling relations. For simplicity we adopt a redshift-independent scaling relation  between black  hole mass and  stellar mass,  $M_{BH} = 0.002  M$  \citep{Marconi_Hunt2003}. The scatter in  this relation is ignored. A  single X-ray  bolometric conversion factor is further adopted, $L_{bol}  = 25 \, L_X(\rm 2 - 10 \,  keV)$ \citep{Elvis1994}.   Under these  assumptions the specific accretion rate is estimated as

\begin{equation}\label{eq:lambda}
\lambda = \frac{25 \,\,  L_X(\rm 2-10\,keV)}{ 1.26 \times 10^{38} \,\, 0.002  \, M}.
\end{equation}

\noindent For the stellar mass function of galaxies in Equation \ref{eq:lambda2phi} we adopt the parametrisation of \cite{Ilbert2013}. They use two Schechter functions \citep{Press1974} with parameters  evolving with redshift to represent the total mass function of galaxies in different redshift slices between $z=0$ and $z=4$. The $\psi(M, z)$ at any given redshift is determined by interpolating the corresponding mass functions in neighboring redshift bins. In Equation \ref{eq:lambda2phi} the X-ray luminosity function is represented by the point estimates of \citet{Aird2015}. They determine the obscuration-corrected space density of AGN in bins of X-ray luminosity and within twelve redshift slices between $z=0$ to $z=7$. Their analysis accounts for observational biases arising from moderate levels of the AGN line-of-sight obscuration. Quantitatively this corresponds to equivalent hydrogen column densities of up to $N_H \approx \rm 10^{24} \, cm^{-2}$, i.e. what is often referred to as the Compton thin limit. For higher levels of obscuration, $N_H > \rm 10^{24} \, cm^{-2}$ (Compton thick), the detectability of AGN in X-ray observations is severely biased and hence, the whereabouts of such sources is still debated. In this work we follow \citet{Aird2015} and assume that the fraction of AGN in the interval  $N_H = \rm 10^{24} - 10^{26}\, cm^{-2}$ is 34\% of the space density of moderately obscured sources ($N_H=\rm 10^{22}-10^{24} \, cm^{-2}$). This is within the range proposed by recent studies  \citep[e.g.][]{Ajello2012, Buchner2015, Akylas2016}. The redshift and luminosity dependence of this fraction is still not well constrained by observations. In our analysis we therefore choose to ignore such dependences and simply add a flat fraction to the point-estimates of the AGN space densities presented by  \citet{Aird2015}. 

\begin{figure*}
\begin{center}
\includegraphics[width=2\columnwidth]{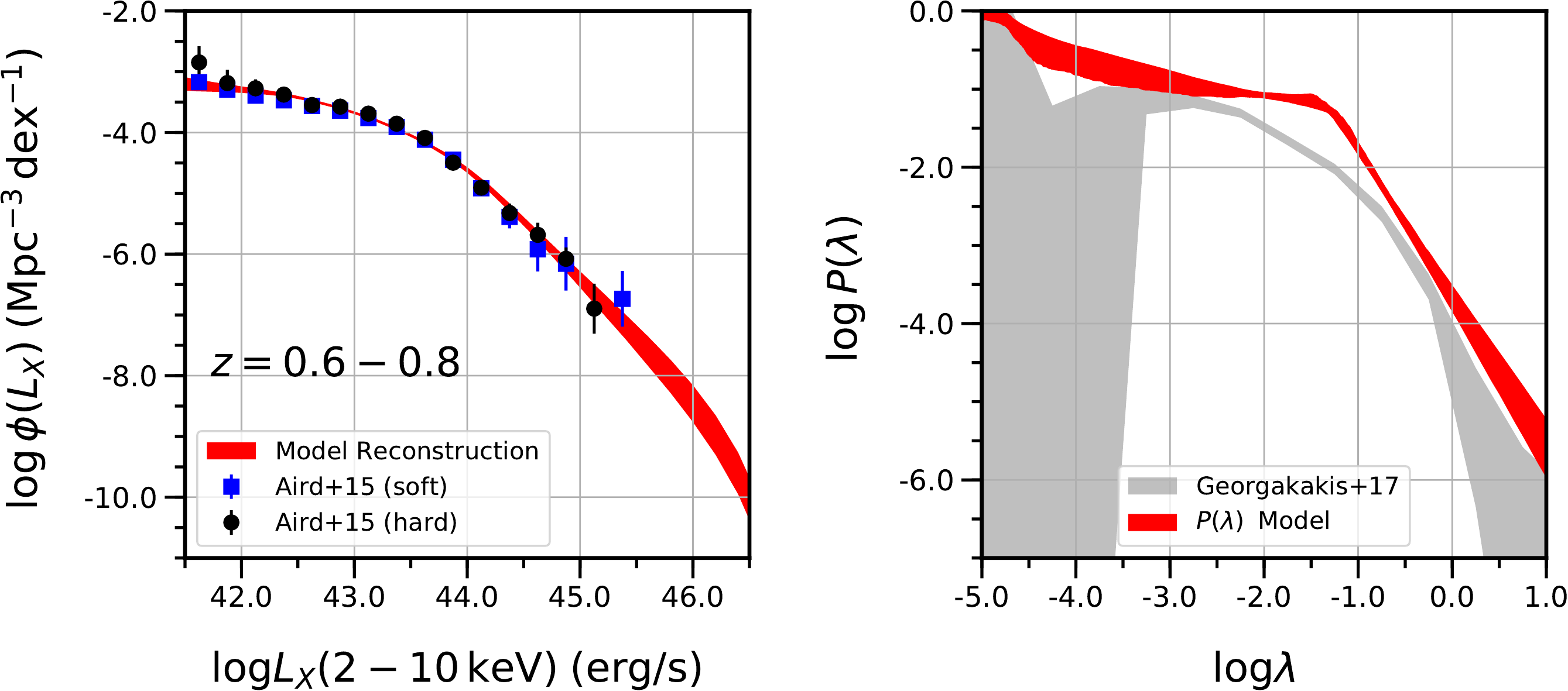}
\end{center}
  \caption{ The left panel shows the obscuration-corrected X-ray luminosity function estimated by \protect\cite{Aird2015} from the hard-band (2-10\,keV; black circles) and soft-band (0.5-2\,keV; blue squares) X-ray selected samples and for the redshift interval $z=0.6-0.8$. The datapoints include corrections for  Compton thick AGN, which are assumed to represent 34\% of the obscured population independent of redshift and X-ray luminosity. These observations are compared with the reconstructed X-ray luminosity function (red shaded region) estimated via Equation \ref{eq:lambda2phi} by convolving the mass function of galaxies with the specific accretion-rate distribution model of Equation \ref{eq:pl}. The width of red-shaded region corresponds to the confidence interval relative to the median that includes 68\% of the reconstructed space density estimates at fixed X-ray luminosity. The corresponding $P(\lambda)$ model is plotted as a function of specific accretion rate on the right panel (red shaded region). The width corresponds to the $1\sigma$ uncertainty. This is compared with the non-parametric estimate of the $P(\lambda)$ of \protect\cite{Georgakakis2017_plz} corresponding to a mean redshift of z=0.75 (grey shaded region). The difference between the two independently estimated specific accretion-rate distributions is that the red shaded region includes corrections for obscured AGN, including Compton thick ones. The \protect\cite{Georgakakis2017_plz} analysis does not account for the impact of obscuration, hence the overall lower normalization.}\label{fig:plz}
\end{figure*}

The specific accretion rate distribution at a given redshift interval is parametrised by a 3-segment broken power-law 

\begin{equation}\label{eq:pl}  
  P(\lambda, z) =  \left\{
  \begin{array}{ll}  K_1 \, \lambda^{\gamma_1}, & \lambda \ge \lambda_1,  \\ 
   & \\ 
K_2 \,  \lambda^{\gamma_2}, & \lambda_2 \leq \lambda <\lambda_1,\\
   & \\ 
K_3 \,  \lambda^{\gamma_3}, & \lambda <\lambda_2,\\
\end{array} \right.
\end{equation}

\noindent where the normalization parameters $K_2$, $K_3$ depend on $K_1$ through Equation \ref{eq:plznorm}. There are therefore six independent parameters, three power-law indices ($\gamma_1$, $\gamma_2$, $\gamma_3$), two breaking points ($\lambda_1$, $\lambda_2$) and one normalization  ($K_1$), that need to be estimated via Equation \ref{eq:lambda2phi}. The $P(\lambda, z)$ is defined in the specific accretion-rate interval ($\lambda_{min}$, $\lambda_{max}$) = ($10^{-5}$, $10$). The parametrisation in Equation \ref{eq:pl}  does not include an explicit redshift dependence. Instead  separate fits are obtained for each of the twelve redshift slices defined by \citet{Aird2015} with redshift edges (0.01, 0.2, 0.4, 0.6, 0.8, 1.0, 1.2, 1.5, 2.0, 2.5, 3.5, 5.0, 7.0). The \citet{Aird2017} luminosity function point-estimates in these intervals are corrected upward for the assumed fraction of Compton  thick AGN (flat 34\%) and then compared with the model predictions (Equation \ref{eq:plznorm}). The Monte Carlo Markov Chain code {\sc emcee} \citep{emcee} is used to infer the free parameters of Equation \ref{eq:pl} by sampling the likelihood function 

\begin{equation}\label{eq:likelhood}
\mathcal{L} = \sum_i \frac{(\phi_i(L_X,z)-\phi_{model}(L_X,z))^2}{\sigma_i^2},
\end{equation}

\noindent where $\phi_i(L_X,z)$ are the luminosity function point-estimates at given X-ray luminosity and redshift intervals, $\sigma_i$ are the corresponding uncertainties, $\phi_{model}(L_X,z)$ are the model predictions. The summation is over all point-estimates at a given redshift bin. Figure \ref{fig:plz} plots the results of the fit for the redshift bin $z=0.6-0.8$. It shows the X-ray luminosity function reconstructed via Equation \ref{eq:lambda2phi} in comparison with the observations of \citet{Aird2015}. Also shown is the corresponding specific accretion-rate distribution in comparison with previous studies \citep{Georgakakis2017_plz} that do not correct for obscuration biases.   

\section{Multiwavelength counterparts of the COSMOS-Legacy X-ray survey}\label{appendix:cosmos}

X-ray sources in the COSMOS Chandra Legacy survey catalogue presented by \cite{Georgakakis2015}  are identified with multiwavelength counterparts in the COSMOS2015 photometric catalogue \citep{Laigle2016}  using the Maximum Likelihood method \citep{Sutherland_and_Saunders1992}. Potential associations are searched for within a radius of 4\,arcsec. Magnitude priors are built in the $r$, $ip$, $Ks$ \citep[UltraVISTA-DR2][]{McCracken2012}, IRAC\,$3.6\mu m$  and IRAC\,$4.8\mu m$ (SPLASH Spitzer legacy program) photometric bands. First the magnitude distribution of all likely counterparts of X-ray sources within the search radius of 4\,arcsec is constructed. From this we subtract the magnitude distribution of sources within the same aperture at random positions within the Chandra Legacy survey footprint. The resulting distribution in 0.5\,magnitude bins in each band is smoothed using an 1-D kernel with weights (0.25, 0.5, 0.25) for magnitudes ($m-0.5$, $m$, $m+0.5$). This smoothed histogram is used as prior of the Maximum Likelihood method. The likelihood ratios (LR) and reliabilities of each counterpart are estimated for all 5 photometric bands adopted in the analysis. From the 5 pairs of likelihood ratio and reliability of a given association the one with the maximum LR is kept and is assigned to the counterpart in question. The near- and mid-infrared photometric bands are typically the ones that maximise the LR of a given association because of the lower density of sources at these longer wavelengths. 

There are 3627 sources in the \cite{Georgakakis2015} COSMOS Chandra Legacy survey catalogue. Of these 1278 are either outside the field of view of the COSMOS2015 photometric catalogue (587) or fall within bright-star masked regions (691). For the remaining sources a likelihood ratio cut of  $LR>1$ is adopted, which yields an identification rate of about 92\% with an expected spurious fraction of less 3\%. 

\section{X-ray WISE associations in the Stripe82X survey}\label{app:stripe82x}

\begin{figure}
\begin{center}
\includegraphics[width=0.9\columnwidth]{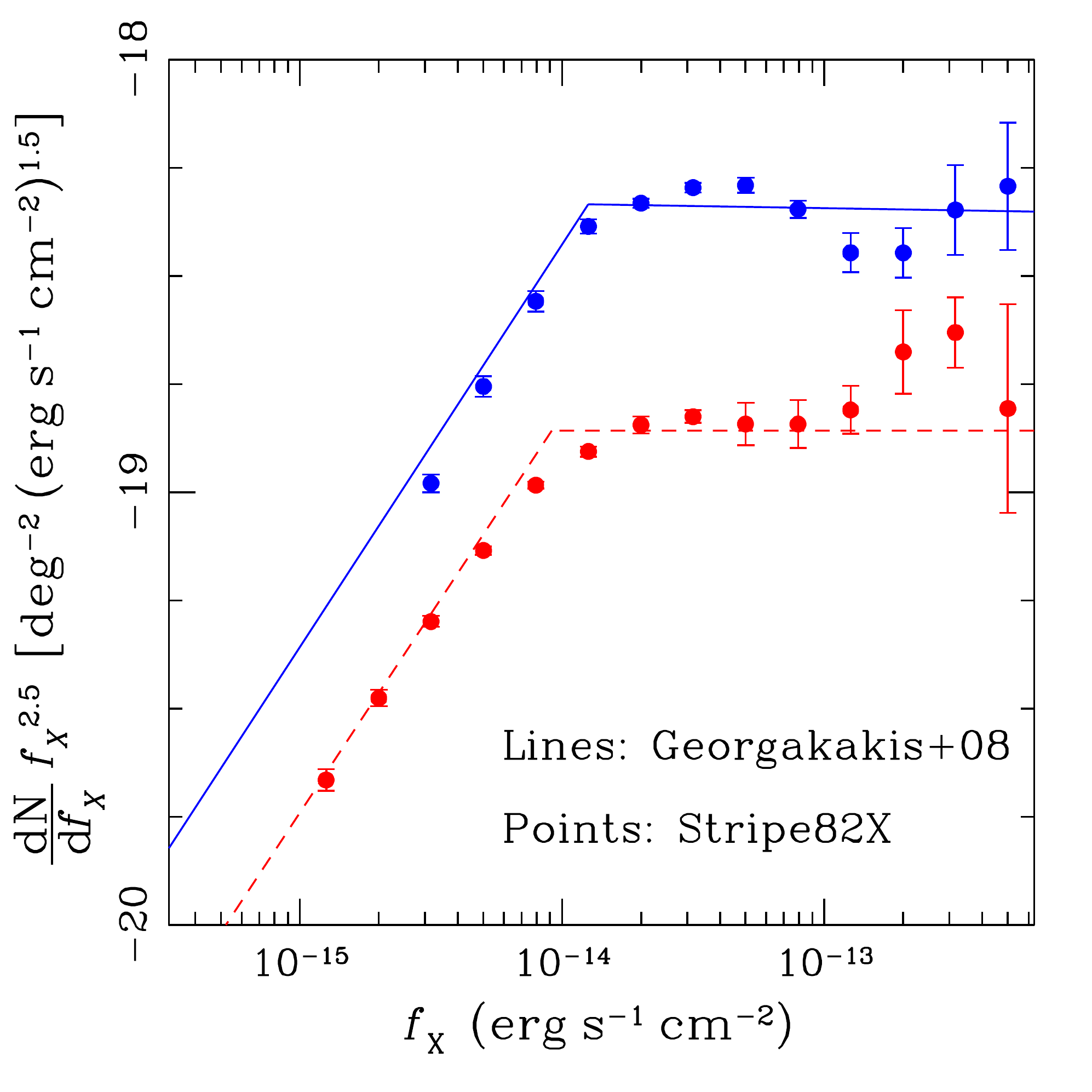}
\end{center}
  \caption{ Differential X-ray numbers counts normalized to the Euclidean slope in the soft (0.5-2\,keV; red dots, red dotted line) and hard (2-10\,keV, blue dots, blue solid line) spectral bands. The data points show the reconstructed number counts using the Stripe82X source catalogue and corresponding sensitivity curves described in the text. The lines are the best-fit double power-law $\log N - \log S$ relation estimated by \protect\cite{Georgakakis2008_sense} using a combination of deep and shallow X-ray survey fields.
  }\label{fig:stripe82x_dnds}
\end{figure}

This work uses a custom X-ray catalogue of the cycle-13 (AO13) XMM-{\it Newton} survey of the Stripe82X area \citep{LaMassa2016}. This is produced using the methods and reduction pipeline presented by \cite{Georgakakis_Nandra2011}. The XMM-{\it Newton} observation identification numbers (OBSIDS) of the analysed data are 0742830101, 0747400101, 0747420101, 0747440101, 0747390101, 0747410101 and 0747430101. These observations were taken in the "Mosaic" mode of the XMM-{\it Newton} and cover a total area of about $\rm 15.9\,deg^2$. The final catalogue consists of 3558 unique X-ray sources detected in at least one of the 0.5-2 (soft), 2-8 (hard) or 0.5-8\,keV (full) spectral bands to the Poisson false detection threshold of $<4 \times 10^{-6}$. The total number of detected sources is larger than the XMM-{\it Newton} A013 catalogue published by the Stripe82X collaboration \citep{LaMassa2016}, which numbers 2862 unique detections. This is because of the more conservative threshold adopted by \citet{LaMassa2016}, which roughly corresponds to a Poisson false detection probability of $<3 \times 10^{-7}$. A detailed comparison between the two source catalogues is beyond the scope of this work. The differential X-ray counts in the soft and hard bands are shown in Figure \ref{fig:stripe82x_dnds} in comparison with literature results.

The Stripe82X AO13 X-ray sources are matched to the ALLWISE data release catalog, which combines data from the {\it WISE} cryogenic and NEOWISE \citep{Mainzer2011} as well as the post-cryogenic survey phases. We avoid spurious ALLWISE detections and regions with poor mid-infrared photometry by requiring that the contamination and confusion flags ({\sc cc\_flags}) of the ALLWISE catalog are zero. The X-ray/ALLWISE cross-matching uses the Maximum Likelihood method \citep{Sutherland_and_Saunders1992} as implemented in \citep{LaMassa2019}. The maximum radius within which potential counterparts are search for is set to 6\,arcsec. This is because the positional accuracy of the  XMM-{\it Newton} is estimated to be about 1.5\,arcsec ($1\sigma$ rms). A likelihood ratio cut of  $LR>0.4$ is adopted for the ALLWISE counterparts of X-ray sources, which yields an identification rate of 65\% with an expected spurious fraction of about 5\%.

\bsp	
\label{lastpage}
\end{document}